\documentclass[prd,twocolumn,aps,groupedaddress,superscriptaddress]{revtex4-2}
\usepackage{amsfonts,amsmath,amssymb} \usepackage{bm}
\usepackage{mathrsfs}
\usepackage{epsfig} \usepackage{latexsym}
\usepackage[dvipsnames]{xcolor}
\usepackage{graphicx}

	\newcommand{\beq}{\begin{eqnarray}}
	\newcommand{\eeq}{\end{eqnarray}}
	\newcommand{\nn}{\nonumber}

	\renewcommand{\v}{{\bf v}}

\newcommand{\bem}{\begin{pmatrix}}
\newcommand{\eem}{\end{pmatrix}}

\newcommand{\A}{{\bar{A}}}
\newcommand{\B}{{\bar{B}}}
\newcommand{\Sz}{{S^z=0}}

\begin{document}
\title{Mesoscopic fluctuations in entanglement dynamics}
\author{Lih-King Lim}		
\address{School of Physics, Zhejiang  University,  Hangzhou, Zhejiang  310027, China}
\author{Cunzhong Lou}		
\address{School of Physics, Zhejiang  University,  Hangzhou, Zhejiang  310027, China}
\author{Chushun Tian}	
\email{ct@mail.itp.ac.cn}	
\address{CAS Key Laboratory of Theoretical Physics and Institute of Theoretical Physics, Chinese Academy of Sciences,  Beijing 100190, China}
 \date{\today}

\begin{abstract}
Understanding fluctuation phenomena plays a dominant role in the development of many-body physics. The time evolution of entanglement is essential to a broad range of subjects in many-body physics, ranging from exotic quantum matter to quantum thermalization. Stemming from various dynamical processes of information, fluctuations in entanglement evolution differ conceptually from out-of-equilibrium fluctuations of traditional physical quantities. Their studies remain elusive. Here we uncover an emergent random structure in the evolution of the many-body wavefunction in two classes of integrable --- either interacting or noninteracting --- lattice models. It gives rise to out-of-equilibrium entanglement fluctuations which fall into the paradigm of mesoscopic fluctuations of wave interference origin. Specifically, the entanglement entropy variance obeys a universal scaling law, in each class, and the full distribution displays a sub-Gaussian upper and a sub-Gamma lower tail. These statistics are independent of both the system's microscopic details and the choice of entanglement probes, and broaden the class of mesoscopic universalities. They have practical implications for controlling entanglement in mesoscopic devices.
\end{abstract}
\maketitle

When an isolated many-body system evolves, entanglement tends to spread. Owing to the diversity of the fate of the wavefunction evolution (e.g., localized or delocalized, thermalized or not thermalized), a wealth of entanglement patterns develop \cite{Cardy05,Prosen08,Moore12,Abanin13,Greiner19,Kaufman16,Huse13}. These patterns are the building blocks of the physics of recently discovered exotic phases of matter \cite{Abanin13,Greiner19,Fisher23}, and are central to the foundations of statistical mechanics \cite{Kaufman16,Greiner19}.  Understanding the long-time evolution of entanglement, and especially its universal aspects, is indispensable in the study of pattern formation.

To address this issue, one often investigates mesoscopic rather than macroscopic systems. Recent advancements in quantum simulation platforms, ranging from cold atoms, trapped ions to superconducting qubits, have made possible the measurement of information-theoretic observables and the experimental study of entanglement evolution \cite{Kaufman16,Greiner19,Brydges19}. In these investigations, quantum coherence is maintained across the entire sample, as required also for mesoscopic electronic and photonic devices \cite{Sheng05,Akkermans07}. At the same time, the relationship between the evolution of entanglement and quantum thermalization in isolated systems is currently under investigations \cite{Kaufman16,Greiner19}. Since various scenarios for the latter \cite{Popescu06,Lebowitz06,Deutsch91,Srednicki94,Rigol08,Izrailev16} are built upon a basis of wavefunctions with finite spatial extent, emphasis has naturally been placed on the dynamics of entanglement on the mesoscopic scale.

\begin{figure*}
\includegraphics[width=1.8\columnwidth]{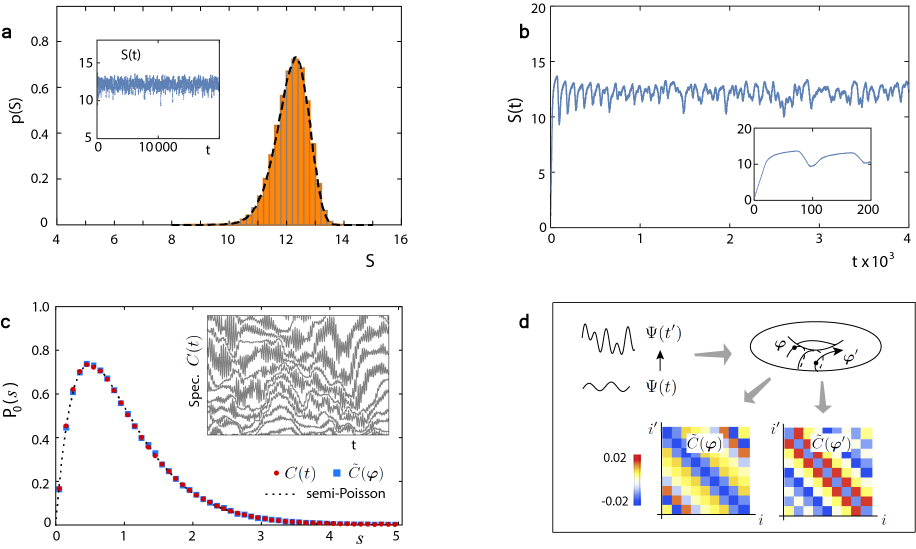}
\caption{{\bf Emergence of mesoscopic fluctuations in entanglement evolution.} {\bf a.} We simulate entanglement entropy evolution $S(t)$ of Rice-Mele model up to $t=10^4$ in unit of $\hbar/J$ ($L_A=25,\,L=124$); see the inset. Its fluctuation statistics (histograms) is shown to be equivalent to the statistics of entanglement entropy fluctuations in an ensemble of virtual disordered samples (dashed line, for $5\times 10^5$ disorder realizations $\boldsymbol{\varphi}$). {\bf b.} These  long-time fluctuations differ from the profile of $S(t)$ at early time. Inset: At early time $S(t)$ exhibits growth followed by damped oscillations. {\bf c.} Simulations show that the nearest-neighbor spacing distribution characterizing spectral fluctuations of the correlation matrix $C(t)$ (inset) is indistinguishable from that for an ensemble of truly random matrices $\tilde{C}(\boldsymbol{\varphi})$, and is semi-Poissonian. {\bf d.} Physically, as system's wavefunction evolves, the dynamical phases $\boldsymbol{\varphi}=(\omega_1t,\ldots,\omega_Nt)$ sweeps out an ensemble of mesoscopic samples $\tilde{C}(\boldsymbol{\varphi})$. { The quench protocol is $(J,J',M): (1,0.5,0.5)\rightarrow (1, 1.5, 1.5)$.}}
\label{fig1}
\end{figure*}

A prominent feature of mesoscopic systems is the occurrence of unique fluctuation phenomena, when randomness due to quenched disorders \cite{Sheng05,Akkermans07} or chaos \cite{Casati06,Izrailev90} is present. Notably, the conductance -- a basic probe of mesoscopic transport -- fluctuations have a universal variance, independent of sample size and the strength of randomness \cite{Altshuler85,Lee85}. Mesoscopic fluctuations are of wave interference origin and conceptually different from thermodynamic fluctuations. They are related to various entanglement properties \cite{Levitov09,Huse19}. The universality of these fluctuations is at the heart of mesoscopic physics.

In fact, there is a rapid increase of interests in entanglement fluctuations. In particular, understanding out-of-equilibrium entanglement fluctuation properties is a key to the statistical physics of isolated systems \cite{Lebowitz10,Deutsch20}. So far studies have focused on the kinematic case \cite{Popescu06,Page93,Hayden06,Nadal10,Rigol22}, where fluctuations arise from random sampling of some pure state ensemble, initiated by Page \cite{Page93}. On the one hand, since kinematic theories cannot describe wave effects and dynamical properties of the Schr{\"o}dinger evolution \cite{Tian21}, out-of-equilibrium entanglement fluctuations are beyond the framework of those theories. On the other hand, there have been big efforts on out-of-equilibrium fluctuations in isolated quantum systems \cite{Sun06,Reimann08,Popescu09,Zanardi10,Zanardi11,Santos13,Haque15,Marquardt17}. But so far focus has been traditional physical quantities, and little has been known about information-theoretic observables such as the entanglement entropy and the R$\acute{\rm e}$nyi entropy \cite{Deutsch20,Nakagawa18}.

Here, we develop an analytical theory for long-time dynamics of entanglement in two classes of integrable lattice models. One class of models, including the Rice-Mele model and the transverse field Ising chain, can be mapped to noninteracting fermions; the other class includes interacting spin chains, with the spin-$1/2$ Heisenberg XXZ model as a representative. Our theory relies crucially on the uncovering of a random structure emergent from the dynamical phases in the wavefunction evolution. Treating various information-theoretic observables as unconventional mesoscopic probes, we show that their out-of-equilibrium fluctuations fall into the paradigm of universal mesoscopic fluctuations in disordered or chaotic systems. Our findings have immediate implications for controlling entanglement in quantum simulation platforms.\\

{\noindent{\large {\bf Results}}}

{\noindent{\bf  Description of main results}}

{\noindent{\small{\bf Emergent mesoscopic fluctuations.}}} We find that the many-body wavefunction evolution endows the correlation matrix (the reduced density matrix) a random structure for noninteracting (interacting) models, even though the system is neither chaotic nor disordered. Specifically, for noninteracting models the time dependence enters through $N\approx {L\over 2}$ dynamical phases $(\omega_1t,\ldots,\omega_Nt)\equiv\boldsymbol{\omega} t$, with $L$ being the number of unit cells, so that the instantaneous correlation matrix $C(t)$ is given by some $N$-variable (matrix-valued) function $\tilde{C}(\boldsymbol{\varphi})$ for ${\boldsymbol{\varphi}=\boldsymbol{\omega}t}$; due to the incommensurality of $\boldsymbol{\omega}$ an ensemble of random matrices $\tilde{C}(\boldsymbol{\varphi})$ then results. Each $\tilde{C}(\boldsymbol{\varphi})$ is determined by $\boldsymbol{\varphi}$, the virtual disorder realization uniformly distributed over a $N$-dimensional torus (Fig.~\ref{fig1}). It describes a virtual disordered sample, and determines entanglement properties of that sample in the same fashion as $C(t)$ determines system's instantaneous entanglement properties. For interacting models, $C(t)$ and $\tilde{C}(\boldsymbol{\varphi})$ are replaced by the instantaneous reduced density matrix $\rho_A(t)$ and its $N$-variable counterpart $\tilde{\rho}_A(\boldsymbol{\varphi})$, respectively, and $N$ grows exponentially with $L$. So, when system's wavefunction evolves, the trajectory $\boldsymbol{\varphi}=\boldsymbol{\omega}t$ sweeps out the entire disorder ensemble, trading the temporal fluctuations of various information-theoretic observables to mesoscopic {\it sample-to-sample} fluctuations \cite{Altshuler85,Lee85}. In particular, we find that these out-of-equilibrium entanglement fluctuations arise from wave interference, similar to mesoscopic fluctuations. Interestingly, this kind of trajectories play important roles in Chirikov's studies of the relations between mesoscopic physics and quantum chaos \cite{Chirikov97}.

However, there are important differences between ordinary quenched disorders and the randomness emergent from entanglement evolution. As shown below, the latter has a strength $\sim 1/\sqrt{L}$ for noninteracting and $\sim {\rm e}^{-L}$ for interacting models, and thus diminishes for $L\rightarrow\infty$. This situation renders canonical mesoscopic theories based on diagrammatical \cite{Sheng05,Akkermans07} and field-theoretical \cite{Kamenev11} methods inapplicable, since they require the disorder strength to be independent of the sample size. In addition, because $C(t)$ is a (block-)Toeplitz matrix and very little \cite{Boglomony20} is known about the spectral statistics of random Toeplitz matrices, mesoscopic theories based on random matrix methods \cite{Beenakker97} are inapplicable either. Here we develop a different approach based on the modern nonasymptotic probability theory \cite{Boucheron13}, that relies merely on the statistical independence of the components of $\boldsymbol{\varphi}$ and applies to any $L$. A related approach has recently been used to find novel universalities in mesoscopic transport \cite{Tian18}.\\

\begin{figure*}[t]
\includegraphics[width=2\columnwidth]{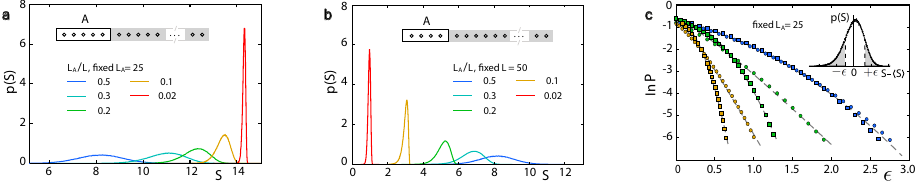}
\caption{{\bf Entanglement entropy distribution of Rice-Mele model.} We perform statistical analysis of the temporal fluctuations in the simulated entanglement entropy evolution. {\bf a.} Variation of the distribution with increasing $L$ at fixed $L_A$. {\bf b.} Same as a., but with increasing $L_A$ at fixed $L$. {\bf c.} The large deviation probability $\mathbf{P}(|S-\langle S\rangle|\geq \epsilon)$, with upper and lower tail respectively, is well fitted by Eq.~(\ref{meq:2}) (dashed lines), implying that the upper (squares) tail distribution is sub-Gaussian and the lower (circles) is sub-Gamma. The ratio $L_A/L$ is $0.1$ (yellow), $0.2$ (green) and $0.5$ (blue){, with the same quench protocol as Fig. 1}.}
\label{fig2}
\end{figure*}

{\noindent{\small{\bf Fluctuation statistics.}}} Uncovering the random structure, we show that fluctuations in entanglement evolution exhibit intriguing universal behaviors for each class of models, independent of microscopic details. First, when the variance ${\rm Var}(S)$ of the entanglement entropy $S$ as well as $L$ and $L_A$ (the subsystem size) are rescaled by appropriate microscopic quantities, the universal scaling law:
\begin{eqnarray}
{\rm Var} (S)=\left\{
\begin{array}{ll}
  1/L+L_A^3/L^2,& {\rm for\,\, noninteracting}\\
  L_A^\beta {\rm e}^{-L},& {\rm for\,\, interacting}
\end{array}
\right.
\label{meq:1}
\end{eqnarray}models follows, where $\beta$ is between $2$ and $4$, depending on both the initial state and system's parameters. Second, the distribution of $S$ has a universal shape, and is asymmetric with respect to its mean $\langle S\rangle$, displaying a sub-Gaussian upper and a sub-Gamma lower tail. In particular, for both classes, the probability for large deviation $\epsilon$ is
\begin{eqnarray}
\mathbf{P}(|S-\langle S\rangle|\geq\epsilon)=\left\{
\begin{array}{ll}
  {\rm e}^{-{\epsilon^2\over 2\mathfrak{b}_+}}, & {\rm for}\, S-\langle S\rangle>0 \\
  {\rm e}^{-{\epsilon^2\over 2(\mathfrak{b}_- + \mathfrak{c} \epsilon)}}, & {\rm for}\, S-\langle S\rangle<0
\end{array}
\right.,
\label{meq:2}
\end{eqnarray}
where $\mathfrak{b}_\pm\propto{\rm Var}(S)$ and $\mathfrak{c}>0$ depends on the ratio $L_A/L$. Third, Eqs.~(\ref{meq:1}) and (\ref{meq:2}) hold for other probes, e.g., the R$\acute{\rm e}$nyi entropy. These universal fluctuation behaviors are irrespective of the location of $\langle S\rangle$ in Page's curve \cite{Page93}. By Eq.~(\ref{meq:1}) at fixed $L_A$ the variance vanishes in the limit $L\rightarrow\infty$ (cf.~Fig.~\ref{fig2}a), implying the full suppression of temporal fluctuations beyond some critical time, in agreement with a benchmark result of entanglement evolution \cite{Cardy05} { and for the first time generalizing that result to interacting spin models analytically. How the entanglement entropy saturates in interacting systems is crucial to understanding experiments on entanglement dynamics \cite{Kaufman16,Greiner19}.} In contrast, at fixed $L$, as $L_A$ increases the variance { displays a power-law growth (cf.~Figs.~\ref{fig2}b and \ref{fig5}c), which is faster} than $\sim L_A$ displayed by typical extensive quantities. We shall see below this enhanced growth results from quantum interference.\\

{\noindent  {\bf Theory and numerical verifications}}

{ \noindent{\small{\bf Noninteracting models.}}} Having summarized the main results, we outline the derivations and present numerical verifications. A complete description is given in Supplementary Notes 1-8 for noninteracting models and Note 9 for interacting models. We start from the free fermion case, and focus on the Rice-Mele model with the Hamiltonian (see Supplementary Note 1)
\beq
H_{RM}&=&-\sum_{i=1}^{L} (J c_{i\bar{A}}^{\dag} \,c^{}_{i\bar{B}}+J'c_{i\bar{B}}^\dag\, c^{}_{(i+1)\bar{A}}+\rm{h.c.}) \nonumber\\
&&+M \sum_{i=1}^{L} (c_{i\bar{A}}^\dag\, c^{}_{i\bar{A}}-c_{i\bar{B}}^\dag \,c^{}_{i\bar{B}}).
\eeq
Here $J,J'$ are the hopping amplitudes, $M$ is the staggered onsite mass, $c^\dag_{i\sigma }\,, c^{}_{i\sigma }$ ($\sigma=\bar{A}\,,\bar{B}$) are, respectively, fermionic creation and annihilation operators at the $\sigma$-sublattice sites belonging to the $i$-th unit cell. The system has a total of $L$ unit cells, and is subjected to the periodic boundary condition. Generalizations to other models { mappable to free fermions are straightforward.} Let the system be at the half-filling ground state $\Psi(0)$. At $t=0$ we suddenly change parameters of the Hamiltonian. So the pre-quench state $\Psi(0)$ evolves unitarily under the new Hamiltonian to state $\Psi(t)$ at later time $t$. Because $\Psi(0)$ is a Gaussian state and the system is fermionic, the instantaneous entanglement entropy can be expressed as
\begin{equation}\label{meq:3}
  S(t) =\int d\lambda\,e(\lambda)\, \textrm{Tr}_A\,\delta (\lambda-C (t))
\end{equation}
using the method in \cite{Peschel03,Kitaev03,Korepin04,Peschel09}. Here $e(\lambda)=-\lambda\ln \lambda-(1-\lambda) \ln (1-\lambda)$ is the binary entropy function. $\textrm{Tr}_A\,\delta (...)$ gives the spectral density of the { correlation matrix $C(t)$ with element $C_{i\sigma,i'\sigma'}(t)=\langle \Psi(t)| c_{i\sigma}^\dag c_{i'\sigma'} |\Psi(t)\rangle$.} The trace is restricted to the subsystem A. When replacing $e(\lambda)$ by an appropriate function of $\lambda$, we obtain other entanglement probes such as the R$\acute{\rm e}$nyi entropy. This kind of expressions indicate that the evolving spectral density underlies out-of-equilibrium behaviors of different entanglement probes. They are analogous to the expressions for probes of mesoscopic transport. Indeed, if we replace $C(t)$ by the product of transmission matrix and its hermitian conjugate, we transform Eq.~(\ref{meq:3}) to the Landauer formula for conductance with $e(\lambda)$ changed to $\lambda$, and to formulaes for other transport probes with $e(\lambda)$ changed to appropriate functions of $\lambda$ \cite{Beenakker97}.

Because the eigenenergy spectrum displays a reflection and a particle-hole symmetry, when particle eigenenergies ${\omega_m\over 2}$ (Planck's constant set to unity) at Bloch momenta $k_m={2\pi(m-1)\over L},\, m=1,...,N=[{L\over 2}]+1$, are given, all other particle and all hole eigenenergies are known. Due to the translational invariance of the system, the time parameter enters the correlation matrix through the dynamical phases $\boldsymbol{\omega}t$ associated with $\boldsymbol{\omega}\equiv ({\omega_1},...,{\omega_N})$. Specifically, we can define a function $\tilde{C}(\boldsymbol{\varphi})=C_0+C_1(\boldsymbol{\varphi})$ on the $N$-dimensional torus. Leaving its detailed form for Supplementary Note 2, here we only expose its key properties. First, $C_{0,1}$ are block-Toeplitz matrices, with elements $(C_{0,1})_{ii'}$ being $2\times 2$ blocks defined in the sublattice sector and depending on the unit cell indexes $i,i'$ via $(i-i')$, i.e., $(C_{0,1})_{ii'}\equiv (C_{0,1})_{i-i'}$. Second, $C_0$ is $\boldsymbol{\varphi}$-independent, whereas $C_1$ is not and its elements take the form of
\begin{eqnarray}
({C}_1)_{l}\equiv \frac{1}{L}\sum_{m=1}^{N}
\big(R_{l}(k_m)\cos \varphi_m+I_{l}(k_m)\sin \varphi_m\big),
\label{meq:4}
\end{eqnarray}
where the elements of blocks, $R$'s, $I$'s, are complex and depend on $k_m$ (as well as post-quench Hamiltonian parameters). Then $C(t)$ is given by $\tilde{C}(\boldsymbol{\varphi})$ at ${\boldsymbol{\varphi}=\boldsymbol{\omega}t}$. Similarly, with the introduction of $S(\boldsymbol{\varphi})\equiv \int
d\lambda\,e(\lambda)\, \textrm{Tr}_A\,\delta (\lambda-\tilde{C} (\boldsymbol{\varphi}))$ in parallel to Eq.~(\ref{meq:3}) (for notational simplicity we use the same symbol $S$ despite differences in the arguments.), $S(t)$ is given by $S(\boldsymbol{\varphi})$ at $\boldsymbol{\varphi}=\boldsymbol{\omega}t$. This implies that, like $C(t)$, an evolving entanglement probe depends on $t$ through the dynamical phases $\boldsymbol{\omega}t$. Such dependence has an immediate consequence (see Supplementary Note 3 for details). That is, because in general the components of $\boldsymbol{\omega}$ are incommensurate, after initial growth \cite{Cardy05} and damped oscillations \cite{Calabrese20} due to the traversal of quasiparticle pairs or the incomplete revival of wavefunction (Fig.~\ref{fig1}b), an entanglement probe displays quasiperiodic oscillations (Fig.~\ref{fig1}a, inset), which are reproducible under the same initial conditions.

To understand fluctuation properties of quasiperiodic oscillations we note that the trajectory ${\boldsymbol{\varphi}=\boldsymbol{\omega}t}$ generates an ensemble of random matrices $\tilde{C}(\boldsymbol{\varphi})$, each of which is determined by the disorder realization, $\boldsymbol{\varphi}$, and thus is separated into two parts: nonrandom $C_0$ and random $C_1(\boldsymbol{\varphi})$. The probability measure of this ensemble is induced by the uniform distribution of $\boldsymbol{\varphi}$ via Eq.~(\ref{meq:4}). This ensemble has some prominent features (see Supplementary Note 4 for detailed discussions): First, since $\varphi_m$'s are statistically independent, Eq.~(\ref{meq:4}) implies that each element randomly fluctuates around its mean, with a magnitude $\sim 1/\sqrt{L}$. Thus for fixed ratio $L_A/L$ the randomness diminishes in the limit of large matrix size. Second, the elements of two distinct blocks are statistically independent. Third, the average elements decay rapidly with their distance to the main diagonal. These features lead to a semi-Poissonian nearest-neighbor spacing distribution \cite{Montambaux98,Boglomony00},
\begin{equation}\label{meq:13}
  P_0(s)=4s {\rm e}^{-2s},
\end{equation}
as shown in simulations (Fig.~\ref{fig1}c){; this kind of universal intermediate statistics were originally found for Anderson transitions \cite{Shklovskii93}}. Strikingly, despite that the Rice-Mele model is integrable and has no extrinsic randomness, the evolving correlation matrix can exhibit level repulsion: $P_0(s\rightarrow 0)\,{\sim}\,s$, which is a distinctive property of quantum chaos \cite{Casati06,Izrailev90}. We can demonstrate that the statistical equivalence of the ensemble of $\tilde{C}(\boldsymbol{\varphi})$ and the time series $C(t)$ (Fig.~\ref{fig1}c) hinges only on the incommensurabilty of $\boldsymbol{\omega}$ (see Supplementary Note 8 when this condition is not met). Furthermore, much like that a transmission matrix determines transport properties of a mesoscopic sample, a matrix $\tilde{C}(\boldsymbol{\varphi})$ determines $S(\boldsymbol{\varphi})$ and other entanglement probes of a virtual mesoscopic sample at the disorder realization $\boldsymbol{\varphi}$; consequently, the statistical equivalence between $C(t)$ and $\tilde{C}(\boldsymbol{\varphi})$ leads to the statistical equivalence between out-of-equilibrium and sample-to-sample fluctuations of various entanglement probes, in agreement with simulation results (Fig.~\ref{fig1}a).

Exploiting this equivalence, we proceed to study the statistics of entanglement entropy fluctuations (see Supplementary Notes 5 and 6 for full details). To overcome the difficulties discussed in the introduction with the unusual disorder structure, below we combine the continuity properties of the $N$-variable function ${S}(\boldsymbol{\varphi})$ with the nonasymptotic probabilistic method, so-called concentration inequality \cite{Boucheron13}. This allows us to work out a statistical theory for mesoscopic sample-to-sample fluctuations of $S(\boldsymbol{\varphi})$ at total system size $L$, which is {\it finite} so that the disorder strength does not vanish.

In order to study the distribution of $S(\boldsymbol{\varphi})$, we introduce the logarithmic moment-generating function $G(u)\equiv \ln \langle {\rm e}^{u (S-\langle S\rangle)}\rangle$, with $u$ being real and $\langle\cdot\rangle$ denoting the average over $\boldsymbol{\varphi}$. Consider the downward fluctuations (i.e., $S-\langle S\rangle<0$) first. Because the $N$ components of $\boldsymbol{\varphi}$ are statistically independent, we can apply the so-called modified logarithmic Sobolev inequality \cite{Boucheron13} to obtain
\begin{equation}\label{meq:9}
  \frac{d}{du}\frac{G}{u}\leq {1\over u^2}\frac{\left\langle\left[\sum_{m=1}^{N} {\rm e}^{u (S-\langle S\rangle)}    \,\phi(-u(S-S_m^-))\right]\right\rangle}{\left\langle {\rm e}^{u (S-\langle S\rangle)}\right\rangle}
\end{equation}
with $\phi(x)={\rm e}^x-x-1$ and $u\leq 0$. Here $S_m^-$ is the maximal values of $S(\boldsymbol{\varphi})$, when $\varphi_{m}$ varies and other arguments are fixed. Observing that the leading $u$-expansion of the right-hand side is ${b_-\over 2}$, with
\begin{equation}\label{meq:7}
  b_-\equiv \sum_{m=1}^{N}\left\langle(S-S_m^-)^2\right\rangle,
\end{equation}
we separate the right-hand side of the inequality into two terms, ${b_-\over 2}$ and the remainder. Then, we show that the latter is bounded by $c_-\frac{d G}{du}$ with $c_-$ being a negative constant. So we cast the inequality (\ref{meq:9}) to
\begin{equation}\label{meq:10}
  \frac{d}{du} \frac{(1+|c_-|u)G}{u}\leq  \frac{b_-}{2},
\end{equation}
which can be readily integrated to give
$G$$\leq$$\frac{b_-}{2}\frac{u^2}{1+|c_-|u}$. Such bound holds also for Gamma random variables. It generalizes the tail behaviors of the Gamma distribution, giving the so-called sub-Gamma tail \cite{Boucheron13}. Specifically, following standard procedures, we can use Markov's inequality to turn this bound for $G$ into a bound for the probability of downward fluctuations. The result is
\begin{equation}\label{meq:11}
  \mathbf{P}(S<\langle S\rangle -\epsilon)\leq {\rm e}^{-\epsilon^2/2(b_- + |c_-| \epsilon)}
\end{equation}
for any $\epsilon$$>$$0$. This gives a sub-Gamma lower tail, which crosses over from a Gaussian to an exponential form at $\epsilon\sim b_-/|c_-|$.
\begin{figure*}[t]
\includegraphics[width=1.9\columnwidth]{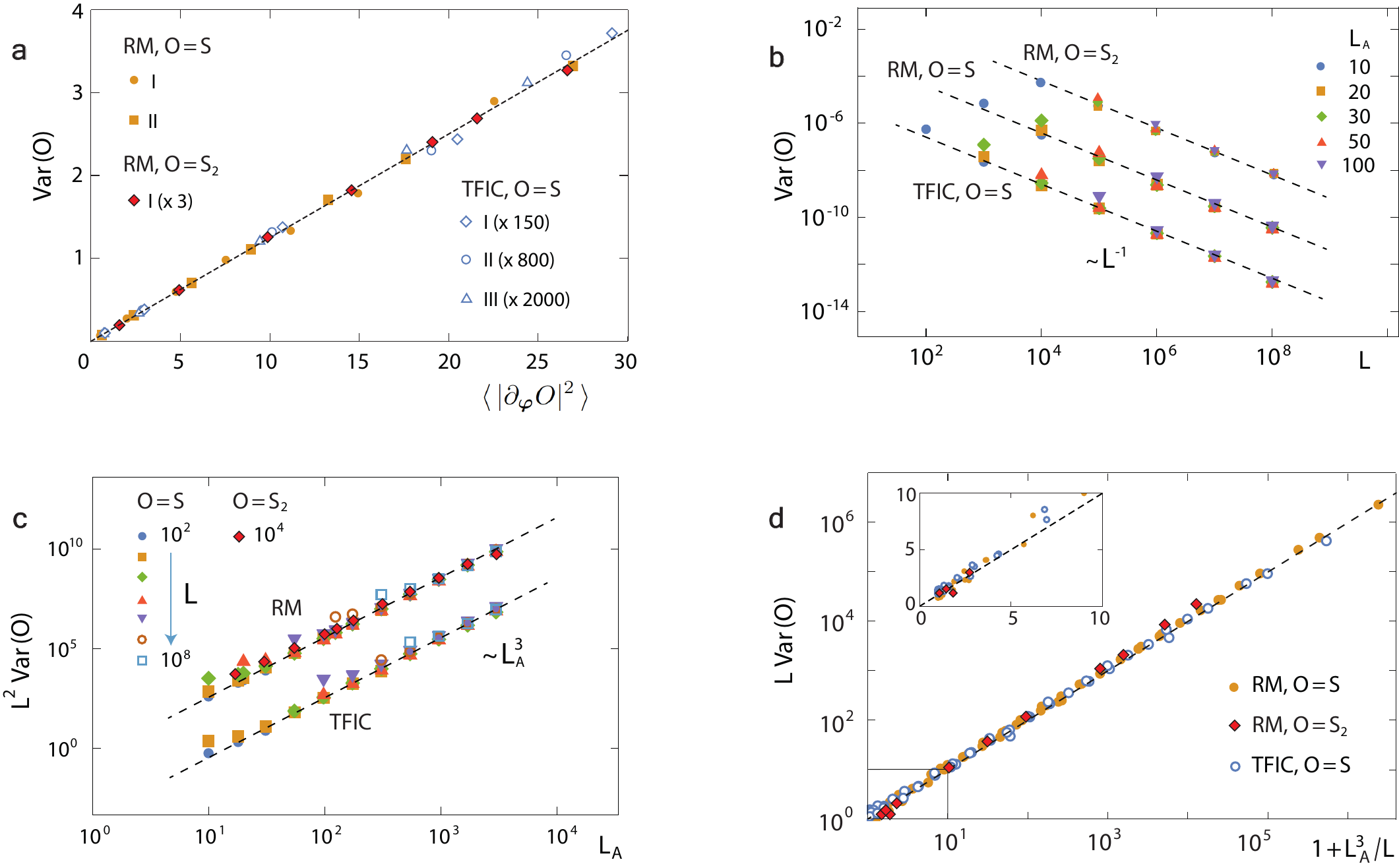}
\caption{{\bf Universal scaling behaviors of the variance in noninteracting  models.} We perform simulations for both Rice-Mele (RM) model and transverse field Ising chain (TFIC), whose Hamiltonian $H_{TFIC}=(-1/2)\sum_{i=1}^{L} (\sigma_i^x \sigma_{i+1}^x+h \sigma_i^z)$ with $\sigma_i^\alpha$ the Pauli matrices and $h$ the external magnetic field, for different sizes and different quench protocols to study the variance of two entanglement probes $O=S,\,S_2$. {\bf a.} For both $O$ the data confirm the relation (\ref{meq:6}). {\bf b-c.} They also confirm the limiting scaling behavior described by the first (second) term of the first line of Eq.~(\ref{meq:1}) for sufficiently small (large) $L^3_A/L$.  {\bf d.} For different models, after rescaling ${\rm Var}(O),\,L,\, L_A$ all data collapse to the universal curve described by Eq.~(\ref{meq:1}) for noninteracting  models. Inset: Zoom-in view of the regime near the origin. All theoretical predictions are presented by dashed lines. The quench protocols of $(J,\,J',\,M)$ for RM are $I$: $(1,0.7,0.3)\rightarrow (0.3,1.0,0.001)$ and $II$: $(0.6,1.0,0.8)\rightarrow (1.0,0.2,10)$, and for TFIC are $I$: $h=3\rightarrow 2$; $II$: $h=3\rightarrow 5$ and $III$: $h=5\rightarrow 2$.}
\label{fig3}
\end{figure*}

Similarly, we can study the upward fluctuations (i.e., $S-\langle S\rangle>0$). We replace $S_m^-$ in the inequality (\ref{meq:9}) by $S_m^+$, which is the minimal $S(\boldsymbol{\varphi})$ when $\varphi_{m}$ varies and other arguments are fixed, and consider $u>0$. Upon separating ${b_+\over 2}$, with $b_+\equiv\sum_{m=1}^{N}\langle(S-S_m^+)^2\rangle$, from the right-hand side of the inequality, the remainder is negative. As a result, $c_-$ is replaced by $0$ and $G\leq\frac{b_+u^2}{2}$, giving
\begin{equation}\label{meq:12}
  \mathbf{P}(S>\langle S\rangle+\epsilon)\leq {\rm e}^{-\epsilon^2/(2b_+)}
\end{equation}
for any $\epsilon>0$, which is a sub-Gaussian upper tail.

The inequalities (\ref{meq:11}) and (\ref{meq:12}) show that $S(\boldsymbol{\varphi})$ concentrates around $\langle S\rangle$ albeit with different bounds for upward and downward fluctuations. Simulations further show that the exact deviation probability for large downward (upward) fluctuations
agrees with the form given by the right-hand side of the corresponding concentration inequality, with $b_\pm$ and $c_-$ as fitting parameters (Fig.~\ref{fig2}c). Therefore, for large deviation, the upper (lower) tail distribution has the universal form given by the first (second) line in Eq.~(\ref{meq:2}), and the parameters $\mathfrak{b}_\pm$ and $\mathfrak{c}$ in Eq.~(\ref{meq:2}) are proportional to $b_\pm$ and $c_-$, respectively. So for large $\epsilon$ the upper tail is always Gaussian ${\rm e}^{-\epsilon^2/(2\mathfrak{b}_+)}$ while the lower is always exponential ${\rm e}^{-\epsilon/(2\mathfrak{c})}$, different from the distribution tails of thermodynamic fluctuations which are symmetric and Gaussian.

With Eq.~(\ref{meq:2}) we find that the variance ${\rm Var}(S)$ is given by $b_\pm$. To calculate the latter, note that by the mean value theorem there exists ${\bar \varphi}_m$ between $\varphi_m$ and $\varphi_m^\pm$ (at which $S^\pm_m$ is reached), so that $(S-S_m^\pm)^2$ is given by $(\varphi_m-\varphi_m^\pm)^2$$(\partial_{\bar{\varphi}_m}S)^2$. Then, for large $L$ the Fourier series of $\partial_{{\varphi}_m}S$ with respect to ${\varphi}_m$ are truncated at the second harmonics, giving $(\partial_{{\bar {\varphi}}_m}S)^2$$\sim$$\int\frac{d{\varphi}_m}{2\pi}(\partial_{{\varphi}_m}S)^2$. Applying these analyses to the definitions of $b_\pm$, we obtain
\begin{equation}\label{meq:6}
  {\rm Var} (S)\propto
\left\langle |\partial_{\boldsymbol{\varphi}} S |^2\right\rangle.
\end{equation}
This relation is confirmed numerically (Fig.~\ref{fig3}a), and the proportionality coefficient is found to be $\approx 1/8$. Equation (\ref{meq:6}) uncovers a relation between entanglement entropy fluctuations and continuity properties of the $N$-variable function $S(\boldsymbol{\varphi})$. It resembles the so-called concentration-of-measure phenomenon, a modern perspective of probability theory \cite{Talagrand96,Ledoux01}, where fluctuations of an observable are controlled by its {\it Lipschitz continuity}. This continuity is a key ingredient of universal wave-to-wave fluctuations in mesoscopic transport \cite{Tian18}.

By definition of $S(\boldsymbol{\varphi})$, we have $\partial_{\varphi_m} S=\textrm{Tr}_A (\ln (\tilde{C}^{-1}-\mathbb{I})\partial_{\varphi_m}C_1)$. Because of $C_1={\cal O}(1/\sqrt{L})$, we expand the logarithm in $C_1$ up to the first order. Taking into account that $C_0$ is short-ranged, we obtain
\begin{equation}\label{meq:8}
  \partial_{\varphi_m}S= -\textrm{Tr}_A\left[(H_{0}+(\partial_{C_0}H_{0})C_1)\partial_{\varphi_m}C_1\right],
\end{equation}
where $H_{0}=\ln(C_0^{-1}-\mathbb{I})$ is the entanglement Hamiltonian in the absence of disorder. Substituting Eq.~(\ref{meq:8}) into Eq.~(\ref{meq:6}), we find that the two terms in Eq.~(\ref{meq:8}) contribute to the variance separately. The contribution by the first term is $a/L$ and that by the second is $bL_A^3/L^2$, and the former (latter) is found to be a subsystem's edge (bulk) effect. Here the coefficient $a$ is proportional to the square of the size of subsystem's edge, and both $a$ and $b$ have no dependence on $L,\,L_A$. Upon rescaling: $L,\,L_A$ by $\sqrt{a/b}$ and ${\rm Var}(S)$ by $\sqrt{ab}$,  we obtain the scaling law (\ref{meq:1}) for noninteracting  models, which is confirmed by simulations (Fig.~\ref{fig3}b-d). By Eq.~(\ref{meq:1}), one enters the regime ${\rm Var}(S)=L^{-1}$ for $L_A\ll L^{1/3}$ (b) and the regime ${\rm Var}(S)=L_A^3/L^2$ for $L_A\gg L^{1/3}$ (c).

Let us consider other entanglement probes such as the second-order R$\acute{\rm e}$nyi entropy $S_2$. As said above, in this case we have an expression similar to Eq.~(\ref{meq:3}), with $e(\lambda)$ changed (see Supplementary Note 2). Repeating the analysis above, we find for $S_2$ the same relation as (\ref{meq:6}). Furthermore, we can calculate $\langle|\partial_{\boldsymbol{\varphi}}S_2|^2\rangle$ in the same way as $\langle|\partial_{\boldsymbol{\varphi}}S|^2\rangle$. As a result, we find that ${\rm Var}(S_2)$ obeys the same scaling law as Eq.~(\ref{meq:1}) for noninteracting  models. These statistics of $S_2$ are confirmed numerically (Fig.~\ref{fig3}). In Supplementary Notes 5-7 we further show that Eqs.~(\ref{meq:1}), (\ref{meq:2}) and (\ref{meq:6}) hold for more general probes.
\begin{figure}
\includegraphics[width=0.9\columnwidth]{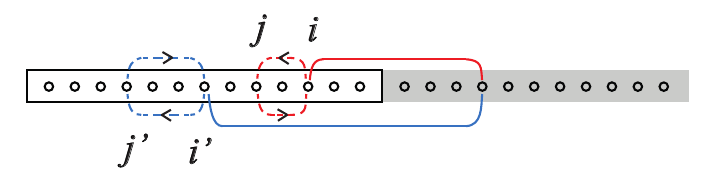}
\caption{{\bf Wave origin of entanglement fluctuations.} The pairing amplitude of the coherent entangled quasiparticle pair (solid lines) fluctuates with time. Constructive interference between two paths due to virtual hopping (blue and red dashed lines), that underlies such fluctuations, leads the variance of a generic entanglement probe to exhibiting the scaling behavior $\sim L_A^3/L^2$ for noninteracting models.}
\label{fig4}
\end{figure}

\begin{figure*}
\includegraphics[width=2\columnwidth]{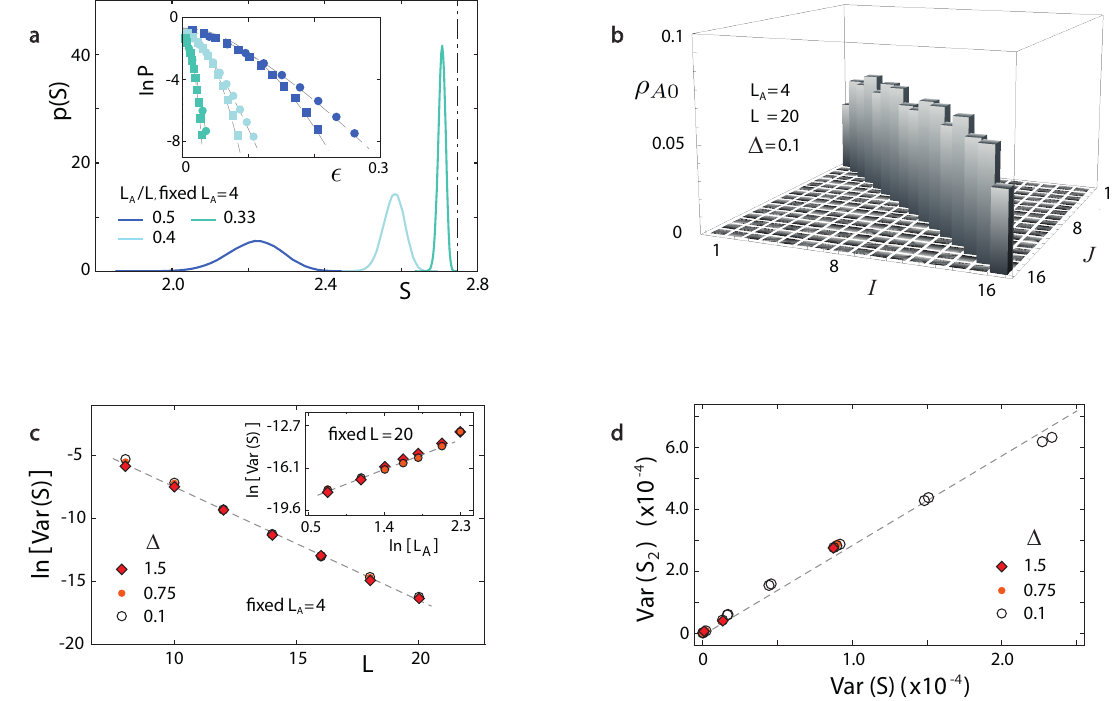}
{\caption{{\bf Entanglement fluctuations of XXZ model.} We perform long-time simulations for this model to study the mesoscopic fluctuations in entanglement evolution. Initially, the system is at an unpolarized random state. {\bf a.} Main panel: The statistical distribution of entanglement entropy for different ratios of $L_A/L$. Inset: The large deviation probability $\mathbf{P}(|S-\langle S\rangle|\geq \epsilon)$, with upper (squares) and lower (circles) tail respectively, is well fitted by Eq.~(\ref{meq:2}) (dashed lines). $\Delta=0.1$. {\bf b.}  Calculating the matrix elements $(\rho_{A0})_{IJ}$ numerically, it is confirmed that $\rho_{A0}$ is proportional to the unit matrix. {\bf c.} The simulated ${\rm Var}(S)$ (symbols) follows the scaling law described by the second line of Eq.~(\ref{meq:1}) (dashed lines). In the main panel (inset), $L_A$ ($L$) is fixed.  {\bf d.} For different $\Delta$ and small $L_A/L$, simulations show that ${\rm Var}(S)\propto{\rm Var}(S_2)$.}
\label{fig5}}
\end{figure*}

To understand physically the scaling behavior we use the concept of coherent entangled quasiparticle pair \cite{Cardy05}. Consider a quasiparticle inside the subsystem A. When pairing with another outside, it contributes to the bipartite entanglement. Due to the Heisenberg uncertainty, this particle's position fluctuates with time, leading to the temporal fluctuation $\Phi(t)$ of the pairing amplitude. In the simplest case, the particle hops virtually from a site $i$ to $j$ (in A as well) and back to $i$. Since the entangled pair is a correlation effect, $\Phi(t)$$\sim$$\sum_{ij}(C_1(\boldsymbol{\omega}t))_{ij}$$(C_1(\boldsymbol{\omega}t))_{ji}$ and thus by Eq.~(\ref{meq:4}) $\Phi(t)$$\sim$${1\over L^2}$$\sum_{ij}$$\sum_{mn}$${\rm e}^{i(k_m-k_n)(i-j)}$${\rm e}^{i(\omega_m+\omega_n)t/2}$, where $k_m,{\omega_m\over 2}$ are respectively the Bloch momentum and the particle eigenenergy associated with the hopping $i$$\rightarrow$$j$, and $k_n,{\omega_n\over 2}$ with $j$$\rightarrow$$i$. The variance of a generic entanglement probe is given by
\begin{eqnarray}\label{meq:14}
  \int dt |\Phi(t)|^2\sim {1\over L^4}\sum_{iji'j'}\sum_{mnm'n'} \delta_{\omega_m+\omega_n,\omega_{m'}+\omega_{n'}}\nonumber\\
  \times {\rm e}^{i((k_m-k_n)(i-j)-(k_{m'}-k_{n'})(i'-j'))},\qquad
\end{eqnarray}
where $(k_m-k_n)(i-j)$ and $(k_{m'}-k_{n'})(i'-j')$ are the phases of the paths: $i$$\rightarrow$$j$$\rightarrow$$i$ and $i'$$\rightarrow$$j'$$\rightarrow$$i'$, respectively. Because $\omega$'s are incommensurate, we obtain $(m,n)$$=$$(m',n')$ or $(n',m')$. So the first sum is dominated by those terms with two phases being identical. As a result, $\int dt |\Phi(t)|^2 \sim L_A^3/L^2$, with the numerator (denominator) given by the first (second) sum: This is the second term in the first line of Eq.~(\ref{meq:1}). We see that it arises from the constructive interference between the two hopping paths (Fig.~\ref{fig4}).\\

{ \noindent{\small{\bf Interacting models.}} For interacting models, the correlation function method and Eq.~(\ref{meq:3}) do not hold in general. What happens then? In this case, we have to retreat back to the more general expression of various information-theoretic observables in terms of the instantaneous reduced density matrix $\rho_A(t)$. Below we generalize the theory above to the spin-1/2 anisotropic Heisenberg XXZ model \cite{Kolezhuk04} defined by the Hamiltonian
\beq
H_{XXZ}=J\sum_{i=1}^L (S_i^x S_{i+1}^x+ S_i^y S_{i+1}^y+\Delta S_i^z S_{i+1}^z),
\eeq
where $S_{i}^\alpha$ are spin-1/2 operators, $\Delta$ is the anisotropy parameter, and the periodic boundary condition is imposed; generalizations to other interacting models are possible, which we will not discuss further. For simplicity here we consider the initial state $\Psi(0)$ to be an unpolarized random state; see Supplementary Note 9 for their detailed description. Other $\Psi(0)$ will be studied in that note.

Let $\Psi(0)$$=$$\sum_{m}$$\chi_m$$\Psi_m$, where $\Psi_m$'s are eigenstates and $\chi_m$'s are superposition coefficients. The number of excited eigenbases, $D$, may be estimated as the participation ratio $(\sum_{m} |\chi_m|^4)^{-1}$, which grows exponentially in $L$ \cite{Santos13}. As the wavefunction evolves, $\rho_A(t)$$=$$\rho_{A0}$$+$$\rho_{A1}(t)$, where $\rho_{A0}$$=$$\sum_{m} |\chi_m|^2$$\text{Tr}_{B}$$\vert\Psi_m\rangle \langle\Psi_m \vert$ and $\rho_{A1}(t)$$=$$\sum_{m \neq n}$${\rm e}^{-i(\omega_m-\omega_n)t}$$\chi_m \chi_n^{*}$$\text{Tr}_{B}$$|\Psi_m\rangle \langle\Psi_n|$ with $\omega_m$'s being eigenenergies, and B is the complement of A. Importantly, all eigenenergy mismatches: $\omega_m$$-$$\omega_n$ here are completely determined by $N$ ($=$$D$$-$$1$) mismatches: $\omega_m$$-$$\omega_1$$\equiv$$\omega_{m1}$ ($m$$=$$2$,...,$D$), because of $\omega_m$$-$$\omega_n$$=$$\omega_{m1}$$-$$\omega_{n1}$. So, similar to the free fermion case, the time parameter enters through the $N$ phases $\boldsymbol{\omega} t$, with $\boldsymbol{\omega}$$\equiv$($\omega_{21}$,...,$\omega_{D1}$). Introducing a function $\tilde\rho_{A}(\boldsymbol{\varphi})$$=$$\rho_{A0}$$+$$\tilde\rho_{A1}(\boldsymbol{\varphi})$ on the $N$-dimensional torus,
\begin{equation}
\label{meq:15}
\tilde\rho_{A1}(\boldsymbol{\varphi}) = \sum_{m\neq n} {\rm e}^{-i(\varphi_m - \varphi_n)} \chi_m \chi_n^*\,\text{Tr}_{B} \vert\Psi_m\rangle \langle\Psi_n\vert,
\end{equation}
and associating with $\tilde\rho_{A}(\boldsymbol{\varphi})$ various entanglement probes, e.g. $S(\boldsymbol{\varphi})$$=$$-$$\text{Tr}_{A}$$(\tilde\rho_{A}\ln \tilde\rho_{A})$ and $S_2(\boldsymbol{\varphi})$$=$$ -$$\ln$$\text{Tr}_A$$(\tilde\rho_{A}^2)$, we then obtain corresponding instantaneous probes from $S(\boldsymbol{\varphi})$, $S_2(\boldsymbol{\varphi})$, etc., by setting $\boldsymbol{\varphi}$$=$$\boldsymbol{\omega} t$.

The components of $\boldsymbol{\omega}$ are incommensurate in general. Thus various probes: $S(t)$, $S_2(t)$, etc. display quasiperiodic oscillations, which are seen in long-time simulations (by using the standard numerical method \cite{Sandvik10}). Most importantly, because the trajectory $\boldsymbol{\varphi}$$=$$\boldsymbol{\omega} t$ is ergodic on torus, the temporal fluctuations of an entanglement probe say $S(t)$ are statistically equivalent to the sample-to-sample fluctuations of $S(\boldsymbol{\varphi})$, with each sample having a disorder realization $\boldsymbol{\varphi}$ and represented by $\tilde\rho_A(\boldsymbol{\varphi})$. Moreover, the sum in Eq.~(\ref{meq:15}) is dominated by $\sim$$L_A^\mu$$D^{1+\nu}$ terms, where $1$$\leq$$\mu$$\leq$$2$, $0$$\leq$$\nu$$<$$1$, and the value of $\mu,\nu$ depends on $\Psi(0)$ and $\Delta$ (see Supplementary Note 9 for details). Taking $\chi_m\chi_n^*$$\sim$$1/D$ into account, we find that $\tilde\rho_{A1}(\boldsymbol{\varphi})$$\sim$$\sqrt{L_A^\mu/D^{1-\nu}}$. So the disorder strength decays exponentially with $L$.

Uncovering this random structure, we can also use the modified logarithmic Sobolov inequality (\ref{meq:9}) to address fluctuation statistics. Repeating previous analysis, we find that the distribution of $S$, $S_2$, etc. all follow Eq.~(\ref{meq:2}), as confirmed numerically (Fig.~\ref{fig5}a). And similar to noninteracting models, the variances of probes $O=$$S$, $S_2$, etc. satisfy Eq.~(\ref{meq:6}). Moreover, as shown in Supplementary Note 9 and confirmed numerically (Fig.~\ref{fig5}b), $\rho_{A0}$$\approx$$\mathbb{I}/D_A$ for $1$$\ll$$L_A$$\ll$$L$, with $D_A$ being the dimension of the subsystem Hilbert space: This belongs to the class of concentration-of-measure phenomena \cite{Talagrand96,Ledoux01,Popescu06,Tian18}, investigations of whose relations to quantum entanglement were initiated in Ref.~\cite{Popescu06}. Substituting this $\rho_{A0}$ into Eq.~(\ref{meq:6}) we find that
\begin{equation}
\label{meq:16}
\text{Var}(O) \propto D_A^2 \langle |\text{Tr}_A ( \tilde\rho_{A1} \partial_{\boldsymbol{\varphi}} \tilde\rho_{A1} )|^2 \rangle,
\end{equation}
where the proportionality coefficient is independent of $L, L_A$. Thus the scaling behaviors of mesoscopic fluctuations of entanglement dynamics must be independent of the choice of $O$, as confirmed numerically (d). Substituting Eq.~(\ref{meq:15}) into Eq.~(\ref{meq:16}), we find $\text{Var}(O)$$\sim$$L_A^\beta$${\rm e}^{-\kappa L}$ ($\kappa$$>$$0$, $\beta$$=$$2$$\mu$$\in$[$2$,$4$]). After rescaling $L$, $L_A$ and $\text{Var}(O)$, we obtain the second line of Eq.~(\ref{meq:1}). In simulations, ${\rm Var}(O)$ is indeed seen to decay exponentially with $L$ for fixed $L_A$ (c, main panel), and to display a power-law increase in $L_A$ for fixed $L$ (c, inset), with a power $2.4$.}\\

{ \noindent{\large{\bf Discussion}}} 
\\Our theory essentially hinges upon the relation between the wavefunction evolution and the trajectory $\boldsymbol{\varphi}=\boldsymbol{\omega}t$ on a high-dimensional torus, and the information-theoretic observable as a function on that torus. So it can be extended to more general contexts. First, it applies to other characteristics of entanglement especially the subsystem complexity, studies of whose temporal fluctuations have been initiated recently \cite{Tonni21}. Secondly, our ongoing studies have shown that there are no principal difficulties to generalize the present theory to nonintegrable interacting spin chains. Thirdly, experiments show that the entanglement dynamics of Bose-Hubbard model exhibits temporal fluctuations \cite{Kaufman16}; it is interesting to generalize the present results to that model and compare them with existing measurements. Fourth, our ongoing studies have also shown that the general theory developed for interacting models, while based on the full reduced density of matrix, should be capable of unifying results for noninteracting and interacting models. Indeed, for noniteracting models, provided the initial state is Gaussian we can use that theory to reproduce Eq.~(\ref{meq:1}); for non-Gaussian initial state, we can show that the emergent disorder carries the same structure as in the Gaussian case, which is a key leading to the scaling law (\ref{meq:1}) and suggests its robustness. Finally, because each virtual disordered sample corresponds to a pure state, our work suggests a simple way of producing a random pure-state ensemble to which great experimental efforts \cite{Choi22} are made. That is, we evolve an initial pure state by a single Hamiltonian and collect states at distinct sufficiently long time.\\
\\
\noindent { \large{\bf Data availability}}
\\
\noindent All data are displayed in the main text and Supplementary Information. Additional data are available from the corresponding author upon request.\\

\noindent  {\large{\bf Code availability}}
\\
\noindent The code that supports the plots in the main text and Supplementary Information is available from the
corresponding author upon request.

\noindent{{\large{\bf Acknowledgements}}}
\noindent We thank Italo Guarneri and Xin Wan for discussions, and Jean-Claude Garreau and Azriel Z. Genack for comments on the manuscript. This work is supported by National Natural Science Foundation of China Grants No. 11925507 (C.T.) and No. 12047503 (C.T.) and No. 11974308 (L.-K.L.).\\
\\
\noindent{{\large{\bf Author contributions}}}
\noindent L.-K.L. and C.T. designed the project and wrote the manuscript. The mathematical ideas were conceived by C.T. and developed into a full theory by C.T. and L.-K.L. together. Numerical analysis was conducted by L.-K.L. and C.L..\\
\\
\noindent{{\large{\bf Competing interests}}}
\noindent The authors declare no competing interests.

\widetext
\newpage

\section*{Supplemental Information: Mesoscopic fluctuations in entanglement dynamics}
\centerline{Lih-King Lim,$^1$ Cunzhong Lou,$^1$ and Chushun Tian$^{2,*}$}
\smallskip
\centerline{\textit{School of Physics, Zhejiang  University,  Hangzhou, Zhejiang  310027, China}}
\centerline{\textit{CAS Key Laboratory of Theoretical Physics and Institute of Theoretical Physics,}}
\centerline{\textit{Chinese Academy of Sciences,  Beijing 100190, China}}
\smallskip
\centerline{$^*$To whom correspondence should be addressed. E-mail: ct@mail.itp.ac.cn}
\setcounter{equation}{0}
\setcounter{figure}{0} 
\bigskip

\begin{center}
{\bf Supplementary Note 0. INTRODUCTION}
\end{center}

In this Supplementary Information (SI) we present a complete description of the theory and details of numerical simulations. We also show that the developed theory applies to generic information-theoretic observables, not limited to the entanglement entropy. For the convenience of readers this SI is written in the self-contained manner. Various definitions and notations are reintroduced, and the results reported in the main text are derived and presented here. Moreover, we discuss the physical implications of the results with substantial extension. It is organized in the following way:
\begin{itemize}
  \item In \ref{sec:pre}, as the preliminary, we introduce the Rice-Mele model and give the overall structure of its correlation matrix.
  \item In \ref{sec:setup} we develop for several lattice fermionic models a formalism for long-time dynamics of entanglement, and demonstrate its connection to classical dynamics on a high-dimensional torus.
  \item In \ref{sec:emergence_randomness} we use the formalism to establish a statistical principle for out-of-equilibrium fluctuations of various information-theoretic observables.
  \item In \ref{sec:emergence}, armed with that principle we show that, despite the models considered are all integrable and not disordered, a random structure emerges from the unitary pure-state evolution, and as a result mesoscopic sample-to-sample fluctuations emerge from entanglement evolution.
  \item In \ref{sec:statistics} we develop a general theory for the statistics of emergent mesoscopic fluctuations, that allows us to study the full distribution and the variance of a generic information-theoretic observable.
  \item In \ref{sec:applications_distribution_EE} we apply the general theory to study entanglement entropy fluctuations. In particular, we uncover a universal scaling law for the variance and derive the full distribution.
  \item In \ref{sec:numerical_tests} we describe the methods of numerical simulations in details. We also show numerically that the $n$th-order R$\acute{\rm e}$nyi entropy exhibits the same universal behaviors as the entanglement entropy.
  \item In \ref{sec:breakdown_ergodicity} we further show that the mesoscopic fluctuations emergent from entanglement evolution disappear for commensurate $\boldsymbol{\omega}$.
   \item In \ref{sec:generalization_to_interacting_models} we generalize the developed analytical theory of mesoscopic fluctuations in entanglement dynamics to an important interacting model, namely, the antiferromagnetic XXZ spin-${1\over 2}$ chain. We further perform numerical experiments to test analytical predictions.
  \item In \ref{sec:TFIC} we describe the dynamics of entanglement in the transverse field Ising chain.
  \item In Supplementary Appendices B-H
  we give some additional technical details.
\end{itemize}
Except Supplementary Notes $1$, $2$ and $8$, we consider incommensurate $\boldsymbol{\omega}$ throughout this SI.

\section{Preliminary: Correlation matrix of the Rice-Mele model}\label{sec:pre}
The Rice-Mele model describes the motion of many fermions on a one-dimensional discrete lattice with two sublattices: $\A,\B$. In the lattice space, its Hamiltonian is given by
\beq
H=-\sum_{i=1}^{L} (J c_{i\A}^{\dag} \,c^{}_{i\B}+J'c_{i\B}^\dag\, c^{}_{(i+1)\A}+h.c.) +M \sum_{i=1}^{L} (c_{i\A}^\dag\, c^{}_{i\A}-c_{i\B}^\dag \,c^{}_{i\B}).
\label{eq:1}
\eeq
Here $J,J'$ are the hopping amplitudes, $M$ is the staggered onsite mass, $c^\dag_{i\sigma }\,, c^{}_{i\sigma }$ ($\sigma=\A\,,\B$) are, respectively, fermionic creation and annihilation operators at the $\sigma$-sublattice sites belonging to the $i$-th unit cell with a total of $L$ unit cells. The periodic boundary condition is imposed. We set the lattice constant to be unity, and set the distance between adjacent $\A\,$- and $\B\,$-sites to be one half.

In this work, we consider generic sudden global quenches of Hamiltonian from $H_{0}$ to $H_{f}$, corresponding to the sudden change of the Hamiltonian parameters in Supplementary Eq.~(\ref{eq:1}):
\beq
(J,J',M)_0\rightarrow (J,J',M)_f
\label{eq:7}
\eeq
at time $t=0$, where the subscript $0$ ($f$) denotes the parameter set in the pre-quench (post-quench) Hamiltonian $H_0$ ($H_f$). Moreover, the initial state $\Psi(0)$ is taken to be the ground state of $H_0$ at half-filling, which corresponds to a fermion number of $L$, and the pure-state evolution results in $\Psi(t) = {\rm e}^{-i H_f t/\hbar} \Psi(0)$. Hereafter the Planck constant is set to unity.

To study the evolution of general entanglement probes, namely information-theoretic observables, after quench, we will see that all information is encoded in the time evolution of the correlation matrix $C(t)\equiv \{C_{i\sigma ,i'\sigma' }(t)\}$ restricted to the subsystem A with $L_A$ contiguous unit cells. Its matrix elements $C_{i\sigma ,i'\sigma' }(t)=\langle \Psi(t)| c^\dag_{i\sigma } c^{}_{i'\sigma' }|\Psi(t) \rangle$, with the indexes $i,i'=1,\ldots, L_A$ and $\sigma,\sigma' =\A,\B$, can be computed to give
\beq
C_{i\sigma ,i'\sigma' }(t)=\frac{1}{L}\sum_k {\rm e}^{i k \,(i-i')} \biggl(\gamma_{\sigma\sigma'} (k) + \alpha_{\sigma\sigma'}(k) \cos (2 E_{fk} t)+ \beta_{\sigma\sigma'}(k) \sin (2 E_{fk} t)\,\biggr),\label{eq:corr}
\eeq
where the coefficients $\alpha$'s, $\beta$'s and $\gamma$'s have no time dependence, but depend only on the Bloch momentum $k$, the sublattice indexes $\sigma,\sigma' $ as well as the parameters of Hamiltonians $H_0$ and $H_f$. $E_{fk}$ is the positive branch of the energy spectrum of $H_f$:
\beq
E_{fk}=\sqrt{M^2 +J^2+J'^2+ 2 J J' \cos k}.\label{eq:4}
\eeq
It shows that the correlation matrix is a block-Toeplitz matrix, sharing the same form as the correlation matrix for the transverse field Ising chain (TFIC) [1,2]. For TFIC, we refer to
\ref{sec:TFIC} for the description. To be specific, its matrix element $C_{i\sigma ,i'\sigma' }$ depends on the unit cell indexes $i,i'$ only through the displacement $l\equiv i-i'\,(=0,\pm 1,\ldots,\pm (L_A-1))$.

We generally have $\gamma_{\B\B}(k)=1-\gamma_{\A\A}(k)$, $\alpha_{\B\B}(k)=-\alpha_{\A\A}(k)$, $\beta_{\B\B}(k)=-\beta_{\A\A}(k)$, $\mathcal{X}_{\B\A}(k)=\mathcal{X}_{\A\B}^*(k)$ for $\mathcal{X}=\gamma,\alpha,\beta$.  So we write the coefficients in the block-matrix form in the sublattice sector as follows
\beq
\check{\gamma}(k)\equiv \left.\left(\begin{array}{cc}
                   \gamma_{\A\A}(k) & \gamma_{\A\B}(k)\\
                   \gamma_{\A\B}^*(k) & 1-\gamma_{\A\A}(k)
                 \end{array}\right)\right.,\,
\check{\alpha}(k)\equiv \left.\left(\begin{array}{cc}
                   \alpha_{\A\A}(k) & \alpha_{\A\B}(k)\\
                   \alpha_{\A\B}^*(k) & -\alpha_{\A\A}(k)
                 \end{array}\right)\right.,\,
\check{\beta}(k)\equiv \left.\left(\begin{array}{cc}
                   \beta_{\A\A}(k) & \beta_{\A\B}(k)\\
                   \beta_{\A\B}^*(k) & -\beta_{\A\A}(k)
                 \end{array}\right)\right.,
\eeq
with six independent complex coefficients. Their explicit forms are given in
\ref{app:rm}.

A closely related model is the Su-Schrieffer-Heeger (SSH) model, by setting $M\rightarrow 0$ in Supplementary Eq.~(\ref{eq:1}). All results pertaining to the Rice-Mele model hold also for the SSH model.

\section{Formalism for long-time dynamics of entanglement}
\label{sec:setup}

In this Supplementary Note we develop a formalism that yields a connection between dynamics of entanglement and classical dynamics on a high-dimensional torus. It is important that throughout this work we consider finite $L$, unlike canonical analytical studies of entanglement evolution [1,3] that address infinite $L$. Of course, the statistical theory developed here can be extended to the case of infinite $L$. We focus on the bipartite entanglement and the condition: $1\ll L_A \leq L/2$ is assumed, so that the subsystem is smaller than its complement but large enough.




\subsection{Relating evolving correlation matrix to classical trajectory on $\boldsymbol{\mathbb{T}^N}$}
\label{sec:time_dependence_correlation}
Due to a particle-hole symmetry and a reflection symmetry of the energy spectrum in Supplementary Eq. (\ref{eq:4}) (see also Supplementary Eqs. (\ref{eq:11a}) and (\ref{eq:12})), we see that the time parameter enters in the instantaneous correlation matrix $C(t)$ in Supplementary Eq. (\ref{eq:corr}) only through
\begin{equation}\label{eq:70}
  N=[L/2]+1
\end{equation}
($[x]$ denotes the integer part of $x>0$.) dynamical phases $(\omega_0t,\ldots,\omega_{N-1}t)\equiv\boldsymbol{\omega} t$.
The frequencies $\omega_m$ are
\beq
\omega_m={2E_{fk_m}}, \quad k_m={2\pi m\over L}, \quad m=0,\ldots,N-1,
\label{eq:15}
\eeq
associated with the energy gaps at the Bloch momenta $k_m$. We actually have for the correlation matrix
\beq
C_{i\sigma ,i'\sigma' }(t)&=&\frac{1}{L}\sum_{m=0}^{N-1} \biggl(1-{1\over 2}\,\delta_{m0}\biggr)\biggl(\left( {\rm e}^{ik_m (i-i')}  \gamma_{\sigma \sigma'}(k_m) +(m\rightarrow -m)\right)\nonumber\\
&&+ \left({\rm e}^{ik_m (i-i')} \alpha_{\sigma \sigma'}(k_m)+(m\rightarrow -m)\right)\cos (\omega_m t)+ \left({\rm e}^{ik_m (i-i')} \beta_{\sigma \sigma'}(k_m)+(m\rightarrow -m)\right) \sin (\omega_m t)\,\biggr)
\label{eq:27a}
\eeq
with the notation: $(m\rightarrow -m)$ denoting the term obtained from that in front of the $+$ sign by making the replacement: $m\rightarrow -m$. Mathematically, we can first introduce a $N$-variable (matrix-valued) function on the $N$-dimensional torus $\mathbb{T}^N$ defined as
\beq
\tilde{C}_{i\sigma ,i'\sigma' }(\boldsymbol{\varphi})&\equiv&\frac{1}{L}\sum_{m=0}^{N-1} \biggl(1-{1\over 2}\,\delta_{m0}\biggr)\biggl(\left( {\rm e}^{ik_m (i-i')}  \gamma_{\sigma \sigma'}(k_m) +(m\rightarrow -m)\right)\nonumber\\
&&+ \left({\rm e}^{ik_m (i-i')} \alpha_{\sigma \sigma'}(k_m)+(m\rightarrow -m)\right)\cos \varphi_m+ \left({\rm e}^{ik_m (i-i')} \beta_{\sigma \sigma'}(k_m)+(m\rightarrow -m)\right) \sin \varphi_m\,\biggr),
\label{eq:27}
\eeq
with $\boldsymbol{\varphi}\equiv(\varphi_0,\ldots,\varphi_{N-1})\in \mathbb{T}^N$. Then, we obtain the instantaneous correlation matrix $C(t)$ from $\tilde{C}(\boldsymbol{\varphi})$ via
\beq
C(t)= \tilde{C}(\boldsymbol{\omega}t).
\label{eq:30}
\eeq
We shall call $\tilde{C}(\boldsymbol{\varphi})$ the correlation matrix as well. Its physical meanings will become clear later.

Note that by definition $\tilde{C}(\boldsymbol{\varphi})$ is periodic in each argument $\varphi_m$,
\begin{equation}\label{eq:87}
  \tilde{C}(\varphi_0,\ldots,\varphi_m,\ldots,\varphi_{N-1})=\tilde{C}(\varphi_0,\ldots,\varphi_m+2\pi,\ldots,\varphi_{N-1}).
\end{equation}
Thus, these phases $\boldsymbol{\varphi}=\boldsymbol{\omega}t$ in $\tilde{C}(\boldsymbol{\varphi})$ completely determine $C(t)$ and entail a classical motion that is a rotation in the $N$-dimensional torus $\mathbb{T}^N$ with constant angular velocity $\boldsymbol{\omega}$ (cf.~Fig.~1(d) in the main text):
\beq
C(t) \leftrightsquigarrow \boldsymbol{\varphi}=\boldsymbol{\omega}t\in \mathbb{T}^N.
\label{eq:16a}
\eeq
In other words, given the initial state $\Psi(0)$, the correlation matrix corresponds one-to-one to a point $\boldsymbol{\omega}t$ along the classical trajectory. This relation plays important roles in investigating fluctuations in long-time entanglement evolution. We remark that this relation makes no reference to the number-theoretical properties of $\boldsymbol{\omega}$, e.g. commensurate or incommensurate, but, as we will demonstrate later, the behaviors of long-time entanglement evolution are very sensitive to these properties. We note that similar correspondence also holds generally for the evolution of quantum expectation value $\langle \Psi(t)|{\cal A}|\Psi(t)\rangle$ of \textit{one-body} operator ${\cal A}$ upon a global quench, see
\ref{app:expectation}.

\subsection{Time evolution of entanglement}
\label{sec:evolution_EE}

\subsubsection{Reduced density of matrix and entanglement entropy}
\label{sec:RDM_EE}

The time evolution of the correlation matrix completely determines the entanglement evolution. Indeed, the instantaneous reduced density matrix provides full information on entanglement evolution. It is given by
\beq
\rho_A(t)={{\rm e}^{-H_A(t)}/Z},\quad Z\equiv {\rm Tr}_A {\rm e}^{-H_A(t)}
\label{eq:62}
\eeq
with the trace restricted to the subsystem A, where $H_A(t)$ is the instantaneous entanglement Hamiltonian. Because the systems considered are noninteracting and fermionic and the evolving state $\Psi(t)$ is Gaussian, $H_A(t)$ takes a quadratic form [4]:
\beq
H_A(t)=\sum_{i,i'=1}^{L_A}\sum_{\sigma,\sigma' =\A,\B}({\cal H}_A(t))_{i\sigma ,i'\sigma' }c^\dagger_{i\sigma }c^{}_{i'\sigma' },
\label{eq:63}
\eeq
and the $2L_A\times 2L_A$ matrix ${\cal H}_A(t)$ is determined by the correlation matrix $C(t)$ via
\beq
{\cal H}_A(t)=\ln(C^{-1}(t)-\mathbb{I}).
\label{eq:64}
\eeq
So we see that the instantaneous reduced density of matrix is a {\it functional} of the instantaneous correlation matrix. Moreover, all $2 L_A$ instantaneous eigenvalues of $C(t)$, denoted as $p_\nu(t)$ ($\nu=1,\ldots, 2L_A$), belong to the interval $[0,1]$. Physically, $p_\nu(t)$ gives the occupation probability of the instantaneous eigenmode $\nu$ of the correlation matrix.

In most of this work, we study the entanglement entropy, which is the von Neumann entropy associated with the reduced density matrix and defined as
\beq
S(t)\equiv-{\rm Tr}_A \left(\rho_A(t)\ln\rho_A(t)\right).
\label{eq:31}
\eeq
Combining it with Supplementary Eqs.~(\ref{eq:62})-(\ref{eq:64}) we find that
\beq
S(t)=-\sum_{\nu=1}^{2L_A}\left(p_{\nu}(t) \ln p_{\nu}(t)+(1-p_{\nu}(t)) \ln (1-p_{\nu}(t))\right)
\label{eq:33}
\eeq
or equivalently,
\beq
S(t)=-{\rm Tr}_A\left(C(t) \ln C(t)+(\mathbb{I}-C(t)) \ln (\mathbb{I}-C(t))\right).
\label{eq:34}
\eeq
So, it is again a functional of $C(t)$, like $\rho_A(t)$. Furthermore, with the introduction of the binary entropy function
\beq
e(\lambda)\equiv-\lambda\ln \lambda-(1-\lambda) \ln (1-\lambda),
\label{eq:35}
\eeq
one can readily express Supplementary Eq.~(\ref{eq:33}) as
\beq
S(t)=\int
d\lambda\, e(\lambda)\, \textrm{Tr}_A\,\delta\left(\lambda-C(t)\right).
\label{eq:36}
\eeq

For the reduced density matrix, we also have a relation similar to Supplementary Eq.~(\ref{eq:30}). Indeed, due to Supplementary Eqs.~(\ref{eq:30}) and (\ref{eq:64}) we have
\beq
{\cal H}_A(t)=\tilde{{\cal H}}_A(\boldsymbol{\omega}t),\quad \tilde{{\cal H}}_A(\boldsymbol{\varphi})\equiv \ln(\tilde{C}^{-1}(\boldsymbol{\varphi})-\mathbb{I}),
\label{eq:158}
\eeq
where $\tilde{\cal H}_A(\boldsymbol{\varphi})$ may be regarded as the entanglement Hamiltonian associated with
$\tilde{C}(\boldsymbol{\varphi})$. Then, upon introducing the following reduced density of matrix associated with
$\tilde{C}(\boldsymbol{\varphi})$:
\beq
\tilde{\rho}_A(\boldsymbol{\varphi})\equiv{{\rm e}^{-\tilde{\cal H}_A(\boldsymbol{\varphi})}\over {\rm Tr}_A {\rm e}^{-\tilde{\cal H}_A(\boldsymbol{\varphi})}},
\label{eq:65}
\eeq
we obtain
\beq
\rho_A(t)=\tilde{\rho}_A(\boldsymbol{\omega}t).
\label{eq:66}
\eeq
By the same token, we can introduce the following entanglement entropy associated with
$\tilde{C}(\boldsymbol{\varphi})$:
\beq
\tilde{S}(\boldsymbol{\varphi})\equiv
\int
d\lambda\, e(\lambda)\, \textrm{Tr}_A\,\delta\left(\lambda-\tilde{C}(\boldsymbol{\varphi})\right),
\label{eq:37}
\eeq
then by Supplementary Eq.~(\ref{eq:30}) the relation:
\beq
S(t)=\tilde{S}(\boldsymbol{\omega}t)
\label{eq:38}
\eeq
follows. Since all entanglement probes associated with $\tilde{C}(\boldsymbol{\varphi})$ such as $\tilde{{\cal H}}_A$, $\tilde{\rho}_A$ and $\tilde{S}$ are obtained from their counterparts associated with $C(t)$ by the replacement: $C(t)\rightarrow\tilde{C} (\boldsymbol{\varphi})$, hereafter we use the same symbols: ${\cal H}_A,\,\rho_A,\, S$, etc. as corresponding probes associated with $C(t)$, keeping in mind the difference in the arguments.

\subsubsection{More general entanglement probes}
\label{sec:general_probe}

Let us consider a larger class of entanglement probes or information-theoretic observables $O(t)$, which are required only to be a functional of $C(t)$, i.e.
\begin{equation}\label{eq:101}
  O(t)\equiv O[C(t)],
\end{equation}
like Supplementary Eqs.~(\ref{eq:64}) and (\ref{eq:34}). In words, $O(t)$ depends on the time parameter only through $C(t)$, albeit its dependence on $C(t)$ is highly nonlinear in general. Therefore, $O(t)$ and $C(t)$ can be diagonalized simultaneously at any $t$. Besides the entanglement entropy and the reduced density of matrix, such entanglement probes include the $n$-th order R$\acute{\rm e}$nyi entropy
\beq
S_n(t)\equiv{1\over 1-n}\ln{\rm Tr}_A \left(\rho_A(t)\right)^n
\label{eq:88}
\eeq
and the entanglement spectrum, namely, the eigenvalues of the reduced density matrix
\beq
\rho_\mathbf{m}(t)=\prod_{\nu=1}^{2L_A}\left(p_\nu(t)\right)^{\delta_{1n_\nu}}\left(1-p_\nu(t)\right)^{\delta_{0n_\nu}}
\label{eq:89}
\eeq
labelled by $\mathbf{m}=\{n_\nu\}$, which is the configuration of occupation numbers $n_\nu\,(=0,1)$ at the eigenmode $\nu$. Owing to the relation Supplementary Eq.~(\ref{eq:30}), $O(t)$ must depend on $t$ via $\boldsymbol{\varphi}=\boldsymbol{\omega}t$, i.e. in the same fashion as Supplementary Eqs.~(\ref{eq:66}) and (\ref{eq:38}). That is, with the introduction of the $N$-variable function: $O(\boldsymbol{\varphi})\equiv O[\tilde{C}(\boldsymbol{\varphi})]$, the relation:
\beq
O(t)= O(\boldsymbol{\varphi})|_{\boldsymbol{\varphi}=\boldsymbol{\omega}t}.
\label{eq:69}
\eeq
then follows. Note that by definition $O(\boldsymbol{\varphi})$ is $2\pi$-periodic in each $\varphi_m$ also.

In order to appreciate more the importance of $C(t)$ to the time evolution of generic entanglement probe more, we proceed to find explicitly the functional dependence of $S_n(t)$ on $C(t)$. First of all, because of $-\ln Z(t)=\sum_{\nu=1}^{2L_A}\ln (1-p_\nu(t))$, we have
\beq
-\ln Z(t)=\int
d\lambda\, \ln (1-\lambda)\, \textrm{Tr}_A\,\delta\left(\lambda-C(t)\right).
\label{eq:90}
\eeq
Moreover, from Supplementary Eqs.~(\ref{eq:63}) and (\ref{eq:64}) we obtain
\beq
\ln{\rm Tr}_A {\rm e}^{-nH_A(t)}&=&\sum_{\nu=1}^{2L_A}\ln \left(1+\left({p_\nu(t)\over 1-p_\nu(t)}\right)^n\right)\nonumber\\
&=&\int
d\lambda\, \ln \left(1+\left({\lambda\over 1-\lambda}\right)^n\right)\, \textrm{Tr}_A\,\delta\left(\lambda-C(t)\right).
\label{eq:91}
\eeq
Then, by substituting Supplementary Eqs.~(\ref{eq:90}) and (\ref{eq:91}) into Supplementary Eq.~(\ref{eq:88}), we obtain
\beq
S_n(t)=\int
d\lambda\, e_n(\lambda)\, \textrm{Tr}_A\,\delta\left(\lambda-C(t)\right),
\label{eq:92}
\eeq
where
\begin{equation}\label{eq:93}
  e_n(\lambda)\equiv {1\over 1-n}\ln \left(\lambda^n+(1-\lambda)^n\right).
\end{equation}
In the limit: $n\rightarrow 1$, $e_n(\lambda)\rightarrow e(\lambda)$. Therefore, we recover the expression for the entanglement entropy Supplementary Eq.~(\ref{eq:36}).

Observing the right-hand sides of Supplementary Eqs.~(\ref{eq:36}) and (\ref{eq:90})-(\ref{eq:92}), we find that they have the common structure:
\beq
O(t)=\int
d\lambda\, \mathfrak{O}(\lambda)\, \textrm{Tr}_A\,\delta\left(\lambda-C(t)\right),
\label{eq:94}
\eeq
and differ only in the function $\mathfrak{O}(\lambda)$. Remarkably, this structure relates the time evolution of different entanglement probes to the same quantity, the instantaneous spectral density of the correlation matrix. Likewise, all $O(\boldsymbol{\varphi})$ can be related to the same quantity, the spectral density of $\tilde{C}(\boldsymbol{\varphi})$, in the way as
\beq
O(\boldsymbol{\varphi})=\int
d\lambda\, \mathfrak{O}(\lambda)\, \textrm{Tr}_A\,\delta\left(\lambda-\tilde{C}(\boldsymbol{\varphi})\right).
\label{eq:95}
\eeq
This expression bears a firm analogy to the expression for canonical quantities in mesoscopic physics [5-10], as we will demonstrate in Supplementary Note $4.3$.
So the evolving correlation matrix $C(t)$ is the building block of our out-of-equilibrium fluctuation theory. Moreover, as we shall see in the next Supplementary Note, the relation Supplementary Eq.~(\ref{eq:69}) allows us to trade out-of-equilibrium entanglement fluctuations to mesoscopic fluctuations from one virtual disordered sample to another, which are seemingly unrelated to the dynamics of entanglement at first glance.

Supplementary Eqs.~(\ref{eq:30}),  (\ref{eq:16a}), (\ref{eq:69}), (\ref{eq:94}) and (\ref{eq:95}) constitute a formalism connecting various information-theoretic observables, that characterize entanglement evolution, to classical rotation on $\mathbb{T}^N$.

\section{Statistical equivalence}
\label{sec:emergence_randomness}

It is well known that dynamical properties of classical rotation on the torus $\mathbb{T}^N$ are extremely sensitive to the number-theoretic properties of the angular velocity $\boldsymbol{\omega}$. Notably, if the equation: $\sum_{m=0}^{N-1} x_m \omega_m=0$ has no nontrivial integer solutions, i.e.
\begin{equation}\label{eq:68}
  \sum_{m=0}^{N-1} x_m \omega_m=0,\quad x_m\in \mathbb{Z}\, \Rightarrow \,x_1=\ldots=x_{N-1}=0,
\end{equation}
then $\omega_0,\ldots,\omega_{N-1}$ are incommensurate or said to  bear the {\it arithmetic linear independence} [11], which should be distinguished from the concept of {\it statistical independence} to be introduced shortly; if the condition Supplementary Eq.~(\ref{eq:68}) is not satisfied, the $N$ frequencies are commensurate. For incommensurate $\boldsymbol{\omega}$ the rotation is ergodic, while for commensurate $\boldsymbol{\omega}$ it is periodic [12]. By the dispersion relation Supplementary Eq.~(\ref{eq:4}) the angular velocity $\boldsymbol{\omega}$ governing the evolution of $C(t)$ (cf.~Supplementary Eq.~(\ref{eq:corr})) is incommensurate. So we will focus on the incommensurate case from now on. In \ref{sec:breakdown_ergodicity}
we will study the consequences of commensurate $\boldsymbol{\omega}$.

By definition $O(\boldsymbol{\varphi})$ introduced in Supplementary Note $2.2$
is $2\pi$-periodic in each argument $\varphi_m$. As a result, $O(t)$ is a {\it quasiperiodic} function of $t$ with {\it frequency basis} $\boldsymbol{\omega}$; in other words, after the initial process the time evolution $O(t)$ transits to quasiperiodic oscillations [13], as exemplified by the entanglement entropy (Fig.~1 of the main text) and the second-order R$\acute{\rm e}$nyi entropy (see Supplementary Figs.~\ref{fig:S1a}(a) and \ref{fig:S1a}(b) below): These quasiperiodic oscillations are conceptually different from the entanglement entropy oscillations due to the traversal of quasiparticle pairs or the incomplete revival of system's wavefunction; the latter oscillations rapidly damp in the course of time [14]. The quasiperiodic oscillations do not exclude the occurrence of a big bump or dip at certain times, and the big dip may signal the recurrence. It should be emphasized that the quasiperiodic oscillations of different probes, all of which can be attributed to those of the correlation matrix, are quantum oscillations (cf.~Supplementary Eq.~(\ref{eq:27a})).

For quasiperiodic $O(t)$ we have
\begin{equation}\label{eq:71}
  \lim_{T\rightarrow \infty}\int_{0}^{T} \frac{dt}{T} O(t)
  =\int {d\boldsymbol{\varphi}\over (2\pi)^N} O(\boldsymbol{\varphi})=\langle O\rangle
\end{equation}
by the ergodic theorem [12]. Here
\begin{equation}\label{eq:20}
  \langle \cdot\rangle\equiv \int {d\boldsymbol{\varphi}\over (2\pi)^N}(\cdot)
\end{equation}
stands for the average with respect to the probability measure $\mathbf{P}$ over $\mathbb{T}^N$, which has a uniform density of $1/(2\pi)^N$. Therefore, $O(t)$ quasiperiodically oscillates around $\langle O\rangle$. The ergodic theorem further gives that for arbitrary interval $\Delta$,
\begin{equation}\label{eq:43}
  \lim_{T\rightarrow \infty}\int_{O(t)\in \Delta} \frac{dt}{T}
  =\int_{O(\boldsymbol{\varphi})\in \Delta} {d\boldsymbol{\varphi}\over (2\pi)^N}.
\end{equation}
Here the left-hand side gives the frequency for the time series $O(t)$ to appear in $\Delta$,  while the right-hand side gives the probability  for  $O(\boldsymbol{\varphi})$ to be in the same interval, i.e. $\mathbf{P}(O(\boldsymbol{\varphi})\in \Delta)$. Consequently, out-of-equilibrium fluctuations displayed by the quasiperiodic oscillations $O(t)$ must have the same statistics as fluctuations of $O(\boldsymbol{\varphi})$ with $\boldsymbol{\varphi}$, provided $\boldsymbol{\varphi}$ is drawn randomly from $\mathbf{P}$. This statistical equivalence holds for any entanglement probes introduced in \ref{sec:evolution_EE}, and is confirmed numerically (Fig.~1a in the main text). More generally, for any set ${\cal A}\subset \mathbb{T}^N$, we have
\begin{equation}\label{eq:96}
  \lim_{T\rightarrow \infty}\int_{\boldsymbol{\omega} t\in {\cal A}} \frac{dt}{T}
  =\int_{\boldsymbol{\varphi}\in {\cal A}} {d\boldsymbol{\varphi}\over (2\pi)^N}\equiv \mathbf{P}(\boldsymbol{\varphi}\in {\cal A}),
\end{equation}
with Supplementary Eq.~(\ref{eq:43}) as a special case.

The right-hand sides of Supplementary Eqs.~(\ref{eq:43}) and (\ref{eq:96}) entail a uniform joint probability density, $1/(2\pi)^N$, for the $N$ components of $\boldsymbol{\varphi}$. Therefore, (with respect to this probability) these components are statistically independent, and each of them is uniformly distributed over the $1$D torus $\mathbb{T}$. That is, the above probability measure $\mathbf{P}$ is a product of $N$ uniform probability measures over $\mathbb{T}$. Such product structure of probability measures allows the applications of the theory of concentration of measure [15-17], and leads to important consequences later on.

\section{Emergence of mesoscopic fluctuations}
\label{sec:emergence}

The Rice-Mele model, the Su-Schriefer-Heeger model, and TFIC are all integrable. Besides, they are deterministic, namely, free of any {\it extrinsic} randomness or stochasticity. Despite of these, armed with the general principle established in the last Supplementary Note we show in this Supplementary Note that some canonical quantum chaotic phenomena can still emerge from the time evolution of the correlation matrix. Most importantly, we show that out-of-equilibrium fluctuations of various entanglement probes can be traded to their fluctuations from one virtual, mesoscopic disordered sample to another.

\subsection{Emergent of ensemble of random correlation matrices}
\label{sec:pseudorandomness}

According to Supplementary Eq.~(\ref{eq:27}), the correlation matrix $\tilde{C}(\boldsymbol{\varphi})$ can be separated into two parts, i.e.
\begin{equation}\label{eq:39}
  \tilde{C}(\boldsymbol{\varphi}) = C_0 + C_1(\boldsymbol{\varphi}).
\end{equation}
Each part inherits the block-Toeplitz matrix structure from $\tilde{C}$, i.e. $(C_{0,1})_{ii'}=(C_{0,1})_{i-i'=l}$ with $(C_{0,1})_l$ being $2\times 2$ blocks defined in the sublattice sector. The first part, $C_0$, has no dependence on $\boldsymbol{\varphi}$ and is well defined for $L\rightarrow\infty$,
\begin{equation}
    (C_0)_{l,\sigma\sigma'} = \frac{1}{L} \sum_k {\rm e}^{i kl} \gamma_{\sigma\sigma'}(k)\stackrel{L\rightarrow\infty}{\longrightarrow} \int_{-\pi}^{\pi}\frac{dk}{2\pi}{\rm e}^{i kl} \gamma_{\sigma\sigma'}(k),
\label{eq:40}
\end{equation}
whereas the second part, $C_1$, depends on $\boldsymbol{\varphi}$,
\beq
     (C_1)_{l,\sigma\sigma'} = \frac{1}{L}\sum_{m=0}^{N-1} \biggl(1-{1\over 2}\,\delta_{m0}\biggr)\biggl(
     \left({\rm e}^{ik_m l}\alpha_{\sigma\sigma'}(k_m)+(m\rightarrow -m)\right)\cos \varphi_m\quad\nonumber\\
+ \left({\rm e}^{ik_m l}\beta_{\sigma\sigma'}(k_m)+(m\rightarrow -m)\right) \sin \varphi_m\,\biggr),
\label{eq:41}
\eeq
and has no limiting behaviors for $L\rightarrow\infty$ because of the incommensurality of $\boldsymbol{\omega}$. Using Supplementary Eqs.~(\ref{eq:22}), (\ref{uv1}), (\ref{eq:24}) and (\ref{uv2}), we find after lengthy but straightforward calculations that the coefficient of $\cos \varphi_m$ ($\sin\varphi_m$) in the summand is real (purely imaginary), which is a function of $k_m$ and denoted as $R_{l,\sigma\sigma'}(k_m)$ ($I_{l,\sigma\sigma'}(k_m)$). So Supplementary Eq.~(\ref{eq:41}) is rewritten as
\beq
     (C_1(\boldsymbol{\varphi}))_{l,\sigma\sigma'} = \frac{1}{L}\sum_{m=0}^{N-1} R_{l,\sigma\sigma'}\left(k_m\right)\,\cos \varphi_m+
     \frac{1}{L}\sum_{m=0}^{N-1} I_{l,\sigma\sigma'}\left(k_m\right)\,\sin \varphi_m\equiv \mathfrak{R}_{l,\sigma\sigma'}(\boldsymbol{\varphi})+i\mathfrak{I}_{l,\sigma\sigma'}(\boldsymbol{\varphi}),
\label{eq:100}
\eeq
where $\mathfrak{R}_{l,\sigma\sigma'}(\boldsymbol{\varphi})$ and $\mathfrak{I}_{l,\sigma\sigma'}(\boldsymbol{\varphi})$ are the real and imaginary part of $(C_1(\boldsymbol{\varphi}))_{l,\sigma\sigma'}$, respectively. It is easy to show that for $l=0$ and $\sigma=\sigma'$ the imaginary part vanishes, i.e. the diagonal of $C_1$ is real. By Supplementary Eqs.~(\ref{eq:30}) and (\ref{eq:39}), we have
\begin{equation}\label{eq:159}
  C(t)= C_0 + C_1(\boldsymbol{\omega}t),
\end{equation}
with
\beq
     (C_1(\boldsymbol{\omega}t))_{l,\sigma\sigma'} = \frac{1}{L}\sum_{m=0}^{N-1} R_{l,\sigma\sigma'}\left(k_m\right)\,\cos (\omega_m t)+
     \frac{1}{L}\sum_{m=0}^{N-1} I_{l,\sigma\sigma'}\left(k_m\right)\,\sin (\omega_m t)\equiv \mathfrak{R}_{l,\sigma\sigma'}(t)+i\mathfrak{I}_{l,\sigma\sigma'}(t).
\label{eq:42}
\eeq
Therefore, $C(t)$ displays quasiperiodic oscillations around $C_0$.

In
\ref{sec:proof} we show that, thanks to Supplementary Eq.~(\ref{eq:96}), both $\mathfrak{R}_{l,\sigma\sigma'}(t)$ and $\mathfrak{I}_{l,\sigma\sigma'}(t)$ or correspondingly $\mathfrak{R}_{l,\sigma\sigma'}(\boldsymbol{\varphi})$ and $\mathfrak{I}_{l,\sigma\sigma'}(\boldsymbol{\varphi})$ are sub-Gaussian centered random variables. More precisely, they have zero mean and their deviation probabilities satisfy the following concentration inequalities:
\begin{eqnarray}
\lim_{T\rightarrow \infty}\int_{|\mathfrak{R}_{l,\sigma\sigma'}(t)|\geq \epsilon} \frac{dt}{T}
=\mathbf{P}\left(|\mathfrak{R}_{l,\sigma\sigma'}(\boldsymbol{\varphi})|\geq \epsilon\right)
&\leq& 2{\rm e}^{-\frac{L}{\eta_{l,\sigma\sigma'}}\, \epsilon^2},\label{eq:72}\\
\lim_{T\rightarrow \infty}\int_{|\mathfrak{I}_{l,\sigma\sigma'}(t)|\geq \epsilon} \frac{dt}{T}
=\mathbf{P}\left(|\mathfrak{I}_{l,\sigma\sigma'}(\boldsymbol{\varphi})|\geq \epsilon\right)
&\leq& 2{\rm e}^{-\frac{L}{\eta'_{l,\sigma\sigma'}}\, \epsilon^2},\label{eq:73}
\end{eqnarray}
for large $L$ and for any positive $\epsilon$. Here
\begin{eqnarray}
  \eta_{l,\sigma\sigma'} &\propto& \int_{-\pi}^{\pi} {dk\over 2\pi} \left({\rm e}^{ikl}\alpha_{\sigma\sigma'}(kl)+{\rm e}^{-ikl}\alpha_{\sigma\sigma'}(-kl)\right)^2, \label{eq:74}\\
  \eta'_{l,\sigma\sigma'} &\propto& -\int_{-\pi}^{\pi} {dk\over 2\pi} \left({\rm e}^{ikl}\beta_{\sigma\sigma'}(kl)+{\rm e}^{-ikl}\beta_{\sigma\sigma'}(-kl)\right)^2, \label{eq:75}
\end{eqnarray}
whose numerical coefficients have no dependence on $L$ and thus are not important. Both $\eta_{l,\sigma\sigma'}$ and $\eta'_{l,\sigma\sigma'}$ are positive. We see that
$C_1(\boldsymbol{\varphi})$ varies around zero with variance $\sim 1/L$, i.e.
\begin{equation}\label{eq:49}
  {\rm Var}(\mathfrak{R}_{l,\sigma\sigma'})\sim {\rm Var}(\mathfrak{I}_{l,\sigma\sigma'})\sim \frac{1}{L}.
\end{equation}
Physically, this scaling can be understood as follows. When $\boldsymbol{\omega}$ is incommensurate, $\cos(\omega_0t)$, $\cos(\omega_1t)$, ..., $\cos(\omega_{N-1}t)$ are $N$ statistically independent random numbers of zero mean [11]. As a result, Supplementary Eq.~(\ref{eq:42}), that corresponds to $\mathfrak{R}_{l,\sigma\sigma'}$ (respectively $\mathfrak{I}_{l,\sigma\sigma'}$), is a sum of $N\sim L$ random numbers. For large $N$ one may expect the central limiting theorem to apply, which then gives $\mathfrak{R}_{l,\sigma\sigma'}\sim \sqrt{N}/L\sim 1/\sqrt{L}$ and likewise for $\mathfrak{I}_{l,\sigma\sigma'}$.

In
\ref{sec:independence} we further show that any random variable $\mathfrak{R}$ or $\mathfrak{I}$ in the block $(C(t))_l$ (respectively $(\tilde{C}(\boldsymbol{\varphi}))_l$) is statistically independent of that in any distinct block $(C(t))_{l'\neq l}$ (respectively $(\tilde{C}(\boldsymbol{\varphi}))_{l'\neq l}$). In contrast, from Supplementary Eq.~(\ref{eq:42}) we see that random variables $\mathfrak{R},\,\mathfrak{I}$ in the same block $(C(t))_l$ (respectively $(\tilde{C}(\boldsymbol{\varphi}))_l$) are {\it not} statistically independent.

So, when the classical trajectory: $\boldsymbol{\varphi}=\boldsymbol{\omega}t$ sweeps out $\mathbb{T}^N$ uniformly, the infinite time series $C(t)$ generates a specific ensemble of random block-Toeplitz matrices $\tilde{C}(\boldsymbol{\varphi})$, denoted as ${\cal E}$. For each member of ${\cal E}$, its $\boldsymbol{\varphi}$-dependent part, $C_1(\boldsymbol{\varphi})$, is composed of $L_A$ statistically independent $2\times 2$ blocks $(C_1(\boldsymbol{\varphi}))_l$ ($l=0,1,\ldots, L_A-1$), the real or imaginary part of whose matrix elements are sub-Gaussian centered random variables, with the variance $\propto \eta_{l,\sigma\sigma'}/L,  \eta'_{l,\sigma\sigma'}/L$. The probability measure over ${\cal E}$ is induced by the uniform probability measure $\mathbf{P}$ over $\mathbb{T}^N$ in the way as follows: Given arbitrary intervals $\Delta_{l,\sigma\sigma'}^{r,\,i}$ for every $l,\sigma,\sigma' $, the probability for $\tilde{C}$ to satisfy: $\textrm{Re}(C_1)_{l,\sigma\sigma'}\in \Delta^r_{l,\sigma\sigma'}$ and $\textrm{Im}(C_1)_{l,\sigma\sigma'}\in \Delta^i_{l,\sigma\sigma'}$ is $\mathbf{P}(\boldsymbol{\varphi}\in {\cal B})$, where the set ${\cal B}$ is defined as
\begin{equation}\label{eq:97}
  {\cal B}\equiv \bigcap_{l,\sigma,\sigma' }\{{\rm Re}(C_1(\boldsymbol{\varphi}))_{l,\sigma\sigma'}\in \Delta_{l,\sigma\sigma'}^r,\,{\rm Im}(C_1(\boldsymbol{\varphi}))_{l,\sigma\sigma'}\in \Delta_{l,\sigma\sigma'}^i\}.
\end{equation}
With this induced probability measure, the ensemble $\mathcal{E}$ is equivalent to the time series $C(t)$ statistically. This ensemble differs from the ensemble of random hermitian Toeplitz matrices studied very recently [18,19] in two aspects. First, each member in $\mathcal{E}$, though being hermitian and Toeplitz-type, carries a block structure; that is, the element of Toeplitz matrix is now a $2 \times 2$ block, instead of a complex number. Second, according to Supplementary Eq.~(\ref{eq:49}) at fixed ratio $L_A/L$ the variance of its matrix elements decays with the matrix dimension and vanishes in the large matrix dimension limit, whereas in canonical studies of random matrices that variance is independent of the matrix dimension.

\subsection{Spectral statistics of correlation matrix}
\label{sec:level_statistics}

We proceed to study the statistical properties of the spectrum of the correlation matrices from either the time series $C(t)$ or $\tilde{C}(\boldsymbol{\varphi})$ drawn randomly from the ensemble $\mathcal{E}$. Since they are statistically equivalent, we will mainly focus on the properties derived from the time series $C(t)$. In this work it suffices to consider the nearest-neighbor spacing distribution, $P_0(s)$, with the eigenvalues rescaled by the mean eigenvalue spacing.

In order to spell out the block-Toeplitz structure in a transparent way,
the correlation matrix $C(t)$ can be rewritten in the following form:
\beq
C(t)=\frac{1}{2}\,\mathbb{I}+{\Gamma}(t),
\label{eq:25}
\eeq
where
\beq
{\Gamma}(t)=\left( \begin{array}{ccccc}
\check{\Pi}_{0}        & \check{\Pi}_{1}        & \ldots & \check{\Pi}_{L_A-2}& \check{\Pi}_{L_A-1} \\
\check{\Pi}_{-1}       & \check{\Pi}_{0}        & \ldots & \check{\Pi}_{L_A-3}& \check{\Pi}_{L_A-2} \\
\vdots                 &                        & \ddots &                    & \vdots\\
\check{\Pi}_{-(L_A-2)} & \check{\Pi}_{-(L_A-3)} & \ldots & \check{\Pi}_{0}    & \check{\Pi}_{1} \\
\check{\Pi}_{-(L_A-1)} & \check{\Pi}_{-(L_A-2)} & \ldots & \check{\Pi}_{-1}   & \check{\Pi}_{0} \end{array}\right)_{L_A\times L_A},\textrm{\ \ \ \ }
\label{eq:28}
\eeq
with
\beq
\check{\Pi}_{l}(t)
=\left( \begin{array}{cc} -\frac{1}{2}\delta_{l0}+f_l & g_l \\ g_{-l}^* & \frac{1}{2}\delta_{l0}-f_l \end{array}\right)
\label{eq:14}
\eeq
defined in the sublattice sector and
\beq
f_l=\frac{1}{L}\sum_{k} \, {\rm e}^{i k l} \,|\tilde{u}_{fk}(t)|^2,\textrm{\ \ \ }g_{l}=\frac{1}{L}\sum_{k} \, {\rm e}^{i k (l+{1\over 2})} \,\tilde{u}^*_{fk}(t) \tilde{v}_{fk}(t).
\eeq
with expressions of $\tilde{u}_{fk}(t)$, $\tilde{v}_{fk}(t)$ given in
\ref{app:rm}.
The statistical property $C(t)$ is therefore determined by $\Gamma(t)$. In parallel, we can generalize the above expressions Supplementary Eqs. (\ref{eq:28}) and (\ref{eq:14}) to define the corresponding $\tilde{C}(\boldsymbol{\varphi})=\frac{1}{2}\,\mathbb{I}+\tilde{\Gamma}(\boldsymbol{\varphi})$ and $\tilde{\Pi}_l(\boldsymbol{\varphi})$.

Owing to the structure described by Supplementary Eqs. (\ref{eq:28}) and (\ref{eq:14}), the eigenvalues of $\Gamma(t)$ come in pair: one positive and the other negative, with the same absolute value. Therefore, its spectrum splits into two subspectra symmetric with respect to zero (cf. Supplementary Fig.~\ref{fig:S3}(b) below): One consists of negative eigenvalues and the other positive. We focus on the former below.

First, we follow the method of Ref.~[19] to estimate small-$s$ behaviors of $P_0$. For this purpose we need to consider only block-Toeplitz matrices of minimal dimension. Such matrices carry the structure of $\Gamma$ but in lower matrix dimension, and are perturbed from the block-diagonal matrix, with each block being ${\rm diag}(\vartheta,-\vartheta)$ and $\vartheta<0$ being some fixed constant. To satisfy the requirements we set $L_A$ in Supplementary Eq.~(\ref{eq:28}) to be two and the diagonal elements of $\check{\Pi}_0$ to be $\pm(\vartheta+\delta\vartheta)$,
with $\delta\vartheta$ being real. Because of $L_A=2$ the subspectra consist of two eigenvalues. By analysis in Supplementary Note $4.1$,
these matrices have only two independent random variables, one from the block $\check{\Pi}_0$ and the other from $\check{\Pi}_1$. So the joint probability of the two eigenvalues can be expressed as an integral of these two variables. Thus the probability for both eigenvalues to be within a small distance of $s$ from $\vartheta$ is $\sim s^2$, with the power $2$ accounting for the number of independent random variables. Since the joint probability is the cumulative spacing distribution $\int_0^s P_0(s')ds'$, we find that
\begin{equation}\label{eq:102}
  P_0(s\rightarrow 0)\,{\sim}\,s.
\end{equation}
Thus $P_0$ vanishes at $s=0$. This is the so-called level repulsion phenomenon --- a defining property of quantum chaos [20]. Strikingly, here the phenomenon occurs in a peculiar system, such that it is integrable and free of any extrinsic randomness and thus has nothing to do with chaos, at first glance.

Next, we estimate large-$s$ behaviors of $P_0$. For this purpose we return to the original matrix, and note that the matrix elements in $\check{\Pi}_l$ and $\tilde{\Pi}_l$ decay rapidly with $l$, namely, the distance to the main diagonal for which $l=0,\,\sigma=\sigma'$. So $\Gamma$ and $\tilde{\Gamma}$ are random band matrices, and their matrix elements are short-ranged correlated since as shown in
\ref{sec:independence} the matrix elements of distinct blocks are statistically independent. Recall that for random band matrices with all their elements being statistically independent, it is generally believed that the eigenvectors may exhibit localization [21]. But it is not clear to us whether this might occur here, because the random Toeplitz band matrix is structured, i.e. all its elements of the same distance to the main diagonal are identical, and additionally the disorder strength has a size dependence shown in Supplementary Eq.~(\ref{eq:49}). From the Toeplitz structure we can expect only that the level repulsion exists up to the scale of mean eigenvalue spacing. So we can follow the arguments of Ref.~[22] to estimate large-$s$ behaviors. Supposing that a band of large width $s$ includes $\langle {\cal N}(s)\rangle$ eigenvalues on the average, by the central limiting theorem we find that the variance of the number of eigenvalues within this band ${\rm Var}({\cal N}(s))\sim \langle {\cal N}(s)\rangle$. Since the probability for the band  to include ${\cal N}(s)$ eigenvalues is $\sim {\rm e}^{-({\cal N}(s)-\langle {\cal N}(s)\rangle)^2/{\rm Var}({\cal N}(s))}$, the probability for the band  to be empty is $\sim {\rm e}^{-\langle {\cal N}(s)\rangle^2/{\rm Var}({\cal N}(s))}\sim {\rm e}^{-const.\langle {\cal N}(s)\rangle}$, with $const.$ being some positive constant. Since $P_0(s)$ is the probability for the band to be empty and $\langle {\cal N}(s)\rangle=s$, a Poissonian tail,
\begin{equation}\label{eq:103}
  P_0(s\gg 1)\,\sim\, {\rm e}^{-const.\langle {\cal N}(s)\rangle}= {\rm e}^{-const.s}
\end{equation}
then follows. The full Poissonian distribution was conjectured for random real symmetric Toeplitz matrices in Ref.~[23].

Finally, the limiting behaviors described by Supplementary Eqs.~(\ref{eq:102}) and (\ref{eq:103}) can be unified via the following simple form:
\begin{equation}\label{eq:104}
  P_0(s)=4s {\rm e}^{-2s}
\end{equation}
with the numerical coefficients fixed by the conditions: $\int_{0}^{\infty}P_0(s) ds=\int_{0}^{\infty}s P_0(s) ds=1$. The distribution Supplementary Eq.~(\ref{eq:104}), that holds for arbitrary $s$, is called semi-Poissonian distribution [24]. It was seen for the first time in Ref.~[18] for truly random hermitian Toeplitz matrices, but without the block structure. This distribution is confirmed in simulations of the spectral statistics of the evolving correlation matrix $C(t)$ (see Fig.~1(c) in the main text).

\begin{table}[b]
\centering
  \caption{Comparisons of different fluctuations.}\label{table:1}
    \begin{tabular}{lll}
     \hline
     \hline
     type of fluctuations $\quad\quad\quad$ & emergent sample-to-sample fluctuations $\quad\quad$ & sample-to-sample fluctuations $\quad\quad$\\
     & in entanglement evolution & in mesoscopic electronics and photonics\\
     \hline
     random matrix representing $\quad$& $\tilde{C}(\boldsymbol{\varphi})$& $T(\boldsymbol{V})$\\
     disordered sample &&\\
     matrix dimension & $2L_A$ & number of wave channels\\
     variance of matrix elements & $\propto 1/L_A$ (at fixed $L_A/L$) & independent of matrix dimension\\
     disorder realization & $\boldsymbol{\varphi}\equiv (\varphi_0,\ldots,\varphi_{N-1})$& $\boldsymbol{V}\equiv (V_{\boldsymbol{r}_1},V_{\boldsymbol{r}_2},\ldots)$\\
     distribution of disorder realizations $\quad$ & uniform in $\mathbb{T}^N$ $\qquad$ & Gaussian in high-dimensional\\
     &&Euclidean space\\
     examples of probes& entanglement entropy, R$\acute{\rm e}$nyi entropy & conductance, shot-noise power\\
     expression of probes & $\int
d\lambda\, \mathfrak{O}(\lambda)\, \textrm{Tr}_A\,\delta(\lambda-\tilde{C}(\boldsymbol{\varphi}))$$\qquad$&
$\int
d\lambda\, \mathfrak{O}(\lambda)\, \textrm{Tr}\,\delta(\lambda-T(\boldsymbol{V}))$\\
     \hline
      \hline
    \end{tabular}
\end{table}

\subsection{A new class of mesoscopic fluctuations}
\label{sec:emergence_mesoscopic_fluctuations}

We have shown that for long time various entanglement probes $O(t)$ quasiperiodically oscillate around their means $\langle O\rangle$. This implies that when the entanglement evolution enters into the stationary state at long time, various probes are {\it not} time independent but display fluctuations around the stationary value. As shown above the statistics of these out-of-equilibrium fluctuations is equivalent to that of the fluctuations of the same probe with $\boldsymbol{\varphi}$, provided $\boldsymbol{\varphi}$ is drawn from $\mathbb{T}^N$ with a uniform probability. Now we show that these fluctuations with $\boldsymbol{\varphi}$ connect the entanglement evolution and mesoscopic physics.

To this end we observe that the fluctuations of entanglement probes with $\boldsymbol{\varphi}$ have following prominent features. First, because a generic probe $O(\boldsymbol{\varphi})$ is a functional of the correlation matrix $\tilde{C}(\boldsymbol{\varphi})$, the fluctuation statistics of different probes can all be attributed to that of $\tilde{C}(\boldsymbol{\varphi})$ or more precisely its spectral density. Second, because the random, namely, $\boldsymbol{\varphi}$-dependent part of $\tilde{C}(\boldsymbol{\varphi})$ vanishes in the limit $L\rightarrow \infty$, these fluctuation phenomena can occur only for finite $L_A$ at fixed ratio $L_A/L\,(>0)$. Third, these phenomena inherit the quantum coherence nature from quasiperiodic oscillations. By these three features the fluctuations of various entanglement probes with $\boldsymbol{\varphi}$ fall into the paradigm of mesoscopic {\it sample-to-sample} fluctuations [5-10]; see Supplementary Table \ref{table:1} for comparisons. Indeed, in mesoscopic electronics or photonics various transport characteristics (e.g. the conductance or transmittance, the shot-noise power, etc.) can be expressed in the same form as the right-hand side of Supplementary Eq.~(\ref{eq:95}). Specifically, upon passing to various mesoscopic transport probes, the hermitian matrix: $\tilde{C}(\boldsymbol{\varphi
})$ is replaced by the hermitian matrix: $T\equiv\mathbf{t}\mathbf{t}^\dagger$, whose dimension gives the number of channels for wave transport in disordered samples and is finite also [25, 26]. Here $\mathbf{t}$ is the transmission matrix that encodes all information of wave propagation from the input to the output end. Similar to that $\tilde{C}(\boldsymbol{\varphi
})$ is determined by $\boldsymbol{\varphi
}$ randomly distributed over $\mathbb{T}^N$, the matrix $\mathbf{t}$ and thereby $T$ are determined by the disordered potential or the dielectric configuration, denoted as $\boldsymbol{V}$; furthermore, since typically the potential and dielectric fluctuations at distinct spatial points are independent and Gaussian, each disorder realization can be viewed as a vector in a high-dimensional Euclidean space, $\boldsymbol{V}\equiv (V_{\boldsymbol{r}_1},V_{\boldsymbol{r}_2},\ldots)$ with $\boldsymbol{r}_1,\boldsymbol{r}_2,\ldots$ labelling distinct spatial points, Gaussian distributed in that space [27].
So, similar to that varying $\boldsymbol{\varphi
}$ gives rise to an ensemble of $\tilde{C}(\boldsymbol{\varphi
})$, varying $\boldsymbol{V}$ gives rise to an ensemble of random matrices $T(\boldsymbol{V})$, and the distribution of $\boldsymbol{V}$ induces the distribution of $T(\boldsymbol{V})$, albeit in a highly complicated manner. With the replacement described above, the right-hand side of Supplementary Eq.~(\ref{eq:95}) gives various transport characteristics, e.g. the conductance or transmittance for $\mathfrak{O}(\lambda)=\lambda$ and the shot-noise power for $\mathfrak{O}(\lambda)=\lambda(1-\lambda)$ [10]. As such, $T(\boldsymbol{V})$ represents a finite disordered sample. In the same fashion, as $\tilde{C}(\boldsymbol{\varphi})$ determines various entanglement probes $O(\boldsymbol{\varphi})$, it mimics a finite disordered sample, with $\boldsymbol{\varphi}$ being the `disorder realization'. So the fluctuations of various entanglement probes with $\boldsymbol{\varphi}$ are traded to mesoscopic sample-to-sample fluctuations. Furthermore, since the former fluctuations are statistically equivalent to the out-of-equilibrium fluctuations displayed by quasiperiodic oscillations, those oscillations have the same nature as the mesoscopic sample-to-sample fluctuations.

However, standard random matrix theories require the variance of matrix elements to be independent of the matrix dimension. This is totally opposed to the present situations: Supplementary Eq.~(\ref{eq:49}) shows that at fixed ratio $L_A/L>0$ the variance of the matrix elements of $\tilde{C}(\boldsymbol{\varphi})$ decays with the matrix dimension $2L_A$ as $\sim 1/L_A$, so that nonrandom $\tilde{C}(\boldsymbol{\varphi})$ results in the limit $L_A\rightarrow\infty
$. Moreover, analytical or mathematical studies of random Toeplitz matrices are highly challenging, as put explicitly by Bogolmony and Giraud: ``seem to be inaccessible to known analytic methods'' [19]; even for the simplest random Toeplitz matrices, only recently have the studies of their spectral statistics been initiated and very few results been reported [18, 19]. For these reasons, we are prohibited to invoke standard mesoscopic theories to study the mesoscopic sample-to-sample fluctuations of entanglement probes. Therefore the long-time dynamics of entanglement brings us a new class of mesoscopic fluctuations.

\section{Statistics of emergent mesoscopic fluctuations: general theory}
\label{sec:statistics}

To overcome the difficulties discussed in the end of last Supplementary Note with the unusual disorder structure, below we develop an approach to study the statistics of mesoscopic fluctuations emergent from entanglement evolution. The idea starts from that a generic evolving entanglement probe $O(t)$ is given by a $N$-variable function, $O(\boldsymbol{\varphi})$, over $\mathbb{T}^N$ according to Supplementary Eq.~(\ref{eq:69}). So, when $\mathbb{T}^N$ is equipped with uniform probability, as imposed by Supplementary Eq.~(\ref{eq:96}), $O(\boldsymbol{\varphi})$ becomes a random variable depending on $N$ --- which is finite but large --- statistically independent angular variables $\varphi_0,\varphi_1,\ldots,\varphi_{N-1}$. Note that, however, the dependence of $O$ on those angular variables is highly nonlinear. Then, taking the advantage of the very fact that $N$ (i.e. $L$) is finite, we can combine the continuity properties of the function $O(\boldsymbol{\varphi})$ with the nonasymptotic probability theory, so-called concentration inequality [28], to study the statistical properties of fluctuations of $O$ with $\boldsymbol{\varphi}$. A similar idea has recently been developed to find the universal {\it wave-to-wave} fluctuations in mesoscopic transport [27].

\subsection{Concentration inequalities}
\label{sec:concentration_inequality}

To implement the idea above, we introduce the logarithmic moment-generating function defined as
\begin{equation}\label{eq:136}
  G(u)\equiv \ln \langle  {\rm e}^{u (O-\langle O \rangle)} \rangle,\quad u\in \mathbb{R}
\end{equation}
for a generic entanglement probe $O$. Because $\varphi_0,\varphi_1,\ldots,\varphi_{N-1}$ are independent and identically distributed random variables, we then have the so-called modified logarithmic Sobolev inequality [28], read
\beq\label{sobo1}
\frac{d}{du}\frac{G(u)}{u}\leq {1\over u^2}\frac{\sum_{m=0}^{N-1} \langle\,  {\rm e}^{u (O-\langle O \rangle)}    \,\phi\bigl(-u(O-O_m)\bigr)          \rangle}{\langle  {\rm e}^{u (O-\langle O \rangle)} \rangle}.
\eeq
Here
\begin{equation}\label{eq:137}
  \phi(x)\equiv {\rm e}^x-x-1.
\end{equation}
$O_m$ is an arbitrary function of $\boldsymbol{\varphi}'_m\equiv (\varphi_0,\varphi_1,\ldots,\varphi_{m-1},\varphi_{m+1},\ldots,\varphi_{N-1}) $, independent of $\varphi_m$. By definition it is also a random variable, but depending on ($N-1$) instead of $N$ independent $\varphi$'s. It is important that the Supplementary inequality (\ref{sobo1}) holds for any $N$. This is fundamentally different from the situations in traditional probability theory, where the number of independent random variables has to be sent to infinity eventually. In other words, the traditional probability theory is a limiting theory, whereas the concentration inequality is not. In this Supplementary Note we need two special classes of $O_m$'s, defined as the infimum (supremum) of $O(\varphi_0,\ldots,\varphi'_m,\ldots,\varphi_{N-1})$ over $\varphi'_m$ with the rest of $(N-1)$ unprimed variables held fixed. The function is denoted as $O^+_m$ ($O^-_m$) correspondingly. For the convenience below we also introduce the following quantity:
\begin{equation}\label{eq:46}
  b_\pm\equiv \sum_{m=0}^{N-1} \left\langle  (O-O_m^\pm)^2\right\rangle.
\end{equation}
Roughly speaking, it is the total variance of the probe $O$ with respect to each argument.

\subsubsection{The upper tail of the distribution}
\label{sec:upper}

Following Ref.~[28], we set $u>0$ and $O_m=O_m^+$ for the Supplementary inequality (\ref{sobo1}) to study the distribution of the upward fluctuations (i.e. $O-\langle O\rangle>0$). However, our subsequent treatments of that inequality differ substantially from those in Ref.~[28]. Observing that $\phi\bigl(-u(O-O_m^+)\bigr)$ can be Taylor expanded as
\begin{equation}\label{eq:138}
  \phi\bigl(-u(O-O_m^+)\bigr)={u^2\over 2}(O-O_m^+)^2+\sum_{n=3}^{\infty}{(-u)^n\over n!}(O-O_m^+)^n,
\end{equation}
we separate from $\phi$ the contribution: ${u^2\over 2}\langle(O-O_m^+)^2\rangle$ to the first term, and rewrite the Supplementary inequality (\ref{eq:138}) as
\begin{equation}\label{eq:139}
  \phi\bigl(-u(O-O_m^+)\bigr)={u^2\over 2}\left\langle(O-O_m^+)^2\right\rangle + {u^2\over 2}\delta(O-O_m^+)^2+\sum_{n=3}^{\infty}{(-u)^n\over n!}(O-O_m^+)^n,
\end{equation}
with $\delta(O-O_m^+)^2\equiv (O-O_m^+)^2-\left\langle(O-O_m^+)^2\right\rangle$ characterizing the deviation of $(O-O_m^+)^2$ from its mean. Substituting Supplementary Eq.~(\ref{eq:139}) into the right-hand side of the Supplementary inequality (\ref{sobo1}), we arrive at
\beq\label{sobo2}
\frac{d}{du} \frac{G(u)}{u}\leq  \frac{b_+}{2}  + \delta F(u),
\eeq
where
\begin{equation}\label{eq:140}
  \delta F(u)  =  \left\langle  {\rm e}^{u (O-\langle O \rangle)} \right\rangle^{-1} \sum_{m=0}^{N-1} \left\langle\,  {\rm e}^{u (O-\langle O \rangle)}    \left(\delta(O-O_m^+)^2+\sum_{n=1}^{\infty}{(-u)^n\over (n+2)!}(O-O_m^+)^{n+2}\right)          \right\rangle.
\end{equation}
It is important to note that although the right-hand side of the Supplementary inequality (\ref{sobo2}) bears some similarities to familiar ones in mathematical literatures [28], there are essential differences. In particular, we do not resort to the bounding property of the function: $\sum_{m=0}^{N-1} (O-O_m^+)^2$, and consequently the first term makes no references to the bounding properties of that function. Whereas in mathematical literatures $b_+$ is determined by the bounding properties of that function.

Next, noting that $O-O_m^+\geq 0$ and the equal sign is taken only at a set of zero (Lebesgue) measure, we find that the right-hand side of the Supplementary inequality (\ref{sobo1})
\beq\label{eq:141}
{1\over u^2}\frac{\sum_{m=0}^{N-1} \langle\,  {\rm e}^{u (O-\langle O \rangle)}    \,\phi\bigl(-u(O-O_m^+)\bigr)          \rangle}{\langle  {\rm e}^{u (O-\langle O \rangle)} \rangle}
\stackrel{u\rightarrow+\infty}{\longrightarrow}{1\over u}\frac{\sum_{m=0}^{N-1} \langle\,  {\rm e}^{u (O-\langle O \rangle)}    \,\bigl(O-O_m^+\bigr)          \rangle}{\langle  {\rm e}^{u (O-\langle O \rangle)} \rangle}.
\eeq
As a result, there must exist some constant $const.$ and some sufficiently large $u^*$, so that
\begin{equation}\label{eq:142}
{1\over u^2}\frac{\sum_{m=0}^{N-1} \langle\,  {\rm e}^{u (O-\langle O \rangle)}    \,\phi\bigl(-u(O-O_m^+)\bigr)          \rangle}{\langle  {\rm e}^{u (O-\langle O \rangle)} \rangle}\leq {const.\over u},\quad {\rm for}\,\, u>u^*.
\end{equation}
This implies that the right-hand side of the modified logarithmic Sobolev inequality is bounded by the curve $\sim {1\over u}$ in the large $u$ regime. On the other hand, it is easy to show that for $u>0$, ${dG(u)\over du}>0$ and grows unboundedly with $u$. Taking this and the Supplementary inequality (\ref{eq:142}) into account, we find that the function $\delta F/{dG\over du}$ must be bounded for $u>0$. Let its supremum be $c_+$. We obtain
\beq\label{sobo3}
\delta F(u)\leq c_+\,  \frac{d G(u)}{du}.
\eeq
As we shall see immediately, that $d G/du$ is positive definite (for $u>0$) entails the sign of $c_+$ important consequences.

Combining the Supplementary inequalities (\ref{sobo2}) and (\ref{sobo3}), we arrive at the following differential inequality:
\beq\label{sobo4}
\frac{d}{du} \frac{G(u)}{u} \leq  \frac{b_+}{2} +  c_+ \,  \frac{d G(u)}{du}.
\eeq
Moreover, by Supplementary Eq.~(\ref{eq:136}) this differential inequality is implemented by the boundary condition: ${G(u)\over u}=0$ at $u=0$. Now, let us study the cases of $c_+<0$ and $c_+>0$ separately:
\begin{itemize}
  \item $c_+<0$: Thanks to ${dG\over du}>0$ the Supplementary inequality (\ref{sobo4}) gives
\beq\label{eq:143}
\frac{d}{du} \frac{G(u)}{u} \leq  \frac{b_+}{2}.
\eeq
It can be solved readily. The result is
\beq\label{eq:144}
G(u)\leq \frac{b_+u^2}{2}.
\eeq
Then it is a standard exercise to use the Markov inequality [28] to obtain the concentration inequality for the upper tail of the distribution. Specifically, by the Markov inequality we have for arbitrary positive $\epsilon$ and $u$,
\beq
\mathbf{P}(O-\langle O \rangle \geq \epsilon)\leq {\rm e}^{-u \epsilon+G(u)}.
\label{eq:145}
\eeq
Combining it with the Supplementary inequality (\ref{eq:144}) we obtain
\beq
\mathbf{P}(O-\langle O \rangle \geq \epsilon)\leq {\rm e}^{-u \epsilon+\frac{b_+u^2}{2}}.
\label{eq:146}
\eeq
Since $u$ is arbitrarily positive, upon minimizing the exponent over positive $u$ it gives
\beq
\mathbf{P}(O-\langle O \rangle \geq \epsilon)\leq {\rm e}^{-{\epsilon^2\over 2b_+}}.
\label{eq:147}
\eeq
This concentration inequality defines a sub-Gaussian upper tail distribution.

\item $c_+>0$: We solve the Supplementary inequality (\ref{sobo4}). The result is
\beq\label{sobo5}
G(u)\leq \frac{b_+}{2}\frac{u^2}{1-c_+u}.
\eeq
This kind of bounds hold also for Gamma random variables, and thus generalize the tail behaviors of the Gamma distribution, giving the so-called sub-Gamma tail distribution [28]. Specifically, by the same method of deriving the Supplementary inequality (\ref{eq:147}), we obtain
\beq
\mathbf{P}(O-\langle O \rangle\geq \epsilon)\leq {\rm e}^{-\frac{\epsilon^2}{2 (b_++c_+ \epsilon)}}.
\label{eq:148}
\eeq
The constants $b_+ $ and $c_+$ are called the {\it variance factor} and {\it scale parameter}, respectively [28]. Of course, it does not mean that $b_+$ gives the exact variance, but gives a bound for the exact variance in general. In the present work, by the choice of $b_\pm$ defined by Supplementary Eq.~(\ref{eq:46}) ($b_-$ is for the lower tail; see the concentration inequalities, namely, Supplementary (\ref{eq:150}) and (\ref{eq:151}) below), which are substantially different from canonical choices in mathematical literatures, we further find that $b_\pm$ are proportional to the exact variance, with the help of numerical simulations (see Supplementary Note $6.5$
for detailed discussions). Finally, we remark that as $\epsilon$ increases, the sub-Gamma bound displays a crossover from the Gaussian form ${\rm e}^{-\epsilon^2/(2b_+)}$ to the exponential form ${\rm e}^{-\epsilon/(2c_+)}$ form at $\epsilon\sim b_+/c_+$.
\end{itemize}

\subsubsection{The lower tail of the distribution}
\label{sec:lower}

The distribution of downward fluctuations (i.e. $\langle O\rangle-O>0$) can be studied in the same way. We set $u<0$ and $O_m=O_m^-$ for the modified logarithmic Sobolev inequality, i.e. Supplementary (\ref{sobo1}). With these two replacements, we have
\beq\label{eq:149}
\delta F(u)\leq c_-\,  \frac{d G(u)}{du}
\eeq
with $d G/du<0$, similar to the bounding described by Supplementary (\ref{sobo3}). The subsequent analysis can be carried out in exactly the same way. But the results are reversed:
\begin{itemize}
  \item $c_-<0$: We obtain
  \beq
\mathbf{P}(\langle O \rangle-O\geq \epsilon)\leq {\rm e}^{-\frac{\epsilon^2}{2 (b_-+|c_-| \epsilon)}},
\label{eq:150}
\eeq
i.e. a sub-Gamma lower tail.
\item  $c_+>0$: We obtain
\beq
\mathbf{P}(\langle O \rangle-O \geq \epsilon)\leq {\rm e}^{-{\epsilon^2\over 2b_-}},
\label{eq:151}
\eeq
i.e. a sub-Gaussian lower tail.
\end{itemize}

\subsection{Calculations of the variance factors $\boldsymbol{b_\pm}$}
\label{sec:fluctuations}

In this part we study the variance factors $b_\pm$ defined by Supplementary Eq.~(\ref{eq:46}) in details. First of all, we keep in mind that $O_m^\pm(\boldsymbol{\varphi}'_m)$ in Supplementary Eq.~(\ref{eq:46}) actually belongs to a special class of (real-valued) functions of $\boldsymbol{\varphi}$, such that they are independent of $\varphi_m$. So $(O-O_m^\pm)^2$ is a function of $\boldsymbol{\varphi}$ also, i.e. $(O-O_m^\pm)^2\equiv(O-O_m^\pm)^2(\boldsymbol{\varphi})$. Let $\varphi_m^\pm$ be the value of $\varphi_m$, at which $O(\boldsymbol{\varphi})=O^\pm_m (\boldsymbol{\varphi}'_m)$. Note that by definition $\varphi_m^\pm$ depends on $\boldsymbol{\varphi}'_m$ as well as $O$. Then, by the mean value theorem there exists ${\bar \varphi}_m^\pm$ between $\varphi_m$ and $\varphi_m^\pm$ which depends on $\boldsymbol{\varphi}$, so that
\begin{equation}\label{eq:48}
  (O-O_m^\pm)^2(\boldsymbol{\varphi})
  =(\varphi_m-\varphi_m^\pm)^2\left(\partial_{{\varphi}_m}O|_{{\varphi}_m=\bar{\varphi}_m^\pm}\right)^2.
\end{equation}
Here the subscript: ${{\varphi}_m=\bar{\varphi}_m^\pm}$ stands for that the arguments of $\partial_{{\varphi}_m}O$ are $(\varphi_0,\ldots,\bar{\varphi}_m^\pm,\ldots,\varphi_{N-1})$, and to make the formula compact we suppress all the arguments on the right-hand side except $\bar{\varphi}_m^\pm$. From Supplementary Eq.~(\ref{eq:48})
\begin{equation}\label{eq:56}
  b_\pm=\sum_{m=0}^{N-1}\left\langle(\varphi_m-\varphi_m^\pm)^2\left(\partial_{{\varphi}_m}O|_{{\varphi}_m=\bar{\varphi}_m^\pm}\right)^2\right\rangle
\end{equation}
follows immediately.

Then we perform the Fourier expansion of $\partial_{{\varphi}_m}O$ with respect to $\varphi_m$. Taking into account the general expression of $O$ (cf.~Supplementary Table \ref{table:1}) and $C_1={\cal O}(1/\sqrt{L})$, we can expand $O$ in $C_1$ for large total system size $L\gg 1$. Keeping the expansion up to the second order (see Supplementary Note $6.1$
for explicit calculations for the entanglement entropy) gives
\begin{equation}\label{eq:50}
  \partial_{{\varphi}_m}O=a_0+a_1\sin(\varphi_m)+
  b_1\cos(\varphi_m)+a_2\sin(2\varphi_m)+
  b_2\cos(2\varphi_m),
\end{equation}
where we have made use of Supplementary Eq.~(\ref{eq:100}) and the coefficients $a$'s, $b$'s are $\varphi_m$ independent. We can rewrite Supplementary Eq.~(\ref{eq:50}) as
\begin{equation}\label{eq:51}
  \partial_{{\varphi}_m}O=A_0+A_1\sin(\varphi_m+\phi_{m1})+A_2\sin(2\varphi_m+\phi_{m2}),
\end{equation}
where $A$'s and $\phi$'s are determined by $a$'s and $b$'s. With the help of this expression it can be readily shown that
\begin{equation}\label{eq:52}
  (\partial_{{\varphi}_m}O)^2\leq (|A_0|+|A_1|+|A_2|)^2\leq 3(A_0^2+A_1^2+A_2^2)
\end{equation}
and
\begin{equation}\label{eq:53}
  \int\frac{d\varphi_m}{2\pi}(\partial_{{\varphi}_m}O)^2=A_0^2+{1\over 2}(A_1^2+A_2^2).
\end{equation}
The Supplementary inequality (\ref{eq:52}) and Supplementary Eq.~(\ref{eq:53}) give
\begin{equation}\label{eq:54}
  (\partial_{{\varphi}_m}O)^2\leq 6\int\frac{d\varphi_m}{2\pi}(\partial_{{\varphi}_m}O)^2.
\end{equation}
Actually the factor $6$ can be improved but this is not important. This bound suggests that (with the components ${\varphi}_{n\neq m}$ treated as parameters) $(\partial_{{{\varphi}}_m}S)^2$ at given ${\varphi}_m$ is the same order as its ${\varphi}_m$-average, i.e.
\begin{equation}\label{eq:57}
  (\partial_{{\varphi}_m}O)^2=r_m(\boldsymbol{\varphi})\int\frac{d\varphi_m}{2\pi}(\partial_{{\varphi}_m}O)^2, \quad r_m(\boldsymbol{\varphi})={\cal O}(1).
\end{equation}
Substituting it into Supplementary Eq.~(\ref{eq:56}) we obtain
\begin{equation}\label{eq:58}
  b_\pm=\sum_{m=0}^{N-1}\left\langle\left((\varphi_m-\varphi_m^\pm)^2
  r_m(\boldsymbol{\varphi}_m^{\pm*})\right)\int\frac{d\varphi_m}{2\pi}(\partial_{{\varphi}_m}O)^2\right\rangle,
\end{equation}
where $\boldsymbol{\varphi}_m^{\pm*}\equiv (\varphi_0,\ldots,\bar{\varphi}^\pm_m,\ldots,\varphi_{N-1})$. Because the first factor characterizes the distance between $\varphi_m$ and $\varphi_m^\pm$ and the second describes the $\varphi_m$ average of $(\partial_{{\varphi}_m}O)^2$, we assume that they are statistically uncorrelated and obtain
\begin{equation}\label{eq:59}
  b_\pm=\sum_{m=0}^{N-1}\left\langle (\varphi_m-\varphi_m^\pm)^2
  r_m(\boldsymbol{\varphi}_m^{\pm*})\right\rangle\left\langle\int\frac{d\varphi_m}{2\pi}(\partial_{{\varphi}_m}O)^2\right\rangle.
\end{equation}
Furthermore, by symmetry $\left\langle (\varphi_m-\varphi_m^\pm)^2
  r_m(\boldsymbol{\varphi}_m^{\pm*})\right\rangle$ must be the same for distinct $m$, at least approximately. As a result, we reduce Supplementary Eq.~(\ref{eq:59}) to
\begin{equation}\label{eq:60}
  b_\pm=R^\pm\left\langle |\partial_{\boldsymbol{\varphi}}O|^2\right\rangle,
\end{equation}
where
\begin{equation}\label{eq:61}
  R^\pm={1\over N}\sum_{m=0}^{N-1}\left\langle (\varphi_m-\varphi_m^\pm)^2
  r_m(\boldsymbol{\varphi}_m^{\pm*})\right\rangle
\end{equation}
is an overall numerical coefficient. As mentioned above, in the present work owing to the special choice of $b_\pm$, which are given by Supplementary Eq.~(\ref{eq:46}), $b_\pm$ are proportional to the variance ${\rm Var}(O)$. As a result,
\begin{equation}\label{eq:135}
  {\rm Var}(O)\propto\left\langle |\partial_{\boldsymbol{\varphi}}O|^2\right\rangle,
\end{equation}
where $|\partial_{\boldsymbol{\varphi}}O|^2=\sum_{m=0}^{N-1}(\partial_{\varphi_m}O(\boldsymbol{\varphi}))^2$. This formula relates fluctuations of the entanglement probe $O$ to the continuity of the $N$-variable function $O({\boldsymbol{\varphi}})$ defined by Supplementary Eq.~(\ref{eq:95}). As such, the mesoscopic fluctuations emergent from entanglement evolution fall into the class of the so-called concentration-of-measure phenomena, where fluctuations of an observable are controlled by its {\it Lipschitz continuity}. The concentration-of-measure phenomena are a new perspective of probability theory [15], and are first reported for mesoscopic electronic and photonic transport in Ref.~[27]. In Fig. 3(a) of the main text, we provide the numerical verification of Supplementary Eq. (\ref{eq:135}) for the entanglement entropy. In Supplementary  Fig. \ref{fig:S2}(a) we provide further numerical verifications for the $n$-th R{\'e}nyi entropy $S_n$.

\section{Statistics of entanglement entropy fluctuations}
\label{sec:applications_distribution_EE}

In this Supplementary Note, we apply the general theory developed in the last Supplementary Note to study the fluctuations of entanglement entropy. In particular, we would like to address their universalities, e.g. to what extent they depend on system's macroscopic and microscopic parameters. The studies in this Supplementary Note can be straightforwardly generalized to $S_n$, and the results do not change.

\subsection{Two contributions to $\boldsymbol{{\rm Var}(S)}$}
\label{sec:contributions}

We first consider one of the most important characteristics, the variance ${\rm Var}(S)$, for which Supplementary Eq.~(\ref{eq:135}) gives
\begin{equation}\label{eq:47}
  {\rm Var}(S)\propto\left\langle |\partial_{\boldsymbol{\varphi}}S|^2\right\rangle.
\end{equation}
Below we calculate explicitly its right-hand side. To this end we need to find the explicit expression of $\partial_{\varphi_m} S(\boldsymbol{\varphi})$. Upon taking the derivative $\partial_{\varphi_m}$ on both sides of Supplementary Eq.~(\ref{eq:37}), we obtain
\beq
\partial_{\varphi_m} S=\textrm{Tr}_A \left({\cal H}_A\partial_{\varphi_m}C_1\right).
\label{eq:107}
\eeq
Here
\begin{equation}\label{eq:160}
  {\cal H}_A(\boldsymbol{\varphi})=\ln ((\tilde{C}(\boldsymbol{\varphi}))^{-1}-\mathbb{I})
\end{equation}
is the coefficient matrix of the entanglement Hamiltonian $H_A(\boldsymbol{\varphi})$, which corresponds to the reduced density matrix Supplementary Eq.~(\ref{eq:65}) and has a quadratic form. We shall call ${\cal H}_A(\boldsymbol{\varphi})$ entanglement Hamiltonian as well.
Because of $C_1={\cal O}(1/\sqrt{L})$ we can expand ${\cal H}_A$ in terms of $C_1$ for $L\gg 1$. Keeping the expansion up to the first order and substituting it into Supplementary Eq.~(\ref{eq:107}), we obtain
\beq
\partial_{\varphi_m} S=(\partial_{\varphi_m} S)_1+(\partial_{\varphi_m} S)_2,
\label{eq:161}
\eeq
where
\beq
(\partial_{\varphi_m} S)_1&=&\textrm{Tr}_A \left(\ln \left(C_0^{-1}-\mathbb{I}\right)\partial_{\varphi_m}C_1\right),\nonumber\\
(\partial_{\varphi_m} S)_2&=&-
\textrm{Tr}_A \left(\left(\left(\mathbb{I}-C_0\right)C_0\right)^{-1}C_1\partial_{\varphi_m}C_1\right)\nonumber\\
&&-{1\over 2}\textrm{Tr}_A \left(\left(\left(\mathbb{I}-C_0\right)^{-1}[\ln \left(\mathbb{I}-C_0\right), C_1]+
C_0^{-1}[\ln C_0, C_1]\right)\partial_{\varphi_m}C_1\right).
\label{eq:109}
\eeq
Taking into account Supplementary Eq.~(\ref{eq:100}) we find that Supplementary Eq.~(\ref{eq:109}) gives the Fourier expansion of $\partial_{\varphi_m} S$ with respect to $\varphi_m$ up to the second harmonic. The square of Supplementary Eq.~(\ref{eq:161}) has three terms. It can be readily seen that upon averaging over $\boldsymbol{\varphi}$ the crossing term vanishes. As a result, ${\rm Var}(S)$ separates into two contributions,
\begin{eqnarray}\label{eq:110}
  {\rm Var}(S)={\rm Var}_1(S)+{\rm Var}_2(S)
  \label{eq:130}
\end{eqnarray}
with
\begin{eqnarray}
  {\rm Var}_1(S)&=&const.\sum_{m=0}^{N-1}\left\langle\left(\textrm{Tr}_A \left(\ln \left(C_0^{-1}-\mathbb{I}\right)\partial_{\varphi_m}C_1\right)\right)^2\right\rangle,\label{eq:111}\\
  {\rm Var}_2(S)&=&const.
  \sum_{m=0}^{N-1}\bigg\langle\bigg[\textrm{Tr}_A\left(\left((\mathbb{I}-C_0)C_0\right)^{-1}C_1\partial_{\varphi_m}C_1\right)\nonumber\\
  &&+{1\over 4}\textrm{Tr}_A \left(\left(\left(\mathbb{I}-C_0\right)^{-1}[\ln \left(\mathbb{I}-C_0\right), \left(\mathbb{I}-C_0\right)^{-1}C_1]+
C_0^{-1}[\ln C_0, C_0^{-1}C_1]\right)\partial_{\varphi_m}C_1\right)\bigg]^2\bigg\rangle,\label{eq:110}
\end{eqnarray}
where $const.$ is the overall numerical proportionality coefficient in Supplementary Eq.~(\ref{eq:47}), and is not important in subsequent discussions. Below we study the two contributions.

\subsection{The contribution $\boldsymbol{{\rm Var}_1(S)}$}
\label{sec:calculations_var_S1}

To calculate Supplementary Eq.~(\ref{eq:111}) we note that the trace operation can be cast to ${\rm Tr}_A(\cdot)=\sum_{i=1}^{L_A}\sum_{\sigma}(\cdot)$. That is, we perform that operation in two steps: In the first step, we sum over the sublattice index $\sigma=\A,\B$; in the second, we sum over the cell index. Now, because the matrix block $(C_0)_{ij}$ decays rapidly with $|i-j|$, we can divide the sum over $i$ into two contributions, one from the bulk of the subsystem A and the other from its edge. The size of the edge, $L_e$, is order of the decay length of $C_0$, and thus is determined merely by the parameters of the Hamiltonian; while the size of the bulk is $\approx L_A$. The bulk and the edge of the subsystem are denoted as $A_b$ and $A_e$, respectively. The cells in $A_b$ are labelled by $i_b$ and in $A_e$ by $i_e$. So,
\begin{eqnarray}\label{eq:119}
  {\rm Tr}_A\left(\ln \left(C_0^{-1}-\mathbb{I}\right)\partial_{\varphi_m}C_1\right)
  =\sum_{i_b\in A_b}\sum_{\sigma}\left(\ln \left(C_0^{-1}-\mathbb{I}\right)\partial_{\varphi_m}C_1\right)_{i_b\sigma,i_b\sigma}+\sum_{i_e\in A_e}\sum_\sigma\left(\ln \left(C_0^{-1}-\mathbb{I}\right)\partial_{\varphi_m}C_1\right)_{i_e\sigma,i_e\sigma}.
\end{eqnarray}
Below we calculate the two terms separately.

For the bulk term, owing to the rapid decay of $C_0$ we can extend all the intermediate cell indexes involved in the matrix product to the total system. Taking this into account, we find that
\beq\label{fluc1}
&&\sum_{i_b\in A_b}\sum_\sigma\left(\ln \left(C_0^{-1}-\mathbb{I}\right)\partial_{\varphi_m}C_1\right)_{i_b\sigma,i_b\sigma}\nonumber\\
&\simeq&\frac{L_A}{L}\biggl(1-{1\over 2}\,\delta_{m0}\biggr)\textrm{tr} \biggl[  -  \biggl(\ln(\check{\gamma}_m^{-1}-\mathbb{I})\,\check{\alpha}_{m}+(m\rightarrow-m)\biggr)\sin\varphi_{m}+ \biggl(\ln(\check{\gamma}_m^{-1}-\mathbb{I})\,\check{\beta}_{m}+(m\rightarrow-m)\biggr)\cos\varphi_{m}\biggr],\quad
\eeq
where the trace $\textrm{tr}$ is restricted to the sublattice sector and $\mathbb{I}$ is the unit matrix in that sector. Moreover, we introduce the matrix defined in the sublattice sector, $\check{\alpha}_m$ with matrix elements $\alpha_{\sigma \sigma'}(k_m)$, and likewise $\check{\beta}_m$ and $\check{\gamma}_m$.

With the introduction of $\check{\Gamma}_m\equiv \check{\gamma}_m-\mathbb{I}/2$, we rewrite Supplementary Eq.~(\ref{fluc1}) as
\beq\label{eq:115}
&&\sum_{i_b\in A_b}\textrm{tr}\left(\ln \left(C_0^{-1}-\mathbb{I}\right)\partial_{\varphi_m}C_1\right)_{i_bi_b}\nonumber\\
&\simeq&\frac{L_A}{L}\biggl(1-{1\over 2}\,\delta_{m0}\biggr)\textrm{tr} \biggl[  -  \biggl(\ln\frac{{\mathbb{I}/2}-\check{\Gamma}_m}{{\mathbb{I}/2}+\check{\Gamma}_m}\,\check{\alpha}_{m}+(m\rightarrow-m)\biggr)\sin\varphi_{m}+ \biggl(\ln\frac{{\mathbb{I}/2}-\check{\Gamma}_m}{{\mathbb{I}/2}+\check{\Gamma}_m}\,\check{\beta}_{m}+(m\rightarrow-m)\biggr)\cos\varphi_{m}\biggr].\,\,\quad
\eeq
Then we make use of the properties (i)-(xii) of $\check{\gamma}_m\,$, $\check{\alpha}_m$ and $\check{\beta}_m\,$ given in
\ref{sec:property_alpha_beta_gamma} to calculate Supplementary Eq.~(\ref{eq:115}).

When we expand the logarithm in $\check{\Gamma}_m$ in Supplementary Eq.~(\ref{eq:115}) and perform the trace, we find that each $\gamma_{\A\B}$ factor is paired with a $T_{\B\A}$ ($T=\gamma, \alpha\, (\textrm{or}\, \beta)$) factor. Therefore, all the phase factors ${\rm e}^{i\delta(k)}$, that are carried by the off-diagonal components of $\check{\gamma}_m\,$, $\check{\alpha}_m$ and $\check{\beta}_m\,$ (cf.~the properties (ii), (iv) and (vi)), cancel out. Therefore, the expansion corresponds, term by term, to the $\check{\Gamma}'_m$-expansion of
\beq
\label{eq:116}
\textrm{tr} \biggl[  -  \biggl(\ln\frac{{\mathbb{I}/2}-\check{\Gamma}'_m}{{\mathbb{I}/2}+\check{\Gamma}'_m}\,\check{\alpha}'_{m}+(m\rightarrow-m)\biggr)\sin\varphi_{m}+ \biggl(\ln\frac{{\mathbb{I}/2}-\check{\Gamma}'_m}{{\mathbb{I}/2}+\check{\Gamma}'_m}\,\check{\beta}'_{m}+(m\rightarrow-m)\biggr)\cos\varphi_{m}\biggr],
\eeq
where $\check{\Gamma}'_m\,$, $\check{\alpha}'_m$ and $\check{\beta}'_m$ differ from $\check{\Gamma}_m\,$, $\check{\alpha}_m$ and $\check{\beta}_m$ only in that their off-diagonal components have no phase factor.
Note that $\check{\Gamma}'_m$ is a real symmetric matrix. Taking this and the properties (v) and (vi) into account, we find that the coefficient of $\cos\varphi_{m}$ vanishes for $m>0$; and for $m=0$ taking this and the properties (xi) and (xii) into account, we find that the coefficient also vanishes. So the second term in Supplementary Eq.~(\ref{eq:116}) vanishes, and thus Supplementary Eq.~(\ref{eq:115}) is simplified as
\beq
\label{fluc3}
\sum_{i_b\in A_b}\textrm{tr}\left(\ln \left(C_0^{-1}-\mathbb{I}\right)\partial_{\varphi_m}C_1\right)_{i_bi_b}
&\simeq& -\frac{L_A}{L}\biggl(1-{1\over 2}\,\delta_{m0}\biggr)\textrm{tr}  \biggl(\ln\frac{{\mathbb{I}/2}-\check{\Gamma}'_m}{{\mathbb{I}/2}+\check{\Gamma}'_m}\,\check{\alpha}'_{m}+(m\rightarrow-m)\biggr)\sin\varphi_{m}\nonumber\\
&=& -\frac{L_A}{L}\biggl(1-{1\over 2}\,\delta_{m0}\biggr)\textrm{tr}  \biggl(\ln\frac{{\mathbb{I}/2}-\check{\Gamma}'_m}{{\mathbb{I}/2}+\check{\Gamma}'_m}\,\breve{\alpha}_{m}\biggr)\sin\varphi_{m}
\eeq
with $\breve{\alpha}_m\equiv \check{\alpha}'_{m}+\check{\alpha}'_{-m}$, where we have used the properties (i) and (ii) to derive the second line.

To calculate Supplementary Eq.~(\ref{fluc3}) we expand the logarithm in $\check{\Gamma}'_{m}$, and consider ${\rm tr}((\check{\Gamma}'_{m})^r\breve{\alpha}_{m})$ for any $r\in \mathbb{N}$. To this end we note that $\check{\Gamma}'_{m}$ and $\breve{\alpha}_m$ have the following general structure:
\beq\label{idgamma}
\check{\Gamma}'_{m}\equiv \left( \begin{array}{cc}\gamma_m  & \tilde{\gamma}_m \\ \tilde{\gamma}_m & -\gamma_m \end{array}\right),\,\,\,\,\,\,\,\breve{\alpha}_m\equiv \left( \begin{array}{cc}\alpha_m  & \tilde{\alpha}_m \\ \tilde{\alpha}_m & -\alpha_m \end{array}\right).
\eeq
Here all matrix elements are real and their explicit expressions can be easily found by using Supplementary Eqs.~(\ref{eq:21}), (\ref{eq:22}), (\ref{eq:23}) and (\ref{eq:24}). In
\ref{ident} we prove the following identity:
\begin{equation}\label{eq:105}
  \gamma_m \alpha_m + \tilde{\gamma}_m \tilde\alpha_m =0.
\end{equation}
With its help one can readily show that
\beq
(\check{\Gamma}'_m)^r \breve{\alpha}_m=\left\{ \begin{array}{lr}(\gamma_m^2+\tilde{\gamma}_m^2)^p\breve{\alpha}_m\,, & \text{for \ } r=2p,\\
i(\gamma_m^2+\tilde{\gamma}_m^2)^p (\gamma_m\tilde{\alpha}_m-\tilde{\gamma}_m\alpha_m)\left(
                                                                                       \begin{array}{cc}
                                                                                         0 & 1 \\
                                                                                         -1 & 0 \\
                                                                                       \end{array}
                                                                                     \right)
\,, & \,\,\text{for\ }r=2p+1.
\end{array}
\right.
\label{eq:117}
\eeq
As a result,
\beq
\sum_{i_b\in A_b}\sum_\sigma\left(\ln \left(C_0^{-1}-\mathbb{I}\right)\partial_{\varphi_m}C_1\right)_{i_b\sigma,i_b\sigma}=0.
\label{eq:121}
\eeq
So, only the edge term contributes to ${\rm Tr}_A\left(\ln \left(C_0^{-1}-\mathbb{I}\right)\partial_{\varphi_m}C_1\right)$.

For the edge term, we have
\beq\label{eq:120}
&&\sum_{i_e\in A_e}\sum_\sigma\left(\ln \left(C_0^{-1}-\mathbb{I}\right)\partial_{\varphi_m}C_1\right)_{i_e\sigma,i_e\sigma}\nonumber\\
&\simeq&\frac{1}{L}\biggl(1-{1\over 2}\,\delta_{m0}\biggr)\sum_{i_e\in A_e}\sum_{i=1}^{L_A}
\textrm{tr} \biggl[  -  \biggl((\ln(C_0^{-1}-\mathbb{I}))_{i_ei}\,\check{\alpha}_{m}{\rm e}^{ik_m(i-i_e)}+(m\rightarrow-m)\biggr)\sin\varphi_{m}\nonumber\\
&&\qquad\qquad\qquad\qquad\qquad\qquad\,+ \biggl((\ln(C_0^{-1}-\mathbb{I}))_{i_ei}\,\check{\beta}_{m}{\rm e}^{ik_m(i-i_e)}+(m\rightarrow-m)\biggr)\cos\varphi_{m}\biggr].
\eeq
Note that the edge size $L_e$ does not scale with $L_A$. Because the matrix block $(\ln(C_0^{-1}-\mathbb{I}))_{ij}$ decays rapidly with $|i-j|$, we can simplify Supplementary Eq.~(\ref{eq:120}) as
\beq\label{eq:219}
&&\sum_{i_e\in A_e}\sum_\sigma\left(\ln \left(C_0^{-1}-\mathbb{I}\right)\partial_{\varphi_m}C_1\right)_{i_e\sigma,i_e\sigma}\nonumber\\
&\simeq&\frac{1}{L}\biggl(1-{1\over 2}\,\delta_{m0}\biggr)\sum_{i_e\in A_e}
\textrm{tr} \biggl[  -  \biggl((\ln(C_0^{-1}-\mathbb{I}))_{i_ei_e}\,(\check{\alpha}_{m}+(m\rightarrow-m))\biggr)\sin\varphi_{m}\nonumber\\
&&\qquad\qquad\qquad\qquad\qquad\,\,+ \biggl((\ln(C_0^{-1}-\mathbb{I}))_{i_ei_e}\,(\check{\beta}_{m}+(m\rightarrow-m))\biggr)\cos\varphi_{m}\biggr].
\eeq
Since $(\ln(C_0^{-1}-\mathbb{I}))_{i_ei_e}$ is independent of $i_e$, we can write it as $\varsigma\otimes \mathbb{I}$ (with $\mathbb{I}$ being the unit matrix in the sublattice sector), and write Supplementary Eq.~(\ref{eq:219}) as
\beq\label{eq:220}
&&\sum_{i_e\in A_e}\sum_\sigma\left(\ln \left(C_0^{-1}-\mathbb{I}\right)\partial_{\varphi_m}C_1\right)_{i_e\sigma,i_e\sigma}\nonumber\\
&\simeq&\frac{L_e}{L}\biggl(1-{1\over 2}\,\delta_{m0}\biggr)\textrm{tr} \biggl[  -  \biggl(\varsigma(\check{\alpha}_{m}+(m\rightarrow-m))\biggr)\sin\varphi_{m}+ \biggl(\varsigma(\check{\beta}_{m}+(m\rightarrow-m))\biggr)\cos\varphi_{m}\biggr].
\eeq
Combining this result with Supplementary Eq.~(\ref{eq:121}) we obtain
\begin{eqnarray}\label{eq:154}
  \left\langle\left(\textrm{Tr}_A \left(\ln \left(C_0^{-1}-\mathbb{I}\right)\partial_{\varphi_m}C_1\right)\right)^2\right\rangle
  ={1\over 2}\left(\frac{L_e}{L}\biggl(1-{1\over 2}\,\delta_{m0}\biggr)\right)^2\left(I_\alpha\left(k_m\right)+I_\beta\left(k_m\right)\right),
\end{eqnarray}
where
\begin{eqnarray}
\label{eq:155}
  I_\alpha\left(k_m\right)&\equiv&\left(\textrm{tr}
  \biggl(\varsigma(\check{\alpha}_{m}+(m\rightarrow-m))\biggr)\right)^2,\nonumber\\
  I_\beta\left(k_m\right)&\equiv&\left(\textrm{tr} \biggl(\varsigma(\check{\alpha}_{m}+(m\rightarrow-m))\biggr)\right)^2.
\end{eqnarray}
Then we substitute Supplementary Eq.~(\ref{eq:154}) into Supplementary Eq.~(\ref{eq:111}). For $L\gg 1$ the sum over $m$ converges to the integral over the momentum. We finally obtain
\beq
{\rm Var}_1(S)=a L^{-1},
\label{eq:134}
\eeq
where the proportionality coefficient
\begin{equation}\label{eq:156}
  a={const.\over 4}\,L^2_e\int_{-\pi}^{\pi}{dk\over 2\pi}\left(I_\alpha\left(k\right)+I_\beta\left(k\right)\right).
\end{equation}
It is important that $a$ depends neither on $L$ nor on $L_A$.

\subsection{The contribution $\boldsymbol{{\rm Var}_2(S)}$}
\label{sec:calculations_var_S2}

For the moment let us ignore the second line in Supplementary Eq.~(\ref{eq:110}) in order to simplify discussions. Because $((\mathbb{I}-C_0)C_0)^{-1}_{ij}$ decays rapidly with $|i-j|$, we have
\begin{eqnarray}\label{eq:122}
  \textrm{Tr}_A\left(\left((\mathbb{I}-C_0)C_0\right)^{-1}C_1\partial_{\varphi_m}C_1\right)
  &\approx& \sum_{i,j=1}^{L_A}{\rm tr}\left(\left((\mathbb{I}-C_0)C_0\right)_{ii}^{-1}(C_1)_{ij}\partial_{\varphi_m}(C_1)_{ji}\right)\nonumber\\
  &=& \sum_{i,j=1}^{L_A}{\rm tr}\left(\kappa\,(C_1)_{i-j}\partial_{\varphi_m}(C_1)_{j-i}\right).
\end{eqnarray}
Here $\kappa\equiv ((\mathbb{I}-C_0)C_0)_{ii}^{-1}$ which is independent of $i$. With the substitution of Supplementary Eq.~(\ref{eq:100}), it is written as
\beq
&&\textrm{Tr}_A\left(\left((\mathbb{I}-C_0)C_0\right)^{-1}C_1\partial_{\varphi_m}C_1\right)={1\over L^2}\left(1-{1\over 2}\delta_{m0}\right)\sum_{i,j=1}^{L_A}\sum_{n=0}^{N-1}\left(1-{1\over 2}\delta_{n0}\right)\nonumber\\
&&\qquad\qquad\times \bigg(-\textrm{tr}\left(\kappa \check{\mathcal{J}}_{cs}(i-j,n,m)\right)  \cos\varphi_n  \sin\varphi_m
+\textrm{tr}\left(\kappa \check{\mathcal{J}}_{sc}(i-j,n,m)\right)  \sin\varphi_n  \cos\varphi_m\nonumber\\
&&\qquad\qquad\qquad+\textrm{tr}\left(\kappa \check{\mathcal{J}}_{cc}(i-j,n,m)\right)  \cos\varphi_n  \cos\varphi_m
-\textrm{tr}\left(\kappa \check{\mathcal{J}}_{ss}(i-j,n,m)\right)  \sin\varphi_n  \sin\varphi_m\bigg),
\label{eq:125}
\eeq
where
\begin{eqnarray}
\check{\cal J}_{cs}(l,n,m)&\equiv&\left({\rm e}^{{ik_n l}}\check{\alpha}_n+(n\rightarrow -n)\right)\left({\rm e}^{-{ik_m l}}\check{\alpha}_m+(m\rightarrow -m)\right),\nonumber\\
\check{\cal J}_{sc}(l,n,m)&\equiv&\left({\rm e}^{{ik_n l}}\check{\beta}_n+(n\rightarrow -n)\right)\left({\rm e}^{-{ik_m l}}\check{\beta}_m+(m\rightarrow -m)\right),\nonumber\\
\check{\cal J}_{cc}(l,n,m)&\equiv&\left({\rm e}^{{ik_n l}}\check{\alpha}_n+(n\rightarrow -n)\right)\left({\rm e}^{-{ik_m l}}\check{\beta}_m+(m\rightarrow -m)\right),\nonumber\\
\check{\cal J}_{ss}(l,n,m)&\equiv&\left({\rm e}^{{ik_n l}}\check{\beta}_n+(n\rightarrow -n)\right)\left({\rm e}^{-{ik_m l}}\check{\alpha}_m+(m\rightarrow -m)\right)
\label{eq:124}
\end{eqnarray}
are matrices in the sublattice sector and are even in $m,n$. Taking the square of Supplementary Eq.~(\ref{eq:125}), we obtain $16$ terms. They have the form as $\sim \mathfrak{g}_1(\varphi_n)\mathfrak{g}_2(\varphi_m)\mathfrak{h}_1(\varphi_{n'})\mathfrak{h}_2(\varphi_m)$, with $\mathfrak{g}_1(\varphi_n)\mathfrak{g}_2(\varphi_m)$ from one trace and $\mathfrak{h}_1(\varphi_{n'})\mathfrak{h}_2(\varphi_m)$ from the other. Here $\mathfrak{g}_{1,2}$ and $\mathfrak{h}_{1,2}$ stand for the symbols of functions: $\cos,\,\sin$. It is easy to see that only the four terms, with $\mathfrak{g}_i=\mathfrak{h}_i\, (i=1,2)$ and $n=n'$, dominate the $\boldsymbol{\varphi}$-averaged square. The average of other terms either vanishes or is smaller by an order of $1/L$. Taking these into account, we have
\beq
&&\sum_{m=0}^{N-1}\left\langle\left[\textrm{Tr}_A\left(\left((\mathbb{I}-C_0)C_0\right)^{-1}C_1\partial_{\varphi_m}C_1\right)\right]^2\right\rangle\nonumber\\
&=&\frac{1}{16 L^4}\sum_{i,j=1}^{L_A}\sum_{i',j'=1}^{L_A}\sum_{\substack{m,n\neq 0, \\ m\neq n}}\, \biggl(\textrm{tr} \Bigl( \kappa \,\check{\mathcal{J}}_{cs}(i-j,n,m) \Bigr) \,\textrm{tr} \Bigl( \kappa \,\check{\mathcal{J}}_{cs}(i'-j',n,m) \Bigr)\nn\\
&&\textrm{\ \ \ \ \ \ \ \ \ \ \ \ \ \ \ \ \ \ \ \  }+(\check{\mathcal{J}}_{cs}\rightarrow \check{\mathcal{J}}_{sc})+(\check{\mathcal{J}}_{cs}\rightarrow \check{\mathcal{J}}_{cc})+(\check{\mathcal{J}}_{cs}\rightarrow
\check{\mathcal{J}}_{ss})\biggr),
\label{eq:126}
\eeq
where the last three terms in the bracket are obtained by replacing $\check{\mathcal{J}}_{cs}$ in the first term by respectively $\check{\mathcal{J}}_{sc}$, $\check{\mathcal{J}}_{cc}$ and $\check{\mathcal{J}}_{ss}$, and we have used the fact that the leading contributions to the sum over $m,n$ come from those terms with $nm\neq 0$.

Next, we perform the sum over the indexes $i,j,i',j'$. Let us observe each trace product in Supplementary Eq.~(\ref{eq:126}).
By Supplementary Eq.~(\ref{eq:124}) the first trace includes four terms. Each term carries a phase factor, and the four factors are different, which are ${\rm e}^{ \pm i{2\pi (n-m)(i-j)\over L}}$ and ${\rm e}^{ \pm i{2\pi (n+m)(i-j)\over L}}$. Similarly, each term in the second trace carries a phase factor, and the  four factors are ${\rm e}^{ \pm i{2\pi (n-m)(i'-j')\over L}}$ and ${\rm e}^{ \pm i{2\pi (n+m)(i'-j')\over L}}$. Therefore, the trace product has $4\times 4=16$ terms, each of which carries a phase factor ${\rm e}^{i\Xi}$, with the phase
\begin{equation}\label{eq:127}
  \Xi\equiv {2\pi\over L}\left(\eta_1(n+\eta_2 m)(i-j)+\eta'_1(n+\eta'_2 m)(i'-j')\right),\quad \eta_1,\,\eta_2,\,\eta'_1,\, \eta'_2=\pm 1.
\end{equation}
Due to $n\pm m\neq 0$ the phase factor ${\rm e}^{i\Xi}$ rapidly oscillates with $i,j,i',j'$. As a result, the sum over $i,j,i',j'$ is dominated by the `stationary phase' configuration, for which $\Xi$ vanishes, requiring $\eta_2=\eta_2'$ and $\eta_1=\eta_1', \,(i-j)+(i'-j')=0$ (or $\eta_1=-\eta_1', \,(i-j)-(i'-j')=0$). This gives
\beq
&&\sum_{m=0}^{N-1}\left\langle\left[\textrm{Tr}_A\left(\left((\mathbb{I}-C_0)C_0\right)^{-1}C_1\partial_{\varphi_m}C_1\right)\right]^2\right\rangle\nonumber\\
&=&\frac{L_A^3}{16 L^4} \sum_{\substack{m,n\neq 0, \\ m\neq n}}\,
\biggl(I_{\alpha\alpha}\left(k_m,k_n\right)+I_{\beta\beta}\left(k_m,k_n\right)+
I_{\alpha\beta}\left(k_m,k_n\right)+I_{\beta\alpha}\left(k_m,k_n\right)\biggr),
\label{eq:129}
\eeq
where
\begin{eqnarray}
\label{eq:128}
  I_{\alpha\alpha}\left(k_m,k_n\right) &\equiv & \left(\textrm{tr} \Bigl[ \kappa (\check{\alpha}_n\check{\alpha}_m+\check{\alpha}_{-n}\check{\alpha}_{-m})\Bigr]\right)^2+
\left(\textrm{tr}\Bigl[ \kappa (\check{\alpha}_n\check{\alpha}_{-m}+\check{\alpha}_{-n}\check{\alpha}_{m})\Bigr]\right)^2, \nonumber\\
  I_{\beta\beta}\left(k_m,k_n\right) &\equiv & \left(\textrm{tr} \Bigl[ \kappa (\check{\beta}_n\check{\beta}_m+\check{\beta}_{-n}\check{\beta}_{-m})\Bigr]\right)^2+
\left(\textrm{tr}\Bigl[ \kappa (\check{\beta}_n\check{\beta}_{-m}+\check{\beta}_{-n}\check{\beta}_{m})\Bigr]\right)^2, \nonumber\\
  I_{\alpha\beta}\left(k_m,k_n\right) &\equiv & \left(\textrm{tr} \Bigl[ \kappa (\check{\alpha}_n\check{\beta}_m+\check{\alpha}_{-n}\check{\beta}_{-m})\Bigr]\right)^2+
\left(\textrm{tr}\Bigl[ \kappa (\check{\alpha}_n\check{\beta}_{-m}+\check{\alpha}_{-n}\check{\beta}_{m})\Bigr]\right)^2, \nonumber\\
I_{\beta\alpha}\left(k_m,k_n\right) &\equiv & \left(\textrm{tr} \Bigl[ \kappa (\check{\beta}_n\check{\alpha}_m+\check{\beta}_{-n}\check{\alpha}_{-m})\Bigr]\right)^2+
\left(\textrm{tr}\Bigl[ \kappa (\check{\beta}_n\check{\alpha}_{-m}+\check{\beta}_{-n}\check{\alpha}_{m})\Bigr]\right)^2.
\end{eqnarray}
For $L\gg 1$ the sum over $m,n$ in Supplementary Eq.~(\ref{eq:129}) can be well approximated by the continuum limit. We obtain
\beq
\sum_{m=0}^{N-1}\left\langle\left[\textrm{Tr}_A\left(\left((\mathbb{I}-C_0)C_0\right)^{-1}C_1\partial_{\varphi_m}C_1\right)\right]^2\right\rangle=
b'L_A^3/L^2,
\label{eq:130}
\eeq
where
\beq
b'=\frac{const.}{16} \int\!\!\!\!\int_{-\pi}^{\pi}{dkdk'\over (2\pi)^2}
\biggl(I_{\alpha\alpha}\left(k,k'\right)+I_{\beta\beta}\left(k,k'\right)+
I_{\alpha\beta}\left(k,k'\right)+I_{\beta\alpha}\left(k,k'\right)\biggr).
\label{eq:133}
\eeq
It is important that $b'$ has no dependence on $L_A$ and $L$, and is completely determined by system's microscopic parameters.

Now let us retrieve the second line in Supplementary Eq.~(\ref{eq:110}). We repeat the derivations above. As a result, we find that only the coefficient $b'$ in Supplementary Eq.~(\ref{eq:130}) is modified and, similar to $b'$, the modified coefficient $b$ has no dependence on $L_A$ and $L$. Finally, we have
\beq
{\rm Var}_2(S)=bL_A^3/L^2.
\label{eq:131}
\eeq
Interestingly, the scaling behavior $\sim L_A^3/L^2$ coincides with a previous result about the entanglement entropy variance in a completely different context [29]. In that context fluctuations are due to randomly drawing a member from a free-fermion eigenstate ensemble; thus that result is kinematic and has nothing to do with the dynamics of entanglement.

\subsection{Universal scaling law for $\boldsymbol{{\rm Var}(S)}$}
\label{sec:scaling}

Combining Supplementary Eqs.~(\ref{eq:134}) and (\ref{eq:131}), we obtain the variance of the entanglement entropy,
\begin{equation}\label{eq:118}
  {\rm Var}(S)=aL^{-1}+bL_A^3/L^2.
\end{equation}
Importantly, from the derivations above we have seen that the first and second term describe a subsystem's edge and bulk effect, respectively. Upon the rescaling: $\tilde{L}\equiv L/\ell,\,\tilde{L}_A\equiv L_A/\ell$ ($\ell\equiv \sqrt{a/b}$), Supplementary Eq.~(\ref{eq:118}) is rewritten as
\begin{eqnarray}
{{\rm Var} (S)/s_0}=1/\tilde{L}+\tilde{L}_A^3/\tilde{L}^2
\label{eq:132}
\end{eqnarray}
with $s_0\equiv\sqrt{ab}$, which is Eq.~(2) in the main text (where we used the same symbols $L,\, L_A$ for notational simplicity). This result is universal with respect to system's detailed constructions and initial states (required to be Gaussian, however), that enter only into the microscopic parameters $\ell$ and $s_0$. It is thus suggested that the behaviors of entanglement entropy fluctuations are completely controlled by two dimensionless macroscopic lengths, $\tilde{L}$ and $\tilde{L}_A$.

Supplementary Eq.~(\ref{eq:132}) implies that in the regime of $\tilde{L}_A\ll \tilde{L}^{1/3}$, the edge term dominates over the bulk term and ${\rm Var}(S)\sim
\tilde{L}^{-1}$. Thus in this regime the fluctuations are very weak. In the special case of $\tilde{L}\rightarrow \infty$ at fixed $\tilde{L}_A$, the variance vanishes. Combined with the fact that $S(t)$ oscillates quasiperiodically, this result implies that fluctuations are fully suppressed and the entanglement entropy is strictly a constant beyond some critical time, in agreement with a celebrated result in entanglement evolution [1, 3]; see
\ref{sec:infinite_L} for details. In the regime of $\tilde{L}_A\gg \tilde{L}^{1/3}$ the fluctuation behaviors are totally opposite. In this case the bulk term dominates over the edge term and ${\rm Var}(S)\sim\tilde{L}_A^3/\tilde{L}^2$. So when the ratio $\tilde{L}_A/\tilde{L}=L_A/L$ is fixed, the variance increases linearly with $\tilde{L}$, and thus enlarging the entire system can drive very strong entanglement fluctuations. This is contrary to the belief [30] that temporal fluctuations observed in experiments on entanglement evolution of finite systems would eventually diminish by increasing the system size. Thus, we see that the behaviors of out-of-equilibrium entanglement entropy fluctuations can be completely different in approaching the limit $L\rightarrow\infty$, depending on how $L_A$ scales with $L$.

The situations above resemble those in the conductance fluctuations of quasi-one-dimensional disordered wires in several aspects. First, the conductance fluctuations are also controlled by some dimensionless macroscopic parameter, namely, the sample length rescaled by the localization length, and the microscopic details of the wire, notably, the disorder strength, enter only into the localization length. Second, varying the rescaled sample length leads to distinct fluctuation behaviors: Increasing the sample length drives the wire from a metallic regime, where the universal conductance fluctuations [5, 6] follow, to a localized regime, where the conductance distribution is very broad so that the variance of logarithmic conductance increases with the sample length linearly [10].

\subsection{The distribution tail}
\label{sec:distribution_EE}

By the general results obtained in Supplementary Note $5.1$,
the upper tail of the distribution is controlled by two parameters $(b_+,c_+)$ and the lower by $(b_-,c_-)$. While to calculate $c_\pm$ analytically is an intractable task, we resort to numerical analysis. We compute the function $\delta F/{dG\over du}$ with the use of {\it ab initio} data of $S(\boldsymbol{\varphi})$ obtained from numerical simulations (see \ref{sec:numerical_tests} for details). Then, according to Supplementary Eqs.~(\ref{sobo3}) and (\ref{eq:149}) the maximal value of $\delta F/{dG\over du}$ over positive $u$ gives $c_+$ while the minimal value over negative $u$ gives $c_-$. We also compute $b_\pm$ numerically by directly using the definitions Supplementary Eq.~(\ref{eq:46}). Supplementary Table \ref{table:2} gives the numerical value of $(b_\pm,\,c_\pm)$ for different $L$ and the ratio $L_A/L$.

We find that for all $L_A/L,\, L$ considered $c_\pm$ are negative. (The only exception is that for $L_A/L=0.5$, $c_+$ is slightly positive, and the reasons are under investigations.) Consequently, by the results obtained in Supplementary Note $5.1$
the distribution displays a sub-Gaussian upper and a sub-Gamma lower tail, described by the concentration inequalities, namely, Supplementary (\ref{eq:147}) and (\ref{eq:150}), respectively. We also see that $b_+\approx b_-$ for all $L_A/L,\,L$, consistent with the analysis in Supplementary Note $5.2$.
Furthermore, Supplementary Table \ref{table:2} shows that $b_-$ decreases rapidly as $L_A/L$ decreases from $0.5$, whereas $c_-$ does not change too much. This implies that as $L_A/L$ decreases the distribution becomes more and more asymmetric. Specifically, for $L_A/L=0.5$ it is near symmetric and near Gaussian; as $L_A/L$ decreases the upper tail is always sub-Gaussian, and with more and more weights transferred from the lower to the upper tail, a heavier and heavier sub-Gamma lower tail develops.

\begin{table}[b]
\centering
  \caption{The value of $(b_\pm,\,c_\pm)$ for different $L$ and $L_A/L$ for the Rice-Mele model with quench parameters $(J,J',M):(1,0.5,0.5)\rightarrow (1,0.5,4.231)$.}\label{table:2}

    \begin{tabular}{c|ccc|ccc}
     \hline
     \hline
             &  &  $L=25$ &  &  & $L=50$  &
\\
     $\quad$$L_A/L\quad$ &   $\quad \langle S \rangle $  & $\quad (b_-,\,c_-)\quad$ & $\quad (b_+,\,c_+)\qquad$ & $\quad$ $\langle S \rangle \quad$  & $\quad (b_-,\,c_-)\quad\quad$ & $\quad (b_+,\,c_+)\quad\quad$ \\
     \hline
     $0.05$&$\quad 21.8$ & $\quad(0.2,\,-0.2)\quad$&$(0.2,\,-0.1)\quad$&$43.6$ &$\quad(0.5,\,-0.2)\quad$&$(0.5,\,-0.1)$\\
     $0.1$ & $\quad20.9$ & $\quad(0.9,\,-0.3)\quad$&$(0.9,\,-0.3)\quad$ &$41.8$ &$(1.8,\,-0.3)$&$(1.8,\,-0.2)$\\
     $0.2$ & $\quad19.0$ & $\quad(3.3,\,-0.5)\quad$&$(3.5,\,-0.4)\quad$&$38.0$ &$(6.6,\,-0.5)$&$(6.9,\,-0.4)$\\
     $0.3$ &$\quad 17.0$ & $\quad(6.5,\,-0.6)\quad$&$(7.0,\,-0.3)\quad$& $34.1$&$(12.8,\,-0.6)$&$(13.9,\,-0.3)$\\
     $0.4$ & $\quad14.7$ & $\quad(9.2,\,-0.5)\quad$&$(10.5,\,-0.1)\quad$& $29.8$&$(17.9,\,-0.5)$&$(20.8,\,-0.1)$\\
     $0.5$ &  $\quad12.5$ & $\quad(9.6,\,-0.3)\quad$&$(11.2,\,0.2) \quad$& $25.0$&$(18.7,\,-0.3)$&$(22.3,\,0.2)$\\
     \hline
      \hline
    \end{tabular}

\begin{tabular}{c}

\\

\end{tabular}

\begin{tabular}{c|ccc}
     \hline
     \hline
          &  & $L=100$  &
\\
     $\quad$$L_A/L\quad$ &   $\quad \langle S \rangle $  & $\quad (b_-,\,c_-)\quad$ & $\quad (b_+,\,c_+)\qquad$ \\
     \hline
     $0.05$&$\quad 87.2$ &$\quad (0.9,\,-0.2) \quad$&$(0.9,\,-0.1)$\\
     $0.1$ & $\quad83.7$ &$(3.6,\,-0.3)$&$(3.6,\,-0.2)$\\
     $0.2$ &$\quad76.1$ &$(13.2,\,-0.5)$&$(13.8,\,-0.3)$\\
     $0.3$ & $\quad68.2$&$(25.6,\,-0.6)$&$(28.1,\,-0.3)$\\
     $0.4$ & $\quad59.4$&$(36.1,\,-0.5)$&$(41.7,\,-0.1)$\\
     $0.5$ &  $\quad50.0$&$(37.0,\,-0.3)$&$(44.5,\,0.1)$\\
     \hline
      \hline
    \end{tabular}
\end{table}

\begin{figure}[t]
\begin{center}
\includegraphics[width=14.cm]{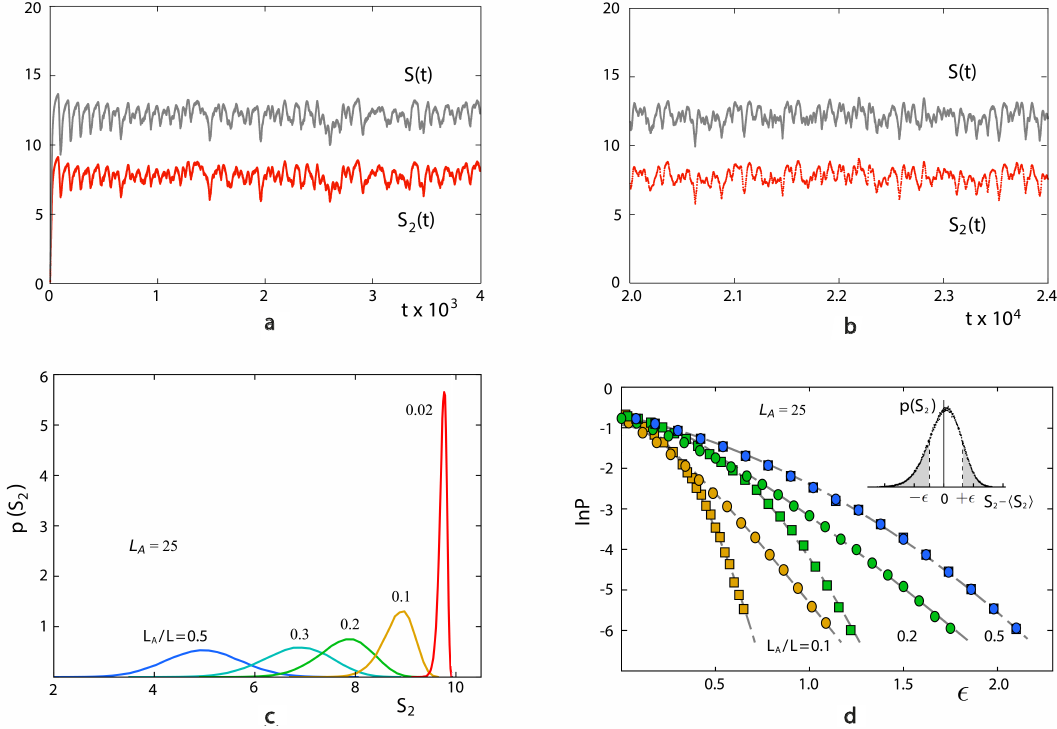}
\end{center}
\caption{{\bf Statistical distribution of out-of-equilibrium fluctuations of R$\acute{\rm e}$nyi entropy.} Numerical simulations of the time evolution of the second-order R$\acute{\rm e}$nyi entropy $S_2(t)$ upon quench. The patterns (red) in different time windows are shown in (a,b), where the patterns of $S(t)$ (grey, same as Fig.~1 in the main text) are also shown for comparison. The evolution of the distribution with $L$ at fixed $L_A$ is shown in (c). For different ratios $L_A/L=0.1$ (yellow), $0.2$ (green) and $0.5$ (blue), numerical data for the upper (squares) and lower (circles) tail are well fitted by the form given by the first and the second line of Supplementary Eq.~(\ref{eq:152}) (dashed lines), respectively (d). The subsystem size $L_A=25$ and for (a,b) the total system size $L=124$.}
\label{fig:S1a}
\end{figure}

The tail behaviors described above can be understood in a simple picture based on the concept of the coherent entangled quasiparticle pair [1] as follows. Upon quench such pairs are locally created across the whole system. Because they have opposite group velocities and, moreover the system is finite, so that a quasiparticle reenters into the system after reaching system's boundary, after long time a stationary configuration of entangled pairs is form. In this configuration the pairs are randomly distributed in the system subjected to Pauli's exclusion principle. Let $\delta$ be the probability for a quasiparticle inside subsystem $A$ to pair with
another outside --- an event contributing to the entanglement entropy $S$, and $(1-\delta)$ be the probability for the exclusive event, namely, an entangled pair staying inside the subsystem A. Then $S$ is given by the total `successful' events for a fixed, but large, number of trials, resulting in a binomial distribution. Now, for $L_A/L=1/2$ an entangled pair has equally the same probability to cross subsystem's boundary or stay inside A, i.e. $\delta=1/2$; it is well known that in this case the binomial distribution reduces to a Gaussian distribution. For small $L_A/L$, $\delta\approx 1$ and thus for an entangled pair to stay in A is a rare event. This results in a heavier lower tail of the $S$ distribution which is exponential, in the same fashion as the binomial distribution approaches its Poissonian limit. So this picture well explains the asymptotic behaviors of the tail bounds and their evolution with $L_A/L$. This suggests that the bounds yield the form of large deviation probability for the upper and the lower tail, respectively. That is, the Supplementary inequalities (\ref{eq:147}) and (\ref{eq:150}) can be promoted to equalities, i.e. for sufficiently large $\epsilon$,
\begin{eqnarray}
\mathbf{P}(|S-\langle S\rangle| \geq  \epsilon)=\left\{
\begin{array}{ll}
  {\rm e}^{-{\epsilon^2\over 2\mathfrak{b}_+}},&{\rm for}\, S-\langle S\rangle>0 \\
  {\rm e}^{-{\epsilon^2\over 2(\mathfrak{b}_- + \mathfrak{c} \epsilon)}},&{\rm for}\, S-\langle S\rangle<0
\end{array}
\right.,
\label{eq:152}
\end{eqnarray}
with the parameters $\mathfrak{b}_\pm \propto b_\pm$ and $\mathfrak{c} \propto |c_-|$ and the proportionality coefficients being absolute constants. This gives Eq.~(1) in the main text, which was confirmed by simulations (see Fig.~2(c) in the main text). In Supplementary Note $7.2$
we provide further simulation results to show that Supplementary Eq.~(\ref{eq:152}) holds also for the second-order R$\acute{\rm e}$nyi entropy $S_2$.

Let us further discuss semi-quantitatively how the rate of the exponential decay $\sim {\rm e}^{-{\epsilon\over 2\mathbf{c}}}\, (\epsilon\gg \mathfrak{b}_-/\mathfrak{c})$ of the lower tail behaves, when $L$ approaches the infinite and the ratio $L_A/L$ is fixed. First of all, Supplementary Eq.~(\ref{eq:118}) implies that ${\rm Var}(S)\sim L$ for sufficiently large $L$ and fixed $L_A/L$. Then, expanding $dG/du$ in $u$ we obtain $dG/du=u{\rm Var}(S)+{\cal O}(u^2)$. So the coefficient of the leading term is $\propto L$, and thus the leading expansion of $L^{-1}dG/du$ is well defined in the limit $L\rightarrow\infty$. It is natural to expect that the entire expansion of $L^{-1}dG/du$ is well defined in such limit, although this is difficult to prove. This implies that all the coefficients of higher order terms cannot grow faster than $L$. Next, according to Supplementary Eq.~(\ref{eq:140}), $\delta F(u)$ is a sum of $N\sim L$ terms. Thus we estimate $\delta F(u)\sim L$. Finally, combining the estimations for $dG/du$ and $\delta F(u)$ with the Supplementary inequality (\ref{eq:149}), we find that $\mathfrak{c}={\cal O}(L^0)$, i.e. depends only on the ratio $L_A/L$. This is in agreement with the data in Supplementary Table \ref{table:2}.


\section{Numerical simulations of noninteracting models}
\label{sec:numerical_tests}

In this Supplementary Note we provide a complete description of numerical simulations and report extended numerical results for noninteracting models.

\subsection{The entanglement entropy $\boldsymbol{S}$}
\label{sec:numerical_EE}

We first diagonalize numerically for each time $t$ the correlation matrix $C(t)$ of Supplementary Eq.~(\ref{eq:corr}) to obtain its eigenvalues $\{p_{\nu}(t)\}_{\nu=1}^{2 L_A}$. We then substitute them into Supplementary Eq. (\ref{eq:33}) to obtain the instantaneous $S(t)$. Upon varying $t$, the pattern $S(t)$ shown in Figs.~1(a) and 1(b) of the main text is obtained. The time is in unit of $\hbar/J$, where $J$ is the amplitude of hopping between nearest $\A$- and $\B$-site.

To obtain the statistics of the random entanglement entropy $S(\boldsymbol{\varphi})$ we simulate an ensemble of random correlation matrices $\tilde{C}(\boldsymbol{\varphi})$. Specifically, we generate $N$ random angles $\varphi_m$ ($m=0,1,\ldots, N-1$) uniformly drawn from the interval $[0,2\pi)$, which gives a disorder realization $\boldsymbol{\varphi}=(\varphi_0,\varphi_1,\ldots,\varphi_{N-1})$, and substitute them into Supplementary Eq.~(\ref{eq:27}). For each $\boldsymbol{\varphi}$ we diagonalize numerically $\tilde{C}(\boldsymbol{\varphi})$ to obtain the eigenvalue spectrum $\{p_{\nu}(\boldsymbol{\varphi})\}_{\nu=1}^{2 L_A}$. Analogous to Supplementary Eq. (\ref{eq:33}), we obtain the corresponding value of $S(\boldsymbol{\varphi})$. The statistics of $S(\boldsymbol{\varphi})$ is obtained for an ensemble consisting of $5\times 10^5$ disorder realizations. The distribution of $S(\boldsymbol{\varphi})$ is shown by the dashed line in Fig.~1(a) of the main text.

When the eigenfrequencies $\omega_0,\omega_1,\ldots,\omega_{N-1}$ are incommensurate, generic for the energy eigenspectrum, the statistics of the time series $S(t)$ and the random function $S(\boldsymbol{\varphi})$ are equivalent, as shown by Supplementary Eq.~(\ref{eq:43}). In Fig.~1(a) of the main text, the dash line and the histogram agree with each other, independent of the time interval used in the sampling of $S(t)$. When $\omega_0,\omega_1,\ldots,\omega_{N-1}$ are commensurate, this statistical equivalence does not hold; see \ref{sec:breakdown_ergodicity}.

\subsection{The $\boldsymbol{n}$th-order R${\acute{\rm {\bf e}}}$nyi entropy $\boldsymbol{S_n}$}
\label{sec:numerical_RE}

As an example for more general entanglement probes discussed in Supplementary Note $2.2$,
we numerically compute the $n$th-order R$\acute{\rm e}$nyi entropy according to Supplementary Eqs.~(\ref{eq:92}) and ~(\ref{eq:93}). We find that its behaviors are similar to those of the entanglement entropy. As shown in Supplementary Fig.~\ref{fig:S1a}, first, after initial growth and damped oscillations, $S_2(t)$ displays quasiperiodic oscillations, fluctuating around its average value, see (a) and (b); second, at fixed $L_A$ the distribution is broadened as $L$ decreases (c), and its upper and lower tail are sub-Gaussian and sub-Gamma, respectively (d). Furthermore, as shown in Supplementary Fig.~\ref{fig:S2}, simulations confirm the relation Supplementary Eq.~(\ref{eq:135}) for $S_n$, for $n=2,3,10$. Even more surprisingly, simulations show that the scaling law Supplementary Eq.~(\ref{eq:118}) holds also for the variance ${\rm Var}(S_n)$ (b-d). These findings confirm that the theory developed in this work applies to general entanglement probes, not limited to the entanglement entropy.

\subsection{Numerical computation of $\boldsymbol{(b^{\pm},c^{\pm})}$}
\label{sec:concentration_inequality_parameters}

In Supplementary Note $5.1$,
we introduce the variance factor and the scale parameter $(b^{\pm},c^{\pm})$ in solving the modified logarithmic Sobolev inequality, i.e. Supplementary (\ref{sobo1}), for the logarithmic moment-generating function $G(u)$ of $S(\boldsymbol{\varphi})$. The $+$ sign corresponds to the upper tail distribution and $-$ to the lower. Here we describe the numerical method for computing these parameters summarized in Supplementary Table \ref{table:2}. For $b^{\pm}$, following Supplementary Eq.~(\ref{eq:46}):
\begin{itemize}
  \item We realize a disorder realization $\boldsymbol{\varphi}$ and compute $S(\boldsymbol{\varphi})$, in precisely the same way as what is described above.
  \item With respect to this $\boldsymbol{\varphi}$, we vary its $m$-th component $\varphi_m$ over $[0,2\pi)$ while keeping other components, i.e. $(\varphi_0,\ldots,\varphi_{m-1},\varphi_{m+1},\ldots,\varphi_{N-1})$ fixed, and subsequently find the minimum (maximum) of $S(\boldsymbol{\varphi})$, which determines $S_{m}^{\pm}$.
  \item  With the same $\boldsymbol{\varphi}$, we repeat the second step for each component and obtain $N$ values of $S_{m}^{\pm}$, and then determine $\sum_{m=0}^{N-1} (S-S_{m}^{\pm})^2$.
  \item We repeat the previous three steps for $2\times 10^3$ disorder realizations $\boldsymbol{\varphi}$ and average the outcome.
\end{itemize}
To obtain the scale parameter $c^{\pm}$, we numerically evaluate the two quantities $\delta F(u)$ and $dG(u)/du$ for a range of $u$  around the origin. $c_+$ is determined by the maximum of $\delta F/(dG/du)$ in the $u>0$ interval, whereas $c_-$ is determined by the corresponding minimum in the $u<0$ interval.

\begin{figure}[t]
\begin{center}
\includegraphics[width=14.cm]{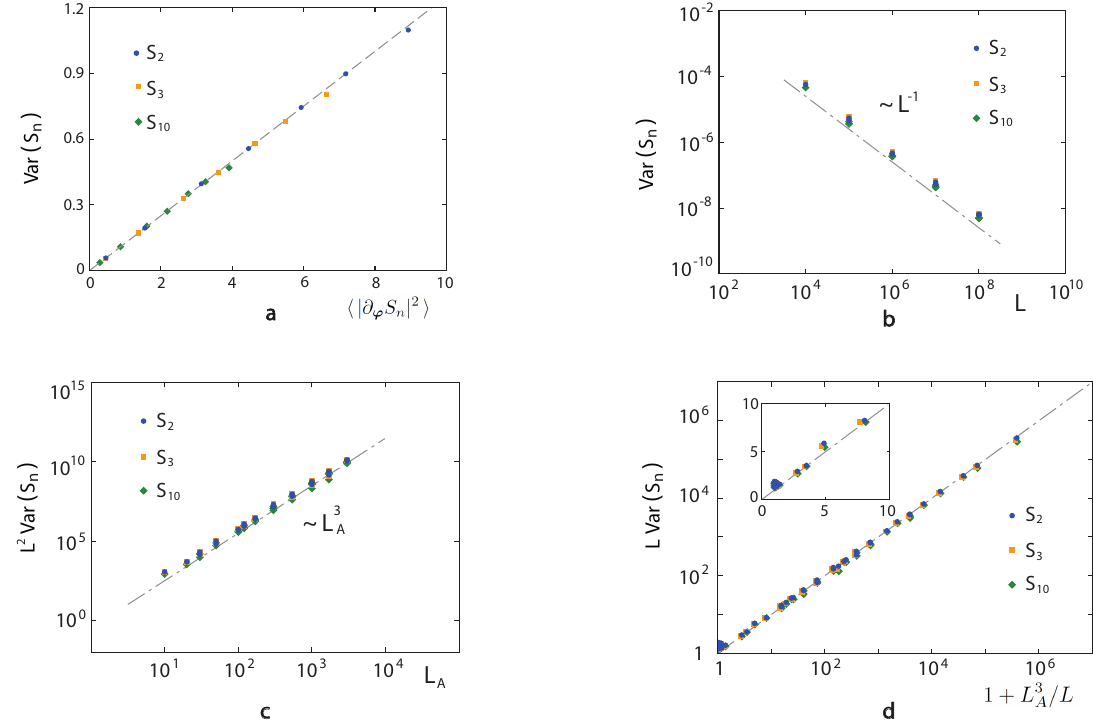}
\end{center}
\caption{{\bf Universal scaling behaviors of the variance of R${\acute{\rm {\bf e}}}$nyi entropies.} Simulation results for the variance of $n$th-order R${\acute{\rm {\bf e}}}$nyi entropy $S_n$, $n=2,3,10$. For each $S_n$, we vary either $L_A$ or $L$, while keeping the other fixed, in the same way as we generate the data for Fig. 3 in the main text. (a) The results confirm the relation Supplementary Eq.~(\ref{eq:135}). (b)-(d) They confirm the first (b) and the second (c) terms, and the overall (d) of the scaling law Supplementary Eq.~(\ref{eq:118}). All theoretical predictions are presented by dashed lines. In (d) ${\rm Var}(S_n),\,L,\, L_A$ are all rescaled in the manner similar to what is performed in the scaling analysis of ${\rm Var}(S)$.}
\label{fig:S2}
\end{figure}

\begin{figure}[t]
\begin{center}
\includegraphics[width=15.cm]{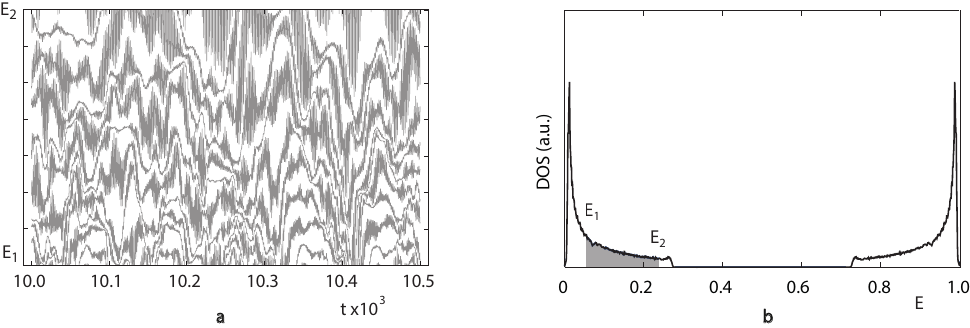}
\end{center}
\caption{{\bf Spectral structures of the instantaneous correlation matrix} $\boldsymbol{C(t).}$ (a) Zoom-out view of the evolving spectrum of $C(t)$. (b) The density of states (DOS) of $C(t)$ shown in arbitrary unit (a.u.) scale. The spectrum lies within the interval $[0,1]$ and it  is particle-hole symmetric. The shaded area is the spectral window where we perform the level statistics shown in Fig.~1(c) of the main text. }
\label{fig:S3}
\end{figure}

\begin{figure}[t]
\begin{center}
\includegraphics[width=16.cm]{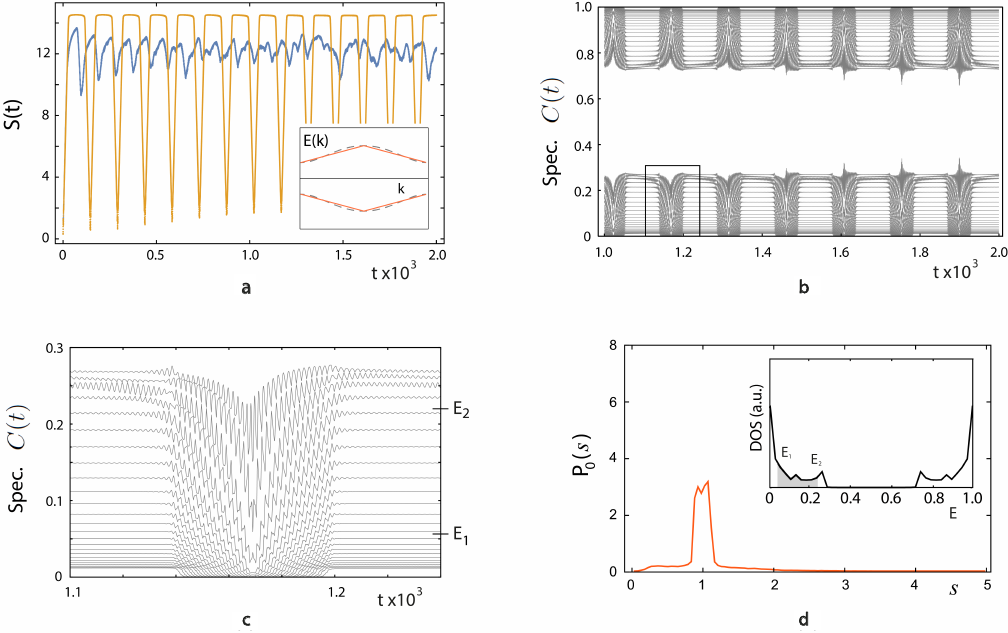}
\end{center}
\caption{{\bf Results for commensurate} $\boldsymbol{\omega.}$ (a) Numerical simulations confirm that, when $\boldsymbol{\omega}$ is commensurate, the pattern of the entanglement entropy is periodic and displays complete revival at multiple periods (yellow curves). The pattern in the incommensurate $\boldsymbol{\omega}$ case is shown for comparison (blue curves, same as Fig.~1 in the main text). Inset of (a) shows the energy spectrum with commensurate energy gap $\omega_m=\omega_0+ m \, r$, $\omega_0=3.162$, $r=0.043$, $m=0,\ldots,62$. The dash lines are the physical energy spectrum Supplementary Eq.~(\ref{eq:11a}). The total system size $L=124$ and the subsystem size $L_A=25$. (b) The spectrum of the corresponding evolving correlation matrix with a zoom-in view in (c). (d) The nearest-neighbor spacing distribution and the DOS (inset). The shaded area in the inset is the spectral window used for statistical analysis.}
\label{fig:S4}
\end{figure}

\subsection{Spectral statistics of the correlation matrix}
\label{sec:level_statistics_correlation_matrix}

We provide more details of the  nearest-neighbor spacing statistics obtained in Fig.~1(c) in the main text. In Supplementary Fig.~\ref{fig:S3}(a) we show the zoom-out view of the time evolving spectrum of the correlation matrix $C(t)$. In Supplementary Fig.~\ref{fig:S3}(b) we show the density of states of the full spectrum, with the shaded area indicating the spectral window used to obtain the nearest-neighbor spacing distribution.

\subsection{Quench parameters}
\label{sec:quench_parameters}

In the generic numerical simulations for the Rice-Mele (RM) model, we use the quench parameters $(J,J',M):(1,0.5,0.5)\rightarrow (1,1.5,1.5)$. For TFIC we use the quench parameter $h=3\rightarrow 5$, with $h$ being the external magnetic field. A detailed account of the dynamics of entanglement in TFIC is given in
\ref{sec:TFIC}. In Fig.~3 of the main text, when obtaining results of various quenches, we use:
\begin{itemize}
\item for RM model $(J,J',M)$ quench I: $(1,0.7,0.3)\rightarrow (0.3,1.0,0.001)$; II: $(0.6,1.0,0.8)\rightarrow (1.0,0.2,10)$.
\item for TFIC quench I: $h=3\rightarrow 2$; II:  $h=3\rightarrow 5$; III:  $h=5\rightarrow 2$.
\end{itemize}


\section{Entanglement evolution with commensurate $\boldsymbol{\omega}$}
\label{sec:breakdown_ergodicity}

So far we have studied the dynamics of entanglement where the frequencies governing the evolution of the correlation matrix $C(t)$ are incommensurate, i.e. $\boldsymbol{\omega}$ is incommensurate. In this Supplementary Note we study the commensurate case. In this case, by Supplementary Eq.~(\ref{eq:27a}), $C(t)$ is periodic in time. Thus all entanglement probes defined by Supplementary Eq.~(\ref{eq:101}) are also periodic in time, and thus display complete revival to the initial values at multiple periods.

Since it is impossible to tune parameters of the physical Rice-Mele model to obtain an energy spectrum with commensurate frequencies, we modify directly the frequencies in the dynamical phases. Specifically, at the Bloch momentum $k_m$ the frequency $\omega_m$ is set to $\omega_0+ m \,r$, with $r$ being a suitably chosen rational number and $\omega_0$ being the zero-momentum eigenfrequency in the incommensurate case. In this way, the resulting spectrum simulates as closely as the physical model (see inset of Supplementary Fig.~\ref{fig:S4}(a)). We assume the eigenstate properties are not modified in any qualitative way.

In Supplementary Fig.~\ref{fig:S4}(a), the result for the entanglement entropy evolution $S(t)$ confirms that it is periodic and displays complete revival to the initial value $S(0)$ at multiple periods (yellow curve). The time profile is completely different from the incommensurate case (blue curve), although in both cases the system has the same degrees of freedom.

We also show the evolving spectrum of $C(t)$ in Supplementary Figs.~\ref{fig:S4}(b) and ~\ref{fig:S4}(c), which clearly show a high degree of regularity. Finally, this regularity results in a very different nearest-neighbor level statistics shown in Supplementary Fig.~\ref{fig:S4}(d), in stark contrast to the incommensurate case giving the semi-Poissonian statistics shown in Fig.~1(c) of the main text.

\section{Generalization to XXZ model}
\label{sec:generalization_to_interacting_models}

In this Supplementary Note we generalize the analytical theory developed above to an important interacting system, namely, the antiferromagnetic XXZ spin-${1\over 2}$ chain. Furthermore, we perform numerical experiments to test analytical predictions obtained from the generalized theory. An XXZ model of size $L$ is described by the following Hamiltonian [36],
\beq
{H}_{\mathrm{xxz}} = \sum_{i=0}^{L-1} H_i,\quad H_i={S}_i^{x}{S}_{i+1}^{x} + {S}_i^{y}{S}_{i+1}^{y} + \Delta\left( {S}_i^{z}{S}_{i+1}^{z} - \frac{1}{4}\right),
\label{ham}
\eeq
where $S_i^{x,y,z}={1\over 2}\sigma_i^{x,y,z}$ are spin-${1\over 2}$ operators, with $\sigma^{x,y,z}_i$ being the Pauli matrices at site $i$, and $\Delta>0$ is the anisotropy parameter. When $\Delta=1$ this system reduces to the isotropic Heisenberg chain (namely, the XXX model), and a $SU(2)$ symmetry arises [37]. But it plays no roles in our work and thus we shall not discuss it further. Below we focus on small but not vanishing $\Delta$, and on even $L\gg 1$ and even $L_A$ with $L_A/L\ll 1$. The system is subjected to the periodic boundary condition $S_{L+1}^\alpha = S_1^\alpha$.

The generalized theory differs from the theory developed in previous Supplementary Notes in two key aspects. First, the correlation function plays no roles; rather, the generalized theory is built upon the reduced density matrix. Secondly, it makes no reference to whether the initial state is Gaussian or not. We can further generalize the theory below to other interacting models, such as the XXZ model with an integrability-breaking term and the antiferromagnetic chain with next-to-nearest neighbor interaction [38]. But we shall not discuss this here for simplicity.

\subsection{Emergence of mesoscopic fluctuations}
\label{sec:emergence_mesoscopic_fluctuations_XXZ}

\subsubsection{Reduced density matrix and entanglement dynamics}
\label{Sec:long-time_dynamics_entanglement}

Let the eigenenergy spectrum of ${H}_{\mathrm{xxz}}$ be $\{\omega_m\}$ and the eigenstates be $\{\Psi_m\}$. A generic initial state $\Psi(0)$ is superposed by $\Psi_m$'s,
\beq
\Psi(0) = \sum_{m}\chi_m\Psi_m,
\label{psi_0}
\eeq
where $\chi_m$'s are superposition coefficients. At later time $t$ the wavefunction is given by
\beq
\Psi(t) = \sum_{m}\chi_m \mathrm{e}^{-i \omega_m t} \Psi_m.
\label{evolution}
\eeq
At this state the density matrix reads
\beq
{\rho}(t) = \left| \Psi(t)\right\rangle\left\langle\Psi(t)\right| = \sum_{m,n}\mathrm{e}^{-i(\omega_{m}-\omega_{n})t}\,\chi_m \chi_n^\ast  \left| \Psi_m\right\rangle\left\langle\Psi_n\right|,
\label{eq:176}
\eeq
and the reduced density matrix  reads
\beq
{\rho}_A(t) = \sum_{m,n}\mathrm{e}^{-i(\omega_{m}-\omega_{n})t}\,\chi_m \chi_n^\ast  {\rm Tr}_{B}\left| \Psi_m\right\rangle\left\langle\Psi_n\right|.
\label{eq:177}
\eeq
The evolution of ${\rho}_A(t)$ provides full information on entanglement evolution. Notably, it gives the evolution of the entanglement entropy $S(t)$ and the $n$-th order R$\acute{\rm e}$nyi entropy $S_n(t)$ defined by Supplementary Eqs.~(\ref{eq:31}) and (\ref{eq:88}), respectively.

To study the evolution Supplementary Eq.~(\ref{eq:177}) it is necessary to understand the number of eigenbases excited by $\Psi(0)$, denoted as $D$. This number may be estimated as the participation ratio in the eigenstate space,
\beq
D \approx \left( \sum_m \left|\chi_m\right|^4 \right)^{-1}.
\eeq
It is much smaller than the dimension of full Hilbert space, i.e. $2^L = \mathrm{e}^{\ln2\,L}\approx\mathrm{e}^{0.7L}$, in general. But similar to the latter, $D$ increases exponentially with $L$, i.e.
\beq
D\sim\mathrm{e}^{\kappa_D L},\quad \kappa_D>0.
\label{dimension}
\eeq
To see this let us first consider a random $\Psi(0)$ without polarization (i.e. the total magnetization $S^z=\langle \Psi(0)|\sum_{i=0}^{L-1}S_i^z|\Psi(0)\rangle=0$); see Supplementary Note $9.5.1$
below for its detailed constructions. So,
\begin{equation}\label{eq:153}
  \langle\Psi(0)|H_\mathrm{xxz}|\Psi(0)\rangle=\sum_{i=0}^{L-1}\langle\Psi(0)|H_i|\Psi(0)\rangle.
\end{equation}
Because $\Psi(0)$ is random $\langle\Psi(0)|H_i|\Psi(0)\rangle$ ($i=0,\cdots, L-1$) are random numbers with nonvanishing mean, implying that $\langle\Psi(0)|H_{\rm xxz}|\Psi(0)\rangle$ is the sum of $L$ independent random number. Since each random number is ${\cal O}(L^0)$, we have
\begin{eqnarray}\label{eq:164}
   \langle\Psi(0)|H_{\rm xxz}^2|\Psi(0)\rangle-\langle\Psi(0)|H_{\rm xxz}|\Psi(0)\rangle^2={\cal O}(L)
\end{eqnarray}
for $L\gg 1$. This scaling also holds, when $\Psi(0)$ is a Neel state
\beq
\left| \mathrm{Neel} \right\rangle \equiv \left|1010\cdots10\right\rangle,
\label{eq:212}
\eeq
with the number $1\,(0)$ denoting an up (down) spin at a lattice site. More precisely, one may readily show that
\begin{eqnarray}\label{eq:175}
   \langle \mathrm{Neel}| {H}^2_{\mathrm{xxz}}| \mathrm{Neel} \rangle - \langle \mathrm{Neel}| {H}_{\mathrm{xxz}}| \mathrm{Neel} \rangle^2=\frac{L}{4}.
\end{eqnarray}
On one hand,
Supplementary Eqs.~(\ref{eq:164}) and (\ref{eq:175}) show that the width of the energy window excited by $\Psi(0)$ increases with $L$ as $\sim \sqrt{L}$. On the other hand, the dimension of Hilbert space grows exponentially with $L$. Moreover, since the spectrum $\{\omega_m\}$ is bounded from both above and below, with the bounds growing as $\sim L$, the typical level spacing must decrease exponentially with $L$. Since the exponential decreasing of the level spacing dominates over the linear growth of the width of the energy window, we have Supplementary Eq.~(\ref{dimension}).

\subsubsection{Relating entanglement dynamics to classical trajectory on $\mathbb{T}^N$}
\label{sec:time_dependence_RDM}

Supplementary Eq.~(\ref{dimension}) shows that the number of frequency mismatches $\omega_{m}-\omega_n$ in Supplementary Eq.~(\ref{eq:176}), where $m,n$ run from $1$ to $D$, is $D(D-1)/2 \sim \mathrm{e}^{2\kappa_D L}$. Noting that
\beq
\omega_{m}-\omega_{n} = \omega_{m1} - \omega_{n1},\quad \omega_{m1}\equiv \omega_m-\omega_{1},
\eeq
we find that all $D(D-1)/2$ frequency mismatches are completely determined by a much smaller set of frequency mismatches, $\{\omega_{m1}\}_{m=2}^{D}$, with a number (for sufficiently large but finite $L$)
\begin{equation}
N=D-1\sim \mathrm{e}^{\kappa_D L}
  \label{eq:187}
\end{equation}
much smaller than $D(D-1)/2$. Identifying this set, we see that similar to the free fermion case, the time parameter in the reduced density matrix enters through the $N$ dynamical phases $\boldsymbol{\omega} t$, with
\begin{equation}\label{eq:183}
  \boldsymbol{\omega}\equiv(\omega_{21},\cdots,\omega_{D1}).
\end{equation}
Now we can generalize Supplementary Eqs.~(\ref{eq:27})-(\ref{eq:16a}). Specifically, we introduce a (matrix-valued) function on the $N$-dimensional torus $\mathbb{T}^N\ni \boldsymbol{\varphi}\equiv (\varphi_0,\varphi_1,\cdots,\varphi_{N-1})$:
\begin{equation}\label{eq:178}
  \tilde\rho_{A}(\boldsymbol{\varphi})=\rho_{A0}+\tilde\rho_{A1}(\boldsymbol{\varphi}),
\end{equation}
where
\beq
{\rho}_{A0} = \sum_{m}|\chi_m|^2\,  {\rm Tr}_B\left| \Psi_m\right\rangle\left\langle\Psi_m\right|
\label{eq:179}
\eeq
has no $\boldsymbol{\varphi}$ dependence, with B being the complement of subsystem A, and
\begin{equation}
\label{eq:180}
\tilde\rho_{A1}(\boldsymbol{\varphi}) = \sum_{m\neq n} {\rm e}^{-i(\varphi_m - \varphi_n)} \chi_m \chi_n^*\,\text{Tr}_B \vert\Psi_m\rangle \langle\Psi_n\vert
\end{equation}
has. Then we obtain the instantaneous reduced density matrix Supplementary Eq.~(\ref{eq:177}) from $\tilde{\rho}_A(\boldsymbol{\varphi})$ via
\beq
\rho_A(t)= \tilde{\rho}_A(\boldsymbol{\omega}t),
\label{eq:181}
\eeq
which is the interacting version of Supplementary Eq.~(\ref{eq:66}). By definition $\tilde{\rho}_A(\boldsymbol{\varphi})$ is periodic in each argument $\varphi_m$. Thus the $N$ phases $\boldsymbol{\varphi}=\boldsymbol{\omega}t$ in $\tilde{\rho}_A(\boldsymbol{\varphi})$ completely determine $\rho_A(t)$ and entail a classical motion --- a rotation in $\mathbb{T}^N$ with constant angular velocity $\boldsymbol{\omega}$:
\beq
\rho_A(t) \leftrightsquigarrow \boldsymbol{\varphi}=\boldsymbol{\omega}t\in \mathbb{T}^N.
\label{eq:182}
\eeq
This is in spirit fully parallel to Supplementary Eq.~(\ref{eq:16a}) in the free fermion case, shown schematically in Fig.~1(d) in the main text. That is, the evolution of the reduced density matrix corresponds to a classical trajectory on $\mathbb{T}^N$.

Now a generic information-theoretic observables $O(t)$ is a functional of $\rho_A(t)$,
\begin{equation}\label{eq:186}
  O(t)\equiv O[\rho_A(t)],
\end{equation}
in contrast to Supplementary Eq.~(\ref{eq:101}). Owing to the relation Supplementary Eq.~(\ref{eq:181}), $O(t)$ must depend on $t$ via $\boldsymbol{\varphi}=\boldsymbol{\omega}t$. That is, with the introduction of the $N$-variable function:
\begin{equation}\label{eq:184}
  O(\boldsymbol{\varphi})\equiv O[\tilde{\rho}_A(\boldsymbol{\varphi})],
\end{equation}
e.g. for the entanglement entropy:
\begin{equation}\label{eq:205}
  S(\boldsymbol{\varphi})=-\text{Tr}_{A}(\tilde\rho_{A}(\boldsymbol{\varphi})\ln \tilde\rho_{A}(\boldsymbol{\varphi})),
\end{equation}
and for the $n$-th order R$\acute{\rm e}$nyi entropy:
\begin{equation}\label{eq:206}
  S_n(\boldsymbol{\varphi})= -{1\over n-1}\ln\text{Tr}_A(\tilde\rho_{A}(\boldsymbol{\varphi}))^n,
\end{equation}
the relation Supplementary Eq.~(\ref{eq:69}) follows again, which we rewrite here,
\beq
O(t)= O(\boldsymbol{\varphi})|_{\boldsymbol{\varphi}=\boldsymbol{\omega}t}
\label{eq:185}
\eeq
for the convenience of readers, although they have completely different meanings. We remark that throughout this Supplementary Note the partial trace $\text{Tr}_A$ is over the subsystem Hilbert space, while in previous Supplementary Notes, namely, Supplementary Eqs.~(\ref{eq:34}), (\ref{eq:92}), (\ref{eq:94}) and (\ref{eq:95}) and ensuing equations, it is over the subsystem lattice.

Observing Supplementary Eqs.~(\ref{eq:205}) and (\ref{eq:206}), we find that all information-theoretic observables $O(\boldsymbol{\varphi})$ can be related to the same quantity, the spectral density of $\tilde{\rho}_A(\boldsymbol{\varphi})$. More precisely, we have
\beq
O(\boldsymbol{\varphi})=\int
d\lambda\, \mathfrak{O}(\lambda)\, \textrm{Tr}_A\,\delta\left(\lambda-\tilde{\rho}_A(\boldsymbol{\varphi})\right),
\label{eq:222}
\eeq
where $\mathfrak{O}(\lambda)=-\lambda\ln\lambda$ for $O=S$ and $\mathfrak{O}(\lambda)=\lambda^n$ for $O={\rm e}^{-(n-1)S_n}$. This expression is an analog of Supplementary Eq.~(\ref{eq:95}).

\subsubsection{Emergence of disordered samples and mesoscopic fluctuations}
\label{sec:mesoscopic_fluctuations}

In general, the $N$ frequencies Supplementary Eq.~(\ref{eq:183}) are incommensurate. In combination with the relation Supplementary Eq.~(\ref{eq:185}), this implies that at long time an entanglement probe $O(t)$ display quasiperiodic oscillations, like in the free fermion case. These quasiperiodic oscillations are conceptually different from the oscillations of conventional [39] and unconventional [40] entanglement probes in interacting models, which arise from the traversal of quasiparticle pairs or the incomplete revival of system's wavefunction, and rapidly damp in the course of time. Most importantly, the statistical equivalence established in \ref{sec:emergence_randomness} applies here also. That is, out-of-equilibrium fluctuations displayed by the quasiperiodic oscillations of $O(t)$ must have the same statistics as fluctuations of $O(\boldsymbol{\varphi})$ with $\boldsymbol{\varphi}$, provided $\boldsymbol{\varphi}$ is drawn randomly from the uniform probability measure $\mathbf{P}$ on $\mathbb{T}^N$.

However, for the XXZ model each virtual disordered sample is represented by a random matrix $\tilde{\rho}_A(\boldsymbol{\varphi})$, namely, Supplementary Eq.~(\ref{eq:178}), instead of $\tilde{C}(\boldsymbol{\varphi})$ for free fermion models. This matrix has two parts: the nonrandom ${\rho}_{A0}$ defined by Supplementary Eq.~(\ref{eq:179}), and the random $\tilde{\rho}_{A1}(\boldsymbol{\varphi})$ defined by Supplementary Eq.~(\ref{eq:180}). The latter depends on the disorder realization, $\boldsymbol{\varphi}$, uniformly distributed over $\mathbb{T}^N$. The entanglement properties of a disordered sample are characterized by various information-theoretic observables: $S(\boldsymbol{\varphi})$, $S_n(\boldsymbol{\varphi})$, etc., which by Supplementary Eq.~(\ref{eq:184}) are functionals of $\tilde{\rho}_A(\boldsymbol{\varphi})$. So the $\boldsymbol{\varphi}$ fluctuations of an information-theoretic observable represent a new class of mesoscopic sample-to-sample fluctuations also, although their detailed mathematical description is very different from that for free fermion models summarized in Supplementary Note $4.3$. 

To understand the scaling behaviors of the disorder strength we first estimate the number of pairs $(m, n)$, denoted as ${\cal W}$, dominate the double sum in Supplementary Eq.~(\ref{eq:180}). Is it $\sim D^2$, as naively expected from that $m,n$ run from $1$ to $D$? Consider an arbitrary local observable ${O}$ in the subsystem A. The time-dependent part of its expectation $\langle\Psi(t)|{O}|\Psi(t)\rangle$ is
\beq
\sum_{m\neq n}
\mathrm{e}^{-i(\omega_m-\omega_n)t} \chi_m \chi_n^\ast  \langle\Psi_n|{O}|\Psi_m\rangle = {\rm Tr}_{A} \left( {O} {\rho}_{A1}(t) \right).
\label{time_dep}
\eeq
From this we see that ${\cal W}$ is the number of pairs $(m, n)$ dominating the sum on the left-hand side of Supplementary Eq.~(\ref{time_dep}). Thus we turn to study the off-diagonal element $\langle\Psi_n|{O}|\Psi_m\rangle$ ($m\neq n$). Because of
\beq
\langle\Psi_n| [O,{H}_{\mathrm{xxz}}] |\Psi_m\rangle = (\omega_m - \omega_n) \langle\Psi_n| O |\Psi_m\rangle,
\eeq
we have
\beq
\omega_m - \omega_n = \frac{\langle\Psi_n| [O,H_{\mathrm{xxz}}] |\Psi_m\rangle}{\langle\Psi_n| O |\Psi_m\rangle}.
\label{energy_diff}
\eeq
A key point is that ${H}_{\mathrm{xxz}}$ is additive and, as a result,
\beq
[{O},{H}_{\mathrm{xxz}}] = [{O},\sum_{i=0}^{L_A-1}{H}_i],
\label{commut}
\eeq
where $\sum_{i=0}^{L_A-1}{H}_i$ may be considered as the XXZ chain restricted to A. Substituting it into Supplementary Eq.~(\ref{energy_diff}), we find that
\beq
\omega_m - \omega_n = \sum_{i=0}^{L_A-1} \frac{\langle\Psi_n| [O,H_i] |\Psi_m\rangle}{\langle\Psi_n| O |\Psi_m\rangle}.
\label{eq:208}
\eeq
The right-hand side cannot grow faster than $\sim L_A$. So
\beq
|\omega_m - \omega_n| \lesssim L_A,
\label{ord}
\eeq
i.e. $|\omega_m - \omega_n|$ must be smaller or comparable to $L_A$. This gives a window $\sim L_A$ of the eigenenergy mismatch, which
is much smaller than the width of the excited energy window $\sim \sqrt{L}$ (for $L_A\ll L$) given by Supplementary Eq.~(\ref{eq:164}) or Supplementary Eq.~(\ref{eq:175}). Then the question is given $m$ how many $n$ there are, such that not only is the condition Supplementary Eq.~(\ref{ord}) satisfied, but also $\langle\Psi_n|O|\Psi_m\rangle$ takes a significant value and thus can dominate the sum in the left-hand side of Supplementary Eq.~(\ref{time_dep}). These two conditions essentially define a sub-spectrum of $H_{\rm xxz}$. Let the typical spacing of two nearest eigenvalues in this sub-spectrum be $s^*$. The desired number of $n$ can be estimated as $L_A/s^*$, and
\beq
{\cal W} \sim DL_A/s^*,
\label{eq:218}
\eeq
where the first factor $D$ accounts for that the index $m$ runs from $1$ to $D$.

We proceed to calculate $s^*$ for fixed $m$. Let us write an unpolarized eigenstate $|\Psi_m\rangle$ as
\beq
| \Psi_m \rangle = \sum_{{\mathfrak{m}}}\,{t}_{{\mathfrak{m}}}^m |{\mathfrak{m}}\rangle,
\label{eq:193}
\eeq
with the sum over all the product basis vectors $|{\mathfrak{m}}\rangle\equiv |s_0s_1\cdots s_{L-1}\rangle$ ($s_i=0,1$) in $\mathscr{H}_{\Sz}$. Here each $|{\mathfrak{m}}\rangle$ is composed of $L/2$ symbols `$0$' (down spins) and $L/2$ symbols `$1$' (up spins). The complex expansion coefficients $t_{\mathfrak{m}}^m$  are constrained by the normalization condition:
\beq
\sum_{{\mathfrak{m}}}\,|{t}_{{\mathfrak{m}}}^m |^2=1.
\label{eq:194}
\eeq
This defines a ($2D_s-1$)-dimensional hypersphere $\mathbb{S}^{2D_s-1}$, where
\begin{equation}\label{eq:221}
  D_s=\Big(
        \begin{array}{c}
          L \\
          L/2 \\
        \end{array}
      \Big)
  ={L!\over ((L/2)!)^2}
\end{equation}
is the dimension of $\mathscr{H}_{\Sz}$. Let
\beq
\langle\Psi_n | = \sum_{\mathfrak{n}}t_{\mathfrak{n}}^{n \ast} \langle \mathfrak{n} |.
\label{eq:215}
\eeq
We substitute Supplementary Eqs.~(\ref{eq:193}) and (\ref{eq:215}) into the left-hand side of Supplementary Eq.~(\ref{time_dep}). Since $\langle\Psi_n|O|\Psi_m\rangle$ takes a significant value, there must exist a large number of $\mathfrak{m},\,\mathfrak{n}$, such that
\beq
\langle \mathfrak{n}|O| \mathfrak{m}\rangle \neq 0.
\eeq
Because $O$ is a local observable in the subsytem A and $\mathfrak{m},\,\mathfrak{n}$ are product states, this can be satisfied only if $\mathfrak{m},\,\mathfrak{n}$ differ only in their spin configurations in A. Thus the spin configuration of $\mathfrak{m},\,\mathfrak{n}$ are almost the same, that can also be readily seen from the definition Supplementary Eq.~(\ref{eq:180}) of $\tilde{\rho}_{A1}(\boldsymbol{\varphi})$. This is a strong condition. Thanks to it $s^*$ must be much larger the level spacing of eigenenergies excited by $\Psi(0)$, which is $\sim D^{-1}$. Furthermore, similar to the latter $s^*$ must decrease exponentially with $L$. Thus $s^*\sim D^{-\nu}$ for fixed $L_A$, with $0\leq \nu <1$. How does $s^*$ scale with $L_A$, when $L$ is fixed? Note that two nearest eigenstates in the subspectrum may be related by the motion of a quasiparticle [37] in A. The volume available for such a particle is $\sim L_A^{\delta}$, with $0\leq \delta \leq 1$, depending on $\Psi(0)$ and the parameter $\Delta$. Thus $s^*\sim L_A^{-\delta}$ for fixed $L$.  Taking all these into account, we have the estimation:
\begin{equation}\label{eq:216}
  s^*\sim L_A^{-\delta} D^{-\nu},
\end{equation}
giving
\begin{equation}\label{eq:217}
  L_A/s^*\sim L_A^{\mu} D^{\nu}, \quad \mu=1+\delta.
\end{equation}
Substituting it into Supplementary Eq.~(\ref{eq:218}) we have
\beq
{\cal W} \sim L_A^\mu D^{1+\nu},\quad 1\leq \mu\leq 2.
\label{number_fre_v1}
\eeq
So the increase of ${\cal W}$ with $D$ is slower than $\sim D^2$. By definition the power $\mu$ depends on both $\Delta$ and $\Psi(0)$.

In Supplementary Table \ref{table:3}, we make comparisons between mesoscopic fluctuations emergent from entanglement dynamics of free fermion and interacting models.

\begin{table}[b]
\centering
{  \caption{Comparisons of mesoscopic entanglement fluctuations in different models.}\label{table:3}
    \begin{tabular}{lll}
     \hline
     \hline
     model $\quad\quad\quad$ & free fermion $\quad\quad$ & interacting $\quad\quad$\\
     \hline
     examples & Rice-Mele, TFIC & XXZ, XXX\\
     random matrix representing $\quad$& $\tilde{C}(\boldsymbol{\varphi})$& $\tilde{\rho}_A(\boldsymbol{\varphi})$\\
     disordered sample &&\\
     matrix dimension & $2L_A$ & $D_A$\\
     variance of matrix elements & $\sim 1/L_A$ (at fixed $L_A/L$) & $\sim L_A^\mu\, {\rm e}^{-(1-\nu)\kappa_D L}$\\
     disorder realization & $\boldsymbol{\varphi}\equiv (\varphi_0,\ldots,\varphi_{N-1})$& $\boldsymbol{\varphi}\equiv (\varphi_0,\ldots,\varphi_{N-1})$\\
     distribution of disorder realizations $\quad$ & uniform in $\mathbb{T}^N$ $\qquad$ & uniform in $\mathbb{T}^N$\\
     growth of torus dimension $N$& linear & exponential \\
     with $L$&&\\
     origin of disorders $\quad$ & incommensurate $\omega_m={2E_{fk_m}}$ & incommensurate $\omega_{m1}=\omega_m-\omega_{1}$\\
     &$m=0,\ldots,N-1$& $m=2,\ldots,D=N+1$\\
     expression of probes & $\int
d\lambda\, \mathfrak{O}(\lambda)\, \textrm{Tr}_A\,\delta(\lambda-\tilde{C}(\boldsymbol{\varphi}))$$\qquad$&
$\int
d\lambda\, \mathfrak{O}(\lambda)\, \textrm{Tr}\,_A\,\delta(\lambda-\tilde{\rho}_A(\boldsymbol{\varphi}))$\\
     \hline
      \hline
    \end{tabular}
}
\end{table}

Now, consider an element of the reduced density matrix $(\tilde\rho_{A1}(\boldsymbol{\varphi}))_{IJ}$, where $I,J$ denote the product basis vectors of the subsystem Hilbert space. Let us perform its $\boldsymbol{\varphi}$ average. With the help of Supplementary Eq.~(\ref{number_fre_v1}) and the estimation: $\chi_m\chi_n^*\sim 1/D$, we obtain from Supplementary Eq.~(\ref{eq:180}) $\langle (\tilde\rho_{A1}(\boldsymbol{\varphi}))_{IJ}\rangle=0$ and
\begin{equation}\label{eq:189}
  {\rm Var}((\tilde\rho_{A1}(\boldsymbol{\varphi}))_{IJ})\sim {\cal W}/D^2\sim L_A^\mu/D^{1-\nu}.
\end{equation}
Thus
\begin{equation}\label{eq:190}
  (\tilde\rho_{A1}(\boldsymbol{\varphi}))_{IJ}\sim \sqrt{L_A^\mu/D^{1-\nu}}\sim L_A^{\mu/2}\, {\rm e}^{-(1-\nu)\kappa_D L/2},
\end{equation}
which implies the disorder strength $\sim L_A^{\mu/2}\, {\rm e}^{-(1-\nu)\kappa_D L/2}$: It displays an exponential decrease in $L$ and a power law increase in $L_A$.

\subsection{Statistical distribution of mesoscopic fluctuations}
\label{sec:statistics_mesoscopic_fluctuations}

With the random structure described above uncovered, we can also use the modified logarithmic Sobolov inequality namely Supplementary (\ref{sobo1}) to study the mesoscopic fluctuations of entanglement probes $O(\boldsymbol{\varphi})=S(\boldsymbol{\varphi}),\,S_n(\boldsymbol{\varphi})$, etc., since ($\varphi_0, \varphi_1, ..., \varphi_{N-1}$) are independent and identically distributed random variables, and the entanglement probe $O(\boldsymbol{\varphi})$ may be regarded as a --- highly nonlinear --- function of these $N$ angular variables. Specially, following the analysis in Supplementary Note $5.1$,
for generic $O$ we still have the concentration inequalities namely Supplementary (\ref{eq:147}) and (\ref{eq:148}) for the upper tail and Supplementary (\ref{eq:150}) and (\ref{eq:151}) for the lower. The difference is that in Supplementary Eq.~(\ref{eq:46}), which we rewrite here:
\beq
b_\pm\equiv \sum_{m=0}^{N-1} \left\langle  (O-O_m^\pm)^2\right\rangle,
\label{bpn}
\eeq
(recall that the average is over the torus $\mathbb{T}^N$ with uniform probability measure $\bf P$.)
the dimension $N$ of the torus increases with $L$ exponentially for sufficiently large but finite $L$ (cf.~Supplementary Eq.~(\ref{eq:187})), whereas in Supplementary Eq.~(\ref{eq:46}) $N$ increases with $L$ linearly (cf.~Supplementary Eq.~(\ref{eq:70})). As we shall see below, this difference is responsible for the arising of completely different scaling behaviors of mesoscopic fluctuations.

For $O=S$, $S_n$, on general grounds we expect that the sign of $c_\pm$ in the concentration inequalities stay the same as that in free fermion models, i.e. $c_\pm<0$ for both upper and lower tails. Thus we have
\beq
\mathbf{P}(O-\langle O \rangle \geq \epsilon)\leq {\rm e}^{-{\frac{\epsilon^2}{2b_+}}}
\label{upper}
\eeq
for the upper tail and
\beq
\mathbf{P}(\langle O \rangle-O\geq \epsilon)\leq {\rm e}^{-\frac{\epsilon^2}{2 (b_-+|c_-| \epsilon)}}
\label{lower}
\eeq
for the lower. Similar to the situations in the free fermion case, simulations further show (see Fig.~5c in the main text and Supplementary Note $9.5$
below) that for large $\epsilon$ the exact deviation probability for upward fluctuations agrees with the form given by the right-hand side of Supplementary (\ref{upper}) and for downward by the right-hand side of Supplementary (\ref{lower}), with $b_\pm$ and $c_-$ as fitting parameters. So we receive Eq.~(2) in the main text for $S$, $S_n$.

\subsection{A general result for the variance of mesoscopic fluctuations}
\label{sec:variance_mesoscopic_fluctuations}

Next, we generalize the method developed in Supplementary Note $5.2$
to study the variance of a generic entanglement probe $O$. From Supplementary Eq.~(\ref{bpn}) we also arrive at Supplementary Eq.~(\ref{eq:56}).
To calculate the latter, we substitute Supplementary Eq.~(\ref{eq:178}) into Supplementary Eq.~(\ref{eq:205}) or Supplementary Eq.~(\ref{eq:206}), and perform the derivative $\partial_{\varphi_m}O$ for $O=S,\, S_n$. In doing so, we encounter an essential technical difference intrinsic to the random structure associated with Supplementary Eq.~(\ref{eq:180}). That is, when we perform the Fourier expansion of $\partial_{{\varphi}_m}O$ with respect to $\varphi_m$, we have
\beq
\begin{aligned}
\partial_{{\varphi}_m}O &= a_0+\sum_{n\neq m}\left( a_{1n}\sin(\varphi_m - \varphi_n)+
b_{1n}\cos(\varphi_m - \varphi_n) \right)\\
&=a_0+\sum_{n\neq m}A_{1n}\sin(\varphi_m - \varphi_n + \phi_{mn1})
\end{aligned}
\label{Fourier}
\eeq
instead of Supplementary Eq.~(\ref{eq:50}). Here $a_{1n}$ and $b_{1n}$ are coefficients independent of $\varphi_m$, $\varphi_n$, and $A_{1n}$ and $\phi_{mn1}$ are determined by $a_{1n}$, $b_{1n}$ and can be easily found. The explicit form of $a_{1n}$, $b_{1n}$, $A_{1n}$ and $\phi_{mn1}$ are not needed for subsequent discussions. For simplicity we have kept the expansion up to the first-order harmonic, since the disorder strength is exponentially small for large $L$. Note that the sum in Supplementary Eq.~(\ref{Fourier}) is over $n$ only, for $m$ is fixed. With the help of Supplementary Eq.~(\ref{Fourier}) we obtain ($\alpha_{mn}\equiv\varphi_n-\phi_{mn1}$)
\beq
\begin{aligned}
(\partial_{{\varphi}_m}O)^2 &\leq (|a_0 |+| \sum_{n\neq m}A_{1n}\sin(\alpha_{mn} - \varphi_m) | )^2
\leq 2(|a_0|^2 + | \sum_{n\neq m}A_{1n}\sin(\alpha_{mn} - \varphi_m) |^2 )\\
&= 2(|a_0|^2 + | \cos{\varphi_m}\sum_{n\neq m}A_{1n}\sin{\alpha_{mn}} - \sin{\varphi_m}\sum_{n\neq m}A_{1n}\cos{\alpha_{mn}} |^2 ).
\end{aligned}
\label{eq:191}
\eeq
Thanks to
\beq
&&| \cos{\varphi_m}\sum_{n\neq m}A_{1n}\sin{\alpha_{mn}} - \sin{\varphi_m}\sum_{n\neq m}A_{1n}\cos{\alpha_{mn}} |^2\nonumber\\
&\leq& 2( \cos^2{\varphi_m}| \sum_{n\neq m}A_{1n}\sin{\alpha_{mn}}|^2 + \sin^2{\varphi_m}|\sum_{n\neq m}A_{1n}\cos{\alpha_{mn}}|^2 )\nonumber\\
&\leq& 2(| \sum_{n\neq m}A_{1n}\sin{\alpha_{mn}}|^2 +|\sum_{n\neq m}A_{1n}\cos{\alpha_{mn}}|^2 ),
\label{eq:192}
\eeq
we cast the Supplementary inequality (\ref{eq:191}) to
\beq
(\partial_{{\varphi}_m}O)^2 \leq 2(|a_0|^2 + 2(|\sum_{n\neq m}A_{1n}\sin{\alpha_{mn}}|^2 + |\sum_{n\neq m}A_{1n}\cos{\alpha_{mn}}|^2 )).
\label{p_ieq}
\eeq
From Supplementary Eq.~(\ref{Fourier}) we also have
\beq
\int\frac{d\varphi_m}{2\pi}(\partial_{{\varphi}_m}O)^2=|a_0|^2+ \frac{1}{2}( |\sum_{n\neq m}A_{1n}\sin{\alpha_{mn}}|^2 + |\sum_{n\neq m}A_{1n}\cos{\alpha_{mn}}|^2 ).
\label{p_inte}
\eeq
The Supplementary inequality (\ref{p_ieq}) and Supplementary Eq.~(\ref{p_inte}) yield
\beq
(\partial_{{\varphi}_m}O)^2\leq 8\int\frac{d\varphi_m}{2\pi}(\partial_{{\varphi}_m}O)^2,
\eeq
which differs from the Supplementary inequality (\ref{eq:54}) only in the numerical factor on the right-hand side. Repeating the analysis from Supplementary Eq.~(\ref{eq:54}), we also arrive at Supplementary Eq.~(\ref{eq:135}) which we rewrite here:
\beq
{\rm Var}(O)\propto \langle |\partial_{\boldsymbol{\varphi}}O|^2\rangle.
\label{var_lip}
\eeq
This general result allows us to study the scaling behaviors of ${\rm Var}(S)$ and ${\rm Var}(S_n)$ in details below. As we shall see immediately, despite this relation holds for the XXZ model also, the scaling behaviors following from it are completely different those for the Rice-Mele model and TFIC.

\subsection{Universal behaviors of variances}
\label{sec:mesoscopic_universality_variance}

We are now ready to study the mesoscopic universality exhibited by the variance of various information-theoretic observables.

\subsubsection{The nonrandom part of the reduced density matrix}
\label{sec:nonrandom_part_RDM}

For $D_s\gg 1$ (recall Supplementary Eq.~(\ref{eq:221}).), by L$\acute{\rm e}$vy lemma the Haar measure on that hypersphere $\mathbb{S}^{2D_s-1}$ displays strong concentration [16, 17] for $L\gg 1$. As a result [42], for almost all states $| \Phi\rangle \in \mathscr{H}_{\Sz}$, the reduced density matrix ${\rm Tr}_B|\Phi\rangle\langle\Phi|$ must be very close to $D_s^{-1}{\rm Tr}_B\, \mathbb{I}$, where $\mathbb{I}$ is the unity matrix in $\mathscr{H}_{\Sz}$, provided the subsystem complement $B$ is sufficiently large, i.e. $L_A/L\ll 1$. Let $\Phi$ be the eigenstate $\Psi_m$ Supplementary Eq.~(\ref{eq:193}). On general ground we expect that most eigenstates Supplementary Eq.~(\ref{eq:193}) satisfy this property, i.e. are {\it typical} with respect to the Haar measure: This expectation is indeed confirmed numerically by computing the trace distance [42] between ${\rm Tr}_B|\Psi_m\rangle\langle\Psi_m|$ and $D_s^{-1}{\rm Tr}_B\, \mathbb{I}$. Thus we have
\beq
{\rm Tr}_B\left| \Psi_m\right\rangle\left\langle\Psi_m\right|\approx D_s^{-1}{\rm Tr}_B\, \mathbb{I}.
\label{eq:195}
\eeq
Now, if the following condition:\\

{\it ($*$) Most weight $|\chi_m|^2$ in the initial state $\Psi(0)$ goes to typical eigenstates.}\\

\noindent is postulated, we can substitute Supplementary Eq.~(\ref{eq:195}) into Supplementary Eq.~(\ref{eq:179}) to obtain
\beq
{\rho}_{A0} \approx D_s^{-1}{\rm Tr}_B\, \mathbb{I}.
\label{eq:196}
\eeq
It implies that ${\rho}_{A0}$ is diagonal.

The remainder is to calculate the diagonal element $({\rho}_{A0})_{II}$. This is essentially the number of bases $|{\mathfrak{m}}\rangle$, whose configuration in A, i.e. $s_0s_1\cdots s_{L_A-1}$ is $I$. Let the magnetization of A be $S_A^z$. Because $L_A$ is an even number, $S_A^z$ must be an integer. In addition, $|S_A^z|\leq L_A/2$. The number of up- and down-spins in $I$ are given respectively by
\beq
L_A^{\uparrow} = \frac{2S_A^z+L_A}{2}, \quad L_A^{\downarrow} = \frac{-2S_A^z+L_A}{2}.
\label{eq:197}
\eeq
So the number of up- and down-spins in the complement $B$ are, respectively,
\beq
L_{B}^{\uparrow} = \frac{L}{2}-L_{A}^{\uparrow} = \frac{L-L_A-2S_A^z}{2},\quad L_{B}^{\downarrow} = \frac{L}{2}-L_{A}^{\downarrow} = \frac{L-L_A+2S_A^z}{2}.
\label{eq:198}
\eeq
Thus
\beq
\begin{aligned}
({\rho}_{A0})_{II} &= {1 \over D_s} \Big(
        \begin{array}{c}
           L_{B}^{\uparrow} \\
          L-L_A \\
        \end{array}
      \Big)
      = {1\over D_s}\frac{(L-L_A)!}{\left((L-L_A+2S_A^z)/2\right)! \left((L-L_A-2S_A^z)/2\right)!}\\
&=\frac{((L-L_A)/2) \left((L-L_A)/2-1\right) \cdots \left((L-L_A)/2-(|S_A^z|-1)\right)}{\left((L-L_A)/2+|S_A^z|\right) \left((L-L_A)/2+(|S_A^z|-1)\right)  \cdots \left((L-L_A)/2+1\right) } {\Big(
        \begin{array}{c}
           L-L_A \\
          (L-L_A)/2 \\
        \end{array}
      \Big)\over D_s}\\
&= \left(1 -\frac{|S_A^z|}{(L-L_A)/2+|S_A^z|} \right) \left(1 -\frac{|S_A^z|}{(L-L_A)/2+(|S_A^z|-1)} \right) \cdots \left(1 -\frac{|S_A^z|}{(L-L_A)/2+1} \right)\,{\Big(
        \begin{array}{c}
           L-L_A \\
          (L-L_A)/2 \\
        \end{array}
      \Big)\over D_s}.
\end{aligned}
\label{eq:199}
\eeq
Note that ${\Big(
        \begin{array}{c}
           L-L_A \\
          (L-L_A)/2 \\
        \end{array}
      \Big)/D_s}\equiv {\rho}_{A0}^*$ gives the diagonal element $({\rho}_{A0})_{II}$ with the configuration $I$ satisfying $S_A^z=0$. According to Supplementary Eq.~(\ref{eq:199}), $({\rho}_{A0})_{II}$ depends on three parameters, i.e. $L,\,L_A$ and $S_A^z$ only. For $L\gg L_A$,
\beq
\left(1 -\frac{|S_A^z|}{(L-L_A)/2+|S_A^z|} \right) \left(1 -\frac{|S_A^z|}{(L-L_A)/2+(|S_A^z|-1)} \right) \cdots \left(1 -\frac{|S_A^z|}{(L-L_A)/2+1} \right)\approx \left(1-\frac{2|S_A^z|}{L}\right)^{|S_A^z|},
\label{eq:201}
\eeq
and Supplementary Eq.~(\ref{eq:199}) reduces to
\beq
({\rho}_{A0})_{II}/{\rho}_{A0}^* \approx\left(1-\frac{2|S_A^z|}{L}\right)^{|S_A^z|}.
\label{eq:200}
\eeq
Because of $|S_A^z|\leq L_A/2\ll L$, we need to study the consequences of Supplementary Eq.~(\ref{eq:200}) in two cases separately. In the first case, $L_A\ll \sqrt{L}$. So Supplementary Eq.~(\ref{eq:200}) can be simplified as $({\rho}_{A0})_{II}/{\rho}_{A0}^* \approx 1-2|S_A^z|^2/L\approx 1$. Therefore, all the diagonal elements have the same value, ${\rho}_{A0}^*$. In the second case, $\sqrt{L}\lesssim L_A$,  i.e. $L_A$ is comparable to or much larger than $\sqrt{L}$. Supplementary Eq.~(\ref{eq:200}) shows that the diagonal elements with $I$ satisfying with $|S_A^z|\ll \sqrt{L}$ have the same value, ${\rho}_{A0}^*$, while the other diagonal elements are negligibly small, compared to ${\rho}_{A0}^*$. Collecting these analysis for both cases, we find that
\beq
{\rho}_{A0} = {\rho}_{A0}^*\,\mathbb{I},
\label{eq:202}
\eeq
with $\mathbb{I}$ being a unit matrix of some subspace of the subsystem Hilbert space with $S_A^z=0$. Let its dimension be $D_A$.

To determine $D_A$ we exploit the normalization condition: ${\rm Tr}_A {\rho}_{A0}=1$. With the substitution of Supplementary Eq.~(\ref{eq:202}), we have
\beq
D_A {\rho}_{A0}^*=1\,\Rightarrow\, D_A = {D_s\over \Big(
        \begin{array}{c}
           L-L_A \\
          (L-L_A)/2 \\
        \end{array}
      \Big)}.
\label{eq:203}
\eeq
$D_A$ may be regarded as the effective dimension associated with ${\rho}_{A0}$. So we can rewrite Supplementary Eq.~(\ref{eq:202}) as
\beq
{\rho}_{A0} = {\mathbb{I}/D_A},
\label{eq:204}
\eeq
which is confirmed numerically (Fig.~5b in the main text).

\subsubsection{Universality with respect to entanglement probes}
\label{sec:universality_entanglement_probes}

Consider the entanglement entropy $S(\boldsymbol{\varphi})$. Applying Supplementary Eq.~(\ref{var_lip}) to Supplementary Eq.~(\ref{eq:205}) we obtain
\beq
{\rm Var}(S)\propto \left\langle
| \operatorname{Tr}_{A} \left( \ln  {\tilde\rho}_A(\boldsymbol{\varphi}) \partial_{\boldsymbol{\varphi}} {\tilde \rho}_A(\boldsymbol{\varphi})
 \right)|^2 \right\rangle,
 \label{var_S}
\eeq
where the proportionality coefficient is a universal numerical constant independent of $L, L_A$. Then we substitute Supplementary Eq.~(\ref{eq:178}) into it, and expand the trace up to the second order in $\tilde{\rho}_{A1}(\boldsymbol{\varphi})$. By further taking Supplementary Eq.~(\ref{eq:204}) into account, we obtain
\beq
\begin{aligned}
\operatorname{Tr}_{A} \left( \ln  {\tilde\rho}_A(\boldsymbol{\varphi}) \partial_{\boldsymbol{\varphi}} {\tilde \rho}_A(\boldsymbol{\varphi})
 \right) &\simeq \operatorname{Tr}_{A} \left( {\rho}_{A0} \partial_{\boldsymbol{\varphi}} \tilde{\rho}_{A1}(\boldsymbol{\varphi})
 \right) + \operatorname{Tr}_{A} \left( {\rho}^{-1}_{A0} \tilde{\rho}_{A1}(\boldsymbol{\varphi}) \partial_{\boldsymbol{\varphi}} \tilde{\rho}_{A1}(\boldsymbol{\varphi})
 \right)\\
 &= D_A \operatorname{Tr}_{A} \left(  \tilde{\rho}_{A1}(\boldsymbol{\varphi}) \partial_{\boldsymbol{\varphi}} \tilde{\rho}_{A1}(\boldsymbol{\varphi})
 \right),
\end{aligned}
\label{S_appro}
\eeq
where in deriving the second line we have used the fact of $\operatorname{Tr}_{A} \tilde{\rho}_{A1}(\boldsymbol{\varphi}) = 0$. Collecting Supplementary Eqs.~(\ref{var_S}) and (\ref{S_appro}) we have
\beq
{\rm Var}(S)\propto D_A^2 \left\langle
| \operatorname{Tr}_{A} \left(  \tilde{\rho}_{A1}(\boldsymbol{\varphi}) \partial_{\boldsymbol{\varphi}} \tilde{\rho}_{A1}(\boldsymbol{\varphi})\right) |^2 \right\rangle.
\label{var_S_v1}
\eeq

Consider the $n$-th order R$\acute{\rm e}$nyi entropy $S_n(\boldsymbol{\varphi})$. By the same token, we obtain
\beq
\begin{aligned}
\partial_{\boldsymbol{\varphi}} S_n(\boldsymbol{\varphi}) &= -\frac{n}{n-1} \operatorname{Tr}_{A} \left(  \left(\tilde{\rho}_{A}(\boldsymbol{\varphi}) \right)^{n-1} \partial_{\boldsymbol{\varphi}} \tilde{\rho}_{A}(\boldsymbol{\varphi})\right) /  \operatorname{Tr}_{A} \left( \left(\tilde{\rho}_{A}(\boldsymbol{\varphi}) \right)^{n} \right)\\
&= -\frac{n}{n-1} \operatorname{Tr}_{A} \left(  \left(\tilde{\rho}_{A}(\boldsymbol{\varphi}) \right)^{n-1} \partial_{\boldsymbol{\varphi}} \tilde{\rho}_{A1}(\boldsymbol{\varphi})\right) /  \operatorname{Tr}_{A} \left( \left(\tilde{\rho}_{A}(\boldsymbol{\varphi}) \right)^{n} \right)\\
&\simeq -n D_A^{-(n-2)} \operatorname{Tr}_{A} \left(  \tilde{\rho}_{A1}(\boldsymbol{\varphi}) \partial_{\boldsymbol{\varphi}} \tilde{\rho}_{A1}(\boldsymbol{\varphi})\right) / {D_A^{-(n-1)}}\\
 &= -n D_A \operatorname{Tr}_{A} \left(  \tilde{\rho}_{A1}(\boldsymbol{\varphi}) \partial_{\boldsymbol{\varphi}} \tilde{\rho}_{A1}(\boldsymbol{\varphi})\right)
\end{aligned}
\label{p_Sn}
\eeq
from Supplementary Eq.~(\ref{eq:206}), which gives
\beq
{\rm Var}(S_n)\propto n^2 D_A^2 \left\langle
| \operatorname{Tr}_{A} \left(  \tilde{\rho}_{A1}(\boldsymbol{\varphi}) \partial_{\boldsymbol{\varphi}} \tilde{\rho}_{A1}(\boldsymbol{\varphi})\right) |^2 \right\rangle.
\label{var_Sn_v1}
\eeq

Comparing Supplementary Eqs.~(\ref{var_S_v1}) and (\ref{var_Sn_v1}), we see that for $O=S,\,S_n$, their variance all obey
\begin{equation}
\label{eq:207}
\text{Var}(O) \propto D_A^2 \langle |\text{Tr}_A ( \tilde\rho_{A1} \partial_{\boldsymbol{\varphi}} \tilde\rho_{A1} )|^2 \rangle,
\end{equation}
where the proportionality coefficient is independent of $L, L_A$. Therefore, like in the free fermion case, different probes $O$ display universal mesoscopic fluctuation behaviors, when $L$, $L_A$ vary. This is confirmed by numerical experiments; see Fig.~5(d) in the main text and Supplementary Fig.~\ref{fig:S5}(d).

\subsubsection{Scaling law of the variance}
\label{sec:scaling_law_variance}

We proceed to study Supplementary Eq.~(\ref{eq:207}) in details. By Supplementary Eq.~(\ref{ord}) we can organize $\tilde{\rho}_{A1}(\boldsymbol{\varphi})$ as
\beq
\tilde{\rho}_{A1}(\boldsymbol{\varphi}) = \widetilde{\sum}_{m,n} \left( \cos(\varphi_m-\varphi_n)\mathscr{C}_{mn} + \sin(\varphi_m-\varphi_n)\mathscr{S}_{mn} \right),
\label{rho_A1_v2}
\eeq
where $\mathscr{C}, \mathscr{S}$ are matrices independent of $\boldsymbol{\varphi}$, and $\widetilde{\sum}_{m,n}$ denotes the double sum over $m,n$ with $m\neq n$ and $\omega_m,\omega_n$ satisfying Supplementary Eq.~(\ref{ord}). Taking the derivative $\partial_{{\varphi}_m}$ gives
\beq
\partial_{\varphi_m} \tilde{\rho}_{A1}(\boldsymbol{\varphi}) = \widetilde{\sum}^m_{n} \left( -\sin(\varphi_m - \varphi_n)(\mathscr{C}_{mn} + \mathscr{C}_{nm}) + \cos(\varphi_m - \varphi_n)(\mathscr{S}_{mn} - \mathscr{S}_{nm}) \right).
\label{p_rho_A1}
\eeq
where $\widetilde{\sum}^m_{n}$ denotes the single sum over $n$ with $n\neq m$ and $\omega_m,\omega_n$ satisfying Supplementary Eq.~(\ref{ord}). With the substitution of Supplementary Eqs.~(\ref{rho_A1_v2}) and (\ref{p_rho_A1}),
\beq
\begin{aligned}
\tilde{\rho}_{A1}(\boldsymbol{\varphi})\partial_{\varphi_m} \tilde{\rho}_{A1}(\boldsymbol{\varphi})
=&\widetilde{\sum}_{p,q}\widetilde{\sum}_{n}^m \left( \cos(\varphi_p - \varphi_q)\mathscr{C}_{pq} + \sin(\varphi_p - \varphi_q)\mathscr{S}_{pq} \right)\\
&\quad\quad\quad\,\,\,\, \times\left( -\sin(\varphi_m - \varphi_n)(\mathscr{C}_{mn} + \mathscr{C}_{nm}) + \cos(\varphi_m - \varphi_n)(\mathscr{S}_{mn} - \mathscr{S}_{nm}) \right),
\label{eq:209}
\end{aligned}
\eeq
where the first sum is over $p, \,q$ and the second over $n$.

When we substitute Supplementary Eq.~(\ref{eq:209}) into the right-hand side of Supplementary Eq.~(\ref{eq:207}), altogether $16$ terms result, each of which takes the following general structure:
\beq
\begin{aligned}
\sum_{m=1}^D\widetilde{\sum}_{p,q}  \widetilde{\sum}^m_{n} \widetilde{\sum}_{p',q'}  \widetilde{\sum}^m_{n'} \operatorname{Tr}_{A}(\cdots) \operatorname{Tr}_{A}(\cdots) \mathfrak{g}(\varphi_p-\varphi_q)\mathfrak{h}(\varphi_m-\varphi_n)\mathfrak{g}'(\varphi_{p'}-\varphi_{q'})\mathfrak{h}'(\varphi_{m}-\varphi_{n'}),
\label{eq:210}
\end{aligned}
\eeq
where $\mathfrak{g},\mathfrak{h},\mathfrak{g}',\mathfrak{h}'$ stand for the symbol of functions: $\cos$, $\sin$, and the trace is completely determined by $\mathscr{C}$, $\mathscr{S}$ and thus is independent of $\boldsymbol{\varphi}$. Now we perform the $\boldsymbol{\varphi}$ average of Supplementary Eq.~(\ref{eq:210}). It can be readily seen that each $\varphi_n$ in $\mathfrak{h}(\varphi_m - \varphi_n)$ must be paired uniquely with $\varphi_{n'}$ in $\mathfrak{h}'(\varphi_m - \varphi_{n'})$, i.e. $n=n'$. So, with the averaging performed the triple sum: $\sum_{m=1}^D \widetilde{\sum}^m_{n}\widetilde{\sum}^m_{n'}$ reduces to a double sum and brings us a factor ${\cal W}$. Furthermore, every $(\varphi_p ,\varphi_q)$ in  $\mathfrak{g}(\varphi_p -\varphi_q)$ must be paired with $(\varphi_{p'} ,\varphi_{q'})$ in  $\mathfrak{g}'(\varphi_{p'} -\varphi_{q'})$, i.e. $(p,q)=(p',q')$ or $(p,q)=(q',p')$. So the quadruple sum $\widetilde{\sum}_{p,q}\widetilde{\sum}_{p',q'}$ reduces to a double sum and brings us another ${\cal W}$ factor. Since each trace is $\sim \frac{1}{D^2}$, we finally have
\beq
\text{Var}(O) \sim \left({\cal W}/D^2\right)^2.
\eeq
With the substitution of Supplementary Eq.~(\ref{number_fre_v1}) it gives
\beq
{\rm Var}(O)\sim L_A^{\beta} \mathrm{e}^{-\kappa L},\quad \beta=2\mu\in [2,4]
,\quad\kappa=2(1-\nu)\kappa_D,
\label{eq:213}
\eeq
which is confirmed numerically; see Fig.~5(b-d) in the main text and more numerical results given below. By their definitions, both $\kappa$ and $\beta$ depend on system's parameters and the initial state in general.

\subsection{Details and more results of numerical simulations}
\label{sec:numerical_tests_interacting_models}

\subsubsection{More details of numerical methods}
\label{sec:random_initial_state}

Because long-time simulations of entanglement dynamics are not well controlled within the framework of density matrix renormalization group, we perform the exact diagonalization of the XXZ model with the use of standard numerical methods [41]. This allows us to find the eigenvalue spectrum $\{\omega_m\}$ and the eigenstates $\{\Psi_m\}$. Given an initial state $\Psi(0)$, we further numerically calculate the overlap $\langle \Psi_m|\Psi(0)\rangle$, which is the superposition coefficient $\chi_m$ in Supplementary Eq.~(\ref{psi_0}). By substituting $\omega_m$, $\Psi_m$ and $\chi_m$ into Supplementary Eq.~(\ref{eq:177}), we can find the instantaneous reduced density of matrix $\rho_A(t)$ for arbitrarily long time, from which we find numerically the evolution of various information-theoretical observables: $S(t)$, $S_n(t)$, etc.. In our simulations, the maximal total system size $L$ available is $20$.

In most part of this work we consider random  initial states. There are two classes of random initial states. In the first class,
\beq
| \Psi_0 \rangle = {1\over \sqrt{D_s}}\sum_{{\mathfrak{m}}}\,{\rm e}^{i\phi_{{\mathfrak{m}}}} |{\mathfrak{m}}\rangle,
\label{eq:211}
\eeq
with the sum over all bases ${\mathfrak{m}}$ in $\mathscr{H}_{\Sz}$. Here all bases ${\mathfrak{m}}$ acquire equal magnitude $1/\sqrt{D_s}$, while the phases $\phi_{{\mathfrak{m}}}$ are statistically independent and randomly drawn from the uniform measure on $[0,2\pi)$ [38]. In the second class,
\beq
| \Psi_0 \rangle = \sum_{{\mathfrak{m}}}\,{t}_{{\mathfrak{m}}} |{\mathfrak{m}}\rangle,\quad
\sum_{{\mathfrak{m}}}\,|{t}_{{\mathfrak{m}}}|^2=1,
\label{eq:214}
\eeq
where the complex coefficients  ${t}_{{\mathfrak{m}}}$ are randomly drawn from the Haar measure on the hypersphere $\mathbb{S}^{2D_s-1}$, and thus their magnitudes are not uniform, unlike Supplementary Eq.~(\ref{eq:211}).

\begin{figure}[t]
\begin{center}
\includegraphics[width=16.cm]{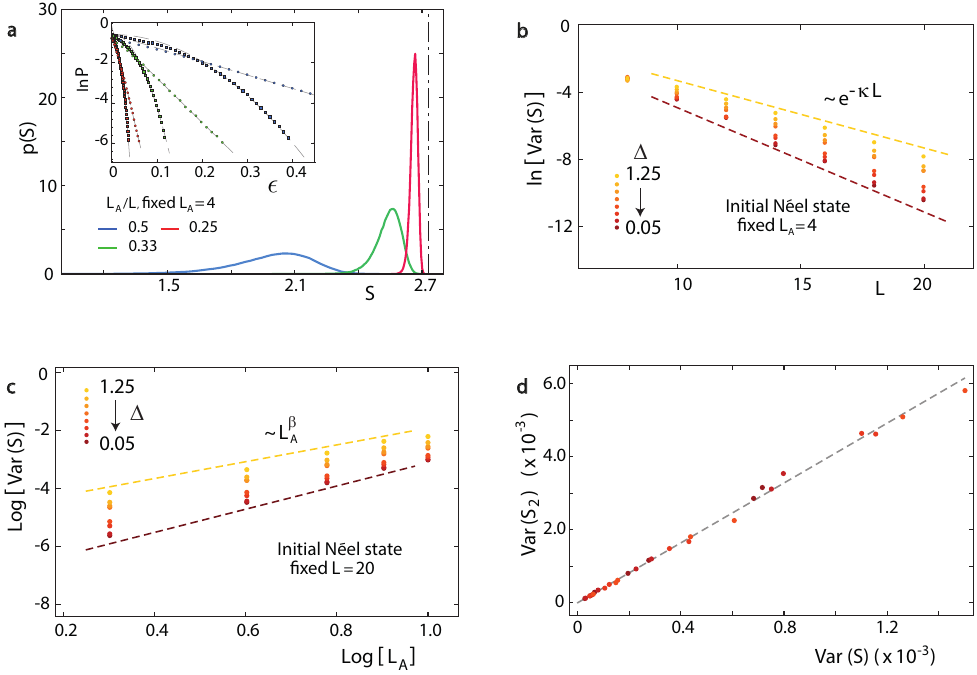}
\end{center}
{
\caption{{\bf Entanglement fluctuations with initial state being Neel.} Simulation results of mesoscopic fluctuations in entanglement dynamics, with $|\Psi(0)\rangle=|{\rm Neel}\rangle$. (a) Variation of the distribution of the entanglement entropy with increasing $L$ at fixed $L_A$ (main panel). The upper (squares) and lower (circles) tail of the deviation probability $\mathbf{P}(|S-\langle S\rangle|\geq \epsilon)$ is well fitted by Eq.~(2) in the main text (dashed lines). $\Delta=0.1$. (b) The simulated ${\rm Var}(S)$ (symbols) for different anisotropy parameter $\Delta$ all display an exponential decay in $L$, when $L_A$ is fixed. The value of $\kappa$ varies from $0.62$ for $\Delta=0.05$ to $0.4$ for $\Delta=1.25$. (c) The simulated ${\rm Var}(S)$ (symbols) for different $\Delta$ all display a power-law increase in $L_A$, when $L$ is fixed. (d) For different $\Delta$ and small $L_A/L$, simulations show that ${\rm Var}(S)\propto{\rm Var}(S_2)$.}
\label{fig:S5}
}
\end{figure}

\subsubsection{More results}
\label{sec:Neel_state}

In the main text (Fig.~5) we report the simulation results for the first class of random $\Psi(0)$. Simulations further show that, provided the random $\Psi(0)$ is changed from the first class to the second while all system parameters, $\Delta,\,L_A,\,L$, are not changed, the fluctuation statistics of entanglement dynamics stay exactly the same. (For this reason, we do not present those simulation results here.) In particular, the scaling behaviors of the variance of $S,S_n$ obey Supplementary Eq.~(\ref{eq:213}) also, and the power $\beta$ and the exponential decay rate $\kappa$ have the same value as those for the first class of $\Psi(0)$.

\begin{figure}[t]
\begin{center}
\includegraphics[width=17.5cm]{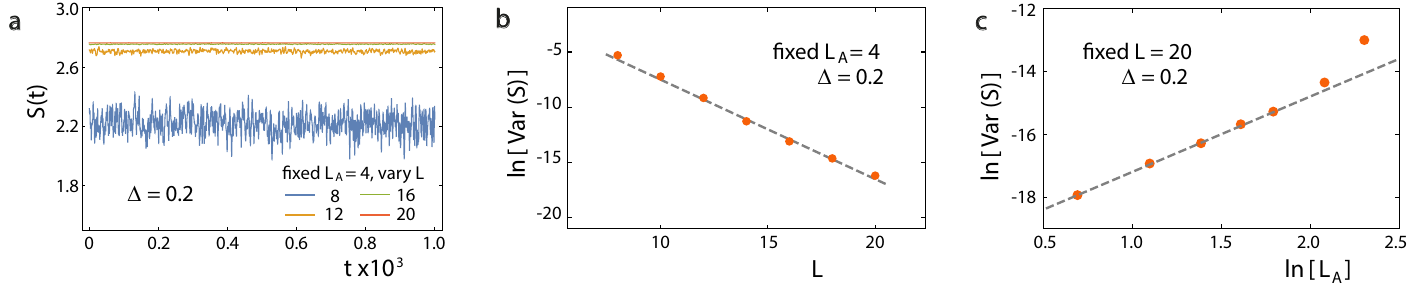}
\end{center}
{
\caption{{\bf Short-time entanglement dynamics of XXZ model.} We perform short-time simulations of entanglement dynamics of the Heisenberg XXZ model up to $t=1000$ in unit of $\hbar/J$ for an initial random state. (a) The patterns of the entanglement entropy $S(t)$ for different $L$ with fixed $L_A$. (b) Statistical analysis of data obtained from short-time evolution shows that ${\rm Var}(S)$ displays an exponential decay in $L$ when $L_A$ is fixed. (c) And it displays a power law increase in $L_A$ when $L$ is fixed.}
\label{fig:S6}
}
\end{figure}

We further perform numerical simulations with $\Psi(0)$ being a Neel state described by Supplementary Eq.~(\ref{eq:212}). The simulation results are shown in Supplementary Fig.~\ref{fig:S5}. From the main panel of (a) we see that, similar to the Rice-Mele model, TFIC, and the XXZ model with random $\Psi(0)$, the distribution of the entanglement entropy $\mathbf{P}(S)$  becomes narrower and narrower, when the ratio $L_A/L$ decreases. The inset of (a) further shows that the distribution of upward (downward) fluctuations is sub-Gaussian (sub-Gamma), with the corresponding large deviation probability described by the first and second line of Eq.~(2) in the main text, respectively. From (b) and (c) we see that the variance of $S$ also follows Supplementary Eq.~(\ref{eq:213}), but with different exponential decay rate $\kappa$ and different power $\beta$. From (d) we see that for small $L_A/L$ the scaling behaviors of different probes: $S,\,S_2$ are the same. So for all initial states considered in this work, the ensuing entanglement dynamics displays universal mesoscopic fluctuations, and the details of the initial state enter only into the parameters $\kappa$ and $\beta$.

So far we have considered long-time evolution. In Supplementary Fig.~\ref{fig:S6} we present the simulation results of short-time evolution of entanglement dynamics up to $t=1000$. It is clear that the scaling law Supplementary Eq.~(\ref{eq:213}) already appears for such short time. This makes present findings particularly relevant to experiments on entanglement dynamics [30], where the evolution time is short but temporal fluctuations of the entanglement entropy were observed.


    \setcounter{section}{0}
    \renewcommand{\thesection}{Supplementary Appendix \Alph{section}}

\section{Entanglement evolution in transverse field Ising chain}
\label{sec:TFIC}

The TFIC Hamiltonian is given by
\beq
H=-\frac{1}{2}\sum_{i=1}^L (\sigma_i^x \sigma_{i+1}^x +h \sigma_i^z),
\eeq
with the periodic boundary condition $\sigma_{L+1}^\alpha = \sigma_1^\alpha$, and the Pauli matrices $\sigma^{x,z}_i$ act on the spin at site $i$ belonging to a spin chain of length $L$. $h$ is the external magnetic field. We also carried a parallel study of the entanglement entropy evolution of TFIC with the use of the numerical solution of the correlation matrix. The method of finding that solution is described in standard textbooks and we simply use analytical expressions given in Ref. \cite{Cardy05}. The key formulae for the numerical computations are Eqs.~(3.3) -- (3.6) in that work. The expressions for finite $L$ simply amount to converting the continuous momentum variable $\varphi$ in those formulae back to a discrete one. That is, for a generic expression $F(\varphi)$, we have
\beq
\frac{1}{2\pi}\int_0^{2\pi}d\varphi \,  F(\varphi)  \,\,\,\rightarrow \,\,\, \frac{1}{L}\sum_{n=-L/2}^{L/2-1}\,F\left(\left(n+{1\over 2}\right){2\pi\over L}\right).
\eeq
Note that on the right-hand side the Bloch momenta are quantized at half-integers. This arises from the antiperiodic boundary condition for the Jordan-Wigner fermions in the even fermion number sector. For studies of the time-dependent correlation matrix, we stay in the paramagnetic phase with $h>1$ when quenching the magnetic field. Moreover, the pre-quench state is the ground state of the pre-quench Hamiltonian.

\section{Formulae for correlation matrix of the Rice-Mele model}\label{app:rm}
Here we give the expressions for the elements of $\check{\gamma}(k), \check{\alpha} (k), \check{\beta}(k)$. In the momentum space, the Hamiltonian Supplementary Eq.~(\ref{eq:1}) reduces to a $2\times 2$ Bloch Hamiltonian which reads
\beq
H(k)=\boldsymbol{B}(k)\cdot \boldsymbol{\sigma}
\label{eq:2}
\eeq
where $\boldsymbol{B}(k)=(B_x(k),B_y(k),B_z(k))$ with
\beq
B_x(k)=-(J+J') \cos(k/2),\,\,\,B_y(k)=(J-J') \sin(k/2),\,\,\,B_z(k)=M,
\label{eq:3}
\eeq
and $\boldsymbol{\sigma}=\{\sigma^x,\sigma^y,\sigma^z\}$, with $\sigma^{x,y,z}$ being the Pauli matrices. The $L$ Bloch momenta are
\beq
k=0,\pm{2\pi m\over L}, \quad m=1,2,\ldots,{L-1\over 2}
\label{eq:10}
\eeq
for odd $L$ and
\beq
k=0,\pi,\pm{2\pi m\over L}, \quad m=1,2,\ldots,{L\over 2}-1
\label{eq:11}
\eeq
for even $L$.

The energy spectrum of the Bloch Hamiltonian Supplementary Eq.~(\ref{eq:2}) is
\beq
\pm E_k=\pm |\boldsymbol{B}(k)| = \pm \sqrt{M^2 +J^2+J'^2+ 2 J J' \cos k}.\label{eq:11a}
\eeq
The $-$ ($+$) sign corresponds to the lower (upper) band. Note that this spectrum exhibits the particle-hole symmetry. Moreover, $E_k$ is symmetric with respect to $k=0$, i.e.
\beq
E_k=E_{-k}.
\label{eq:12}
\eeq
That is, the spectrum bears the reflection symmetry. The eigenstates are $SU(2)$ spinors. The eigenstate corresponding to $E_k$ is
\beq
\left(
  \begin{array}{c}
    u^-_k \\
    v^-_k \\
  \end{array}
\right)=\left(
          \begin{array}{c}
            \sin\frac{\theta_k}{2} \\
            -{\rm e}^{i\phi_k}\cos\frac{\theta_k}{2} \\
          \end{array}
        \right)
\label{eq:5}
\eeq
and to $-E_k$ is
\beq
\left(
  \begin{array}{c}
    u^+_k \\
    v^+_k \\
  \end{array}
\right)=\left(
          \begin{array}{c}
            \cos\frac{\theta_k}{2} \\
            {\rm e}^{i\phi_k}\sin\frac{\theta_k}{2} \\
          \end{array}
        \right),
\label{eq:6}
\eeq
where $\cos\theta_k=M/E_k, \tan\phi_k=B_y(k)/B_x(k)=-{J-J'\over J+J'}\tan({k\over 2})$. It is obvious that
\begin{equation}\label{eq:114}
  \theta_k=\theta_{-k},\qquad \phi_k=-\phi_{-k}.
\end{equation}

With the help of $\Psi(t) = {\rm e}^{-i H_f t/\hbar} \Psi(0)$ as well as Supplementary Eqs. (\ref{eq:5}) and (\ref{eq:6}), the matrix elements of the correlation matrix $C_{i\sigma,i'\sigma'}(t)$ can be computed to give
\beq\label{corrmat1}
&&C_{i\A,i'\A}(t)=\frac{1}{L}\sum_k \, {\rm e}^{i k (i-i')} \,|\tilde{u}_{fk}(t)|^2= \delta_{ii'}-C_{i\B,i'\B}(t),\\\label{corrmat2}
&&C_{i\A,i'\B}(t)=\frac{1}{L}\sum_k \, {\rm e}^{i k (i-i'+{1\over 2})}  \,\tilde{u}^*_{fk}(t) \tilde{v}_{fk}(t) = C^*_{i'\B,i\A}(t)
\eeq
with the summation over all Bloch momenta, since we take the initial state as the half-filling ground state of the pre-quench Hamiltonian. The time-dependent form factors in Supplementary Eqs.~(\ref{corrmat1}) and (\ref{corrmat2}) are given by
\beq
\tilde{u}_{fk}(t) = a_k\, u_{fk}^-\, {\rm e}^{i E_{fk} t} + b_k \,u_{fk}^+ \,{\rm e}^{-i E_{fk} t}
\label{eq:18}
\eeq
and
\beq
\tilde{v}_{fk}(t) = a_k\, v_{fk}^- \,{\rm e}^{i E_{fk} t} + b_k \,v_{fk}^+ \,{\rm e}^{-i E_{fk} t}.
\label{eq:19}
\eeq
The coefficients $a_k$ ($b_k$) are overlaps between the half-filling ground state of the pre-quench Hamiltonian $H_0$ and the lower (upper) band eigenstates of the post-quench Hamiltonian $H_f$, with
\beq
a_k=(u^{-*}_{fk},v^{-*}_{fk}) \left( \begin{array}{c}u^-_{0k}\\v^-_{0k}\end{array} \right),\,\,\,\,
b_k=(u^{+*}_{fk},v^{+*}_{fk}) \left(\begin{array}{c}  u^-_{0k}\\v^-_{0k} \end{array}\right),\label{eq:13}
\eeq
where the subscript $0$ denotes the eigenstate of $H_0$.

For $\sigma=\sigma'$, with the help of Supplementary Eq.~(\ref{eq:18}) we find that
\beq
|\tilde{u}_{fk}(t)|^2=|a_k|^2 (u_{fk}^{-})^2+|b_k|^2 (u_{fk}^{+})^2+u_{fk}^+ u_{fk}^- (a_k b_k^*+a_k^*b_k)\cos(2 E_{fk} t)+i u_{fk}^+ u_{fk}^- (a_k b_k^*-a_k^*b_k)\sin(2 E_{fk} t).
\eeq
Inserting it into Supplementary Eq.~(\ref{corrmat1}) we get
\beq
&&\gamma_{\A\A}(k)=1-\gamma_{\B\B}(k)=|a_k|^2 (u_{fk}^{-})^2+|b_k|^2 (u_{fk}^{+})^2,\label{eq:21}\\
&&\alpha_{\A\A}(k)=-\alpha_{\B\B}(k)=u_{fk}^+ u_{fk}^- (a_k b_k^*+a_k^*b_k),\label{eq:22}\\
&&\beta_{\A\A}(k)=-\beta_{\B\B}(k)=i u_{fk}^+ u_{fk}^- (a_k b_k^*-a_k^*b_k).\label{uv1}
\eeq
For $\sigma\neq \sigma'$, with the help of Supplementary Eqs.~(\ref{eq:18}) and (\ref{eq:19}) we find that
\beq
&&\tilde{u}^*_{fk}(t) \tilde{v}_{fk}(t)={\rm e}^{i\phi_{fk}}   \biggl(    |a_k|^2 u_{fk}^- {v'}_{fk}^- +  |b_k|^2 u_{fk}^+ {v'}_{fk}^++(a_k^* b_k u_{fk}^- {v'}_{fk}^+ + a_k b_k^* u_{fk}^+ {v'}_{fk}^- )\cos(2 E_{fk} t)\nn\\
&&\textrm{ \ \ \ \ \ \ \ \ \ \ \ \ \ \ \ \ \ \ \ \ \ \ \ }-i(a_k^* b_k u_{fk}^- {v'}_{fk}^+ - a_k b_k^* u_{fk}^+ {v'}_{fk}^- )\sin(2 E_{fk} t)\, \biggr),
\eeq
where $v_{fk}^\pm = {\rm e}^{i\phi_{fk}} \,{v'}_{fk}^{\pm}$, and $\phi_{fk}$ is defined in the same way as $\phi_k$ but with post-quench Hamiltonian parameters. Inserting it into Supplementary Eq.~(\ref{corrmat2}) we get
\beq
&&\gamma_{\A\B}(k)=\gamma_{\B\A}(k)^*={\rm e}^{i\delta(k)}\left(|a_k|^2u_{fk}^- {v'}_{fk}^-+ |b_k|^2u_{fk}^+ {v'}_{fk}^+\right),\label{eq:23}\\
&&\alpha_{\A\B}(k)=\alpha_{\B\A}(k)^*= {\rm e}^{i\delta(k)}\left(a_k^* b_k u_{fk}^- {v'}_{fk}^+ + a_k b_k^* u_{fk}^+ {v'}_{fk}^- \right),\label{eq:24}\\
&&\beta_{\A\B}(k)=\beta_{\B\A}(k)^*=-i{\rm e}^{i\delta(k)}\left(a_k^* b_k u_{fk}^- {v'}_{fk}^+ -a_k  b_k^* u_{fk}^+ {v'}_{fk}^-\right),\label{uv2}
\eeq
and $\delta(k)\equiv k/2+\phi_{fk}$. We note that $u_{fk}^{\pm}$, ${v'}_{fk}^{\pm}$ are real, whereas $a_{k}$, $b_{k}$, and $v_{{fk}}^{\pm}$ are generally complex except for $k=0$ when they are also real. Furthermore, $u_{fk}^{\pm}$, ${v'}_{fk}^{\pm}$ are even in $k$, $\delta(k)$ is odd in $k$, and $a_k^*=a_{-k}$, $b_k^*=b_{-k}$.

\section{Some general remarks on quantum expectation value of operators upon a quench}\label{app:expectation}
When a generic isolated many-body system with Hamiltonian $H_f$ is driven out of equilibrium, under Schr{\"o}dinger evolution it evolves from an initial state $\Psi(0)$ to a state $\Psi(t)$ at time $t$, given by
\beq
\Psi(t) = {\rm e}^{-i H_f t} \,\Psi(0).
\label{eq:8}
\eeq
In the basis of many-body eigenstates $\Psi_{\bf m}$ (labelled by {\bf m}), Supplementary Eq.~(\ref{eq:8}) is written as
\beq
\Psi(t)=\sum_{\bf m} {\rm e}^{-i\mathcal{E}_{\bf m} t}\,w_{\bf m}\,\Psi_{\bf m},
\label{eq:162}
\eeq
where $\mathcal{E}_{\bf m}$'s are the many-body eigenenergies and $w_{\bf m}$'s are the superposition coefficients of $\Psi(0)$. Under this evolution, the expectation value of a generic operator ${\cal A}$ at time $t$ is
\begin{eqnarray}
\langle \Psi(t)|{\cal A}|\Psi(t)\rangle=\sum_{{\bf m} {\bf m}'}{\rm e}^{-i(\mathcal{E}_{\bf m}-\mathcal{E}_{{\bf m}'}) t}\,w_{{\bf m}'}^* \,w_{\bf m}\,\langle \Psi_{{\bf m}'}|{\cal A}|\Psi_{\bf m}\rangle.
\label{eq:166}
\end{eqnarray}
So the time parameter enters the evolution $\langle \Psi(t)|{\cal A}|\Psi(t)\rangle$ through a large number $\sim {\cal L}^2$ of dynamical phases: $(\mathcal{E}_{\bf m}-\mathcal{E}_{{\bf m}'})t$ with $\mathcal{E}_{\bf m}\neq \mathcal{E}_{{\bf m}'}$. Here ${\cal L}$ is the number of many-body eigenstates superposing the initial state. Investigations of the emergence of statistical mechanics from such evolving quantum expectation values were initiated in Ref.~[31]. Nowadays it is accepted [32-35] that the quantum expectation can equilibrate after long time, when out-of-equilibrium fluctuations around the equilibrium value are small. However, despite of some progresses it remains a difficult problem to demonstrate under what circumstances this scenario would emerge from quantum dynamics of an isolated system.

Compared to Supplementary Eq.~(\ref{eq:166}), we find that the correlation matrix Supplementary Eq. (\ref{eq:corr}) is simplified substantially. The general reasons underlying this simplification are as follows. First, the Rice-Mele model is composed of free fermions, and many-body effects enter through the so-called exchange interaction, namely, the particle indistinguishability [43]. Second, the system is driven out of equilibrium by global quench. For the first reason each many-body eigenstate ${\bf m}$ corresponds to a configuration, $\{n^{\zeta}_{k}\}$, of the occupation number $n^{\zeta}_{k}=0,1$ at the single-particle eigenstate $(u^\zeta_{fk}, v^\zeta_{fk})^T$ of $H_f$, with $\zeta=+\,(-)$ denoting the particle (hole) band and the superscript $T$ denoting the transpose (see
\ref{app:rm}). Because the quench is global and the pre-quench state $\Psi(0)$ is a half-filling ground state,
the configurations corresponding to the many-body eigenstates of $H_f$ superposing $\Psi(t)$ --- {\it not} all the many-body eigenstates --- must satisfy
\begin{equation}\label{eq:165}
  n^{+}_{k}+n^{-}_{k}=1,\quad {\rm for\,\, every\,\, Bloch\,\,momentum\,}\, k,
\end{equation}
and the corresponding many-body eigenenergies are given by
\begin{equation}\label{eq:163}
  {\cal E}_{\bf m}=\sum_k (n^+_k-n^-_k)E_{fk}=\sum_k (1-2n^-_k)E_{fk},
\end{equation}
where to derive the first equality we have used the reflection symmetry Supplementary Eq.~(\ref{eq:12}) and to derive the second we have used Supplementary Eq.~(\ref{eq:165}). Taking these into account we simplify Supplementary Eq.~(\ref{eq:166}) to
\begin{eqnarray}
\langle \Psi(t)|{\cal A}|\Psi(t)\rangle=\sum_{\{{n_k^-}{}^\prime\}}\sum_{\{{n^-_k}\}}{\rm e}^{it\sum_k({n^-_k}-{n_k^-}{}^\prime) {2E_{fk}}}\,w_{{\bf m}'}^* \,w_{\bf m}\,\langle \Psi_{{\bf m}'}|{\cal A}|\Psi_{\bf m}\rangle.
\label{eq:32}
\end{eqnarray}
Note that due to Supplementary Eq.~(\ref{eq:165}) the many-body eigenstate ${\bf m}$ (or ${\bf m}'$) here is completely fixed by the occupation number configuration of the hole band $\{n_k^-\}\equiv {\bf m}^-$ (or $\{{n_k^-}{}^\prime\}\equiv {\bf m}'{}^-$). In Supplementary Eq.~(\ref{eq:32}), the dynamical phases are governed by the frequencies associated with the energy gaps $2E_{fk}$ of the single-particle band structure and the difference in hole numbers ${n^-_k}-{n_k^-}{}^\prime$ between two states ${\bf m},\,{\bf m}'$.

Then we apply Supplementary Eq.~(\ref{eq:32}) to one-body operators, e.g. the correlation function $C(t)$ which belongs to this class. Such operators take the general form as
\begin{equation}\label{eq:167}
  {\cal A}=\sum_{\zeta\zeta'}\sum_{kk'}{\cal A}_{kk'}^{\zeta\zeta'}{\rm e}^{i(kj-k'j')}d_{k\zeta}^\dagger d_{k'\zeta'},
\end{equation}
where $d^\dag_{k\zeta}\,, d^{}_{k\zeta}$ are, respectively, fermionic creation and annihilation operators at the Bloch momentum $k$ and at the band $\zeta$, the coefficients ${\cal A}_{kk'}^{\zeta\zeta'}$  are complex in general, and $j,j'=1,\ldots, L$ label the unit cells. It is easy to see that
\begin{eqnarray}\label{eq:168}
  \langle\Psi_{{\bf m}'}|d_{k\zeta}^\dagger d_{k'\zeta'}|\Psi_{\bf m}\rangle&=&\delta_{kk'}\langle\Psi_{{\bf m}'}|d_{k\zeta}^\dagger d_{k\zeta'}|\Psi_{\bf m}\rangle\nonumber\\
  &&\times \left\{\begin{array}{ll}
           n_k^\zeta \prod_{k'}\delta_{{n_{k'}^-}{}^\prime{n_{k'}^-}}, &{\rm for}\,\,\zeta=\zeta'\\
           n_k^{\zeta'} \delta_{(1-{n_k^-}{}^\prime)n_k^-}\prod_{k'\neq k}\delta_{{n_{k'}^-}{}^\prime{n_{k'}^-}}, &{\rm for}\,\,\zeta=-\zeta'
         \end{array}
  \right.,
\end{eqnarray}
due to the constraint Supplementary Eq.~(\ref{eq:165}). The product of Kronecker symbols in the second line shows that ${\bf m}'$ is uniquely determined by ${\bf m}$ (more precisely, its occupation number configuration ${\bf m}^-$ at the hole band) for given $\zeta,\,\zeta'$, and thus we denote it as ${\bf m}'({\bf m}^-)$. With the substitution of Supplementary Eqs.~(\ref{eq:167}) and (\ref{eq:168}) into Supplementary Eq.~(\ref{eq:32}), we obtain
\begin{equation}\label{eq:169}
  \langle \Psi(t)|{\cal A}|\Psi(t)\rangle={\cal A}_0+\delta {\cal A}(t),
\end{equation}
where ${\cal A}_0$ has no time dependence,
\begin{eqnarray}\label{eq:170}
 {\cal A}_0=\sum_{k}{\rm e}^{ik(j-j')}\sum_{\zeta} {\cal A}_{kk}^{\zeta\zeta} \sum_{{\bf m}^-} n_k^\zeta |w_{\bf m}|^2,
\end{eqnarray}
and $\delta {\cal A}$ depends on time,
\begin{eqnarray}\label{eq:171}
 \delta {\cal A}(t)=\sum_{k}{\rm e}^{ik(j-j')}\sum_{\zeta} {\cal A}_{kk}^{\zeta(-\zeta)} \sum_{{\bf m}^-} n_k^{-\zeta}\, w_{{\bf m}'({\bf m}^-)}^*\,w_{\bf m} \,{\rm e}^{it(2n_k^--1){2E_{fk}}}.
\end{eqnarray}
The first factors in Supplementary Eqs.~(\ref{eq:170}) and (\ref{eq:171}), ${\rm e}^{ik(j-j')}$, imply that the expectation value is translationally invariant in the unit cell index. Most importantly, we can rewrite Supplementary Eq.~(\ref{eq:171}) as
\begin{eqnarray}\label{eq:172}
 \delta {\cal A}(t)&=&\sum_{k}{\rm e}^{ik(j-j')+i{2E_{fk}t}}\,{\cal A}_{kk}^{+-} \,\sum_{{\bf m}^-} \delta_{1n_{k}^-}\,w_{{\bf m}'({\bf m}^-)}^*\,w_{\bf m}\nonumber\\
  &+&\sum_{k}{\rm e}^{ik(j-j')-i{2E_{fk}t}}\,{\cal A}_{kk}^{-+} \,\sum_{{\bf m}^-} \delta_{0n_{k}^-}w_{{\bf m}'({\bf m}^-)}^*\,w_{\bf m}.
\end{eqnarray}
Taking into account the symmetries Supplementary Eq.~(\ref{eq:11a}) and (\ref{eq:12}), we thus arrive at the same physical picture as in Supplementary Note $2.1$.
That is, for a generic one-body operator ${\cal A}$,
\beq
\langle \Psi(t)|{\cal A}|\Psi(t)\rangle \leftrightsquigarrow \boldsymbol{\varphi}=\boldsymbol{\omega}t\in \mathbb{T}^N,
\label{eq:16}
\eeq
i.e. the quantum expectation $\langle \Psi(t)|{\cal A}|\Psi(t)\rangle$ corresponds one-to-one to a point $\boldsymbol{\omega}t$ along the classical trajectory on $\mathbb{T}^N$.

\section{Proof of two concentration inequalities: Supplementary (\ref{eq:72}) and (\ref{eq:73})}
\label{sec:proof}

We shall prove the Supplementary inequality (\ref{eq:72}) only. The Supplementary inequality (\ref{eq:73}) can be proved in the same way. Since $L$ and thereby $N$ --- the number of independent random variables $\varphi_m$ --- are finite, we adopt the theory of concentration of measure [15-17].

First of all, because $\mathfrak{R}_{l,\sigma\sigma'}(t)$ is a quasiperiodic in $t$, applying Supplementary Eq.~(\ref{eq:71}) gives
\beq
\lim_{T\rightarrow\infty}\int_{0}^{T}\frac{dt}{T}\mathfrak{R}_{l,\sigma\sigma'}(t)=\int {d\boldsymbol{\varphi}\over (2\pi)^N} \mathfrak{R}_{l,\sigma\sigma'}(\boldsymbol{\varphi})=0.
\label{eq:82}
\eeq
Then, by Supplementary Eq.~(\ref{eq:96}) we have
\begin{eqnarray}
\lim_{T\rightarrow \infty}\int_{|\mathfrak{R}_{l,\sigma\sigma'}(t)|\geq \epsilon} \frac{dt}{T}
=\mathbf{P}\left(|\mathfrak{R}_{l,\sigma\sigma'}(\boldsymbol{\varphi})|\geq \epsilon\right).
\label{eq:76}
\end{eqnarray}
The rest of the proof is to bound the probability on the right-hand side. Using Supplementary Eq.~(\ref{eq:42}) we find that
\beq
     \partial_{\varphi_m}\mathfrak{R}_{l,\sigma\sigma'}(\boldsymbol{\varphi}) = -\frac{1}{L}R_{l,\sigma\sigma'}\left(k_m\right)\,\sin \varphi_m.
\label{eq:77}
\eeq
Thus
\beq
     |\partial_{\boldsymbol{\varphi}}\mathfrak{R}_{l,\sigma\sigma'}(\boldsymbol{\varphi})|^2
     &=&\sum_{m=0}^{N-1}(\partial_{\varphi_m}\mathfrak{R}_{l,\sigma\sigma'}(\boldsymbol{\varphi}))^2 = \frac{1}{L^2}\sum_{m=0}^{N-1}\left(R_{l,\sigma\sigma'}\left(k_m\right)\right)^2\,\sin^2 \varphi_m\nonumber\\
     &\leq&\frac{1}{L^2}\sum_{m=0}^{N-1}\left(R_{l,\sigma\sigma'}\left(k_m\right)\right)^2.
\label{eq:78}
\eeq
For $L\gg 1$ the sum converges well to the integral, giving
\beq
     \frac{1}{L^2}\sum_{m=0}^{N-1}\left(R_{l,\sigma\sigma'}\left(k_m\right)\right)^2\approx \frac{1}{2L}\int_{-\pi}^{\pi}\frac{dk}{2\pi}
     \left({\rm e}^{ikl}\alpha_{\sigma\sigma'}(kl)+{\rm e}^{-ikl}\alpha_{\sigma\sigma'}(-kl)\right)^2\equiv \|\mathfrak{R}_{l,\sigma\sigma'}\|_{\rm Lip}^2.
\label{eq:79}
\eeq
Since $\mathbb{T}^N=\mathbb{R}^N/\mathbb{Z}^N$ inherits the Riemann structure from the flat Euclidean space $\mathbb{R}^N$, we have
\begin{equation}\label{eq:80}
  |\mathfrak{R}_{l,\sigma\sigma'}(\boldsymbol{\varphi})-\mathfrak{R}_{l,\sigma\sigma'}(\boldsymbol{\varphi}')|\leq \|\mathfrak{R}_{l,\sigma\sigma'}\|_{\rm Lip} \|\boldsymbol{\varphi}-\boldsymbol{\varphi}'\|.
\end{equation}
That is, the function $\mathfrak{R}_{l,\sigma\sigma'}(\boldsymbol{\varphi})$ is Lipschitz continuous with respect to the Euclidean norm $\|\cdot\|$, and $\|\mathfrak{R}_{l,\sigma\sigma'}\|_{\rm Lip}$ is called the {\it Lipschitz constant}. Combining this continuity property with a well-known theorem in the theory of concentration of measure [17], we find that
\beq
\mathbf{P}\left(|\mathfrak{R}_{l,\sigma\sigma'}(\boldsymbol{\varphi})|\geq \epsilon\right)
\leq 2{\rm e}^{-\frac{\delta'}{\|\mathfrak{R}_{l,\sigma\sigma'}\|_{\rm Lip}^2}\, \epsilon^2},
\label{eq:81}
\eeq
where $\delta'$ is some absolute constant of order unity. With the substitution of Supplementary Eq.~(\ref{eq:79}) into the right-hand side, we justify the Supplementary inequality (\ref{eq:72}) and Supplementary Eq.~(\ref{eq:74}).

\section{Statistical independence of correlation matrix elements}
\label{sec:independence}

In this Supplementary Appendix we show that for large $L$ any random variable $\mathfrak{R}_{l_1,\sigma_1\sigma_1'},\,\mathfrak{I}_{l_1,\sigma_1\sigma_1'}$ in the block $(C(t))_{l_1}$ is statistically independent of $\mathfrak{R}_{l_2,\sigma_2\sigma_2'},\,\mathfrak{I}_{l_2,\sigma_2\sigma_2'}$ in a distinct block $(C(t))_{l_2}$ ($l_1\neq l_2$), or equivalently, any random variable $\mathfrak{R}_{l_1,\sigma_1\sigma_1'},\,\mathfrak{I}_{l_1,\sigma_1\sigma_1'}$ in the block $(\tilde{C}(\boldsymbol{\varphi}))_{l_1}$ is statistically independent of $\mathfrak{R}_{l_2,\sigma_2\sigma_2'},\,\mathfrak{I}_{l_2,\sigma_2\sigma_2'}$ in a distinct block $(\tilde{C}(\boldsymbol{\varphi}))_{l_2}$. Without loss of generality we prove the statistical independence of $\mathfrak{R}_{l_1,\sigma_1\sigma_1'}$ and $\mathfrak{I}_{l_2,\sigma_2\sigma_2'}$ below. The statistical independence of $\mathfrak{R}_{l_1,\sigma_1\sigma_1'}$ and $\mathfrak{R}_{l_2,\sigma_2\sigma_2'}$ and of $\mathfrak{I}_{l_1,\sigma_1\sigma_1'}$ and $\mathfrak{I}_{l_2,\sigma_2\sigma_2'}$ can be proven in the same way.

Consider the characteristic function of those two random variables, defined as
\begin{equation}\label{eq:44}
  \Phi_2(z_1,z_2)\equiv \lim_{T\rightarrow\infty}\int_{0}^{T}{dt\over T}\exp \left(i\left(z_1\mathfrak{R}_{l_1,\sigma_1\sigma_1'}(t)+z_2\mathfrak{I}_{l_2,\sigma_2\sigma_2'}(t)\right)\right).
\end{equation}
Applying Supplementary Eq.~(\ref{eq:71}) we obtain
\begin{equation}\label{eq:45}
  \Phi_2(z_1,z_2)=\int{d\boldsymbol{\varphi}\over (2\pi)^N}\exp \left(i\left(z_1\mathfrak{R}_{l_1,\sigma_1\sigma_1'}(\boldsymbol{\varphi})+z_2\mathfrak{I}_{l_2,\sigma_2\sigma_2'}(\boldsymbol{\varphi})\right)\right).
\end{equation}
With the substitution of $\mathfrak{R}_{l,\sigma\sigma'}(\boldsymbol{\varphi})$ and $\mathfrak{I}_{l,\sigma\sigma'}(\boldsymbol{\varphi})$ given by Supplementary Eq.~(\ref{eq:100}), we obtain
\begin{eqnarray}
\Phi_2(z_1,z_2)&=&\int{d\boldsymbol{\varphi}\over (2\pi)^N}\exp \left({i\over L}\sum_{m=0}^{N-1}\left(z_1R_{l_1,\sigma_1\sigma_1'}(k_m)\cos\varphi_m -iz_2I_{l_2,\sigma_2\sigma_2'}(k_m)\sin\varphi_m\right)\right)\nonumber\\
&=&\prod_{m=0}^{N-1}J_0\left({1\over L}\sqrt{\left(z_1R_{l_1,\sigma_1\sigma_1'}(k_m)\right)^2 -\left(z_2 I_{l_2,\sigma_2\sigma_2'}(k_m)\right)^2}\right),
\label{eq:55}
\end{eqnarray}
where $J_0(x)$ is the zero-order Bessel function. Recall that $I_{l_2,\sigma_2\sigma'_2}$ is purely imaginary. Because of $J_0(x\ll 1)\approx 1-{x^2\over 4}\approx {\rm e}^{-{x^2\over 4}}$,
\begin{eqnarray}
\Phi_2(z_1,z_2)\stackrel{L\gg 1}{\longrightarrow}\prod_{m=0}^{N-1}\exp\left(-{1\over 4L^2}\left(z_1R_{l_1,\sigma_1\sigma_1'}(k_m)\right)^2 -\left(z_2I_{l_2,\sigma_2\sigma_2'}(k_m)\right)^2\right).
\label{eq:67}
\end{eqnarray}

Next, we consider the characteristic function of the single random variable $\mathfrak{R}_{l_1,\sigma_1\sigma_1'}$, defined as
\begin{equation}\label{eq:83}
  \Phi_1(z_1)\equiv \lim_{T\rightarrow\infty}\int_{0}^{T}{dt\over T}\exp \left(iz_1\mathfrak{R}_{l_1,\sigma_1\sigma_1'}(t)\right),
\end{equation}
which is
\begin{equation}\label{eq:98}
  \Phi(z_1)=\int{d\boldsymbol{\varphi}\over (2\pi)^N}\exp \left(iz_1\mathfrak{R}_{l_1,\sigma_1\sigma_1'}(\boldsymbol{\varphi})\right).
\end{equation}
By the same token we find that
\begin{eqnarray}
\Phi_1(z_1)\stackrel{L\gg 1}{\longrightarrow}\prod_{m=0}^{N-1}\exp\left(-{1\over 4L^2}\left(z_1R_{l_1,\sigma_1\sigma_1'}(k_m)\right)^2\right).
\label{eq:84}
\end{eqnarray}
Similarly, we have
\begin{eqnarray}
\Phi_1(z_2)&\equiv& \lim_{T\rightarrow\infty}\int_{0}^{T}{dt\over T}\exp \left(iz_2\mathfrak{I}_{l_2,\sigma_2\sigma_2'}(t)\right)
=\int{d\boldsymbol{\varphi}\over (2\pi)^N}\exp \left(iz_2\mathfrak{I}_{l_2,\sigma_2\sigma_2'}(\boldsymbol{\varphi})\right)\label{eq:99}\\
&\stackrel{L\gg 1}{\longrightarrow}&\prod_{m=0}^{N-1}\exp\left({1\over 4L^2}\left(z_2I_{l_2,\sigma_2\sigma_2'}(k_m)\right)^2\right).
\label{eq:85}
\end{eqnarray}
for the characteristic function of $\mathfrak{I}_{l_2,\sigma_2\sigma_2'}$. From Supplementary Eqs.~(\ref{eq:67}), (\ref{eq:84}) and (\ref{eq:85}) we obtain
\begin{eqnarray}
\Phi_2(z_1,z_2)\stackrel{L\gg 1}{\longrightarrow}\Phi_1(z_1)\Phi_1(z_2)
\label{eq:86}
\end{eqnarray}
That is, $\Phi_2(z_1,z_2)$ factorizes into $\Phi_1(z_1)\Phi_1(z_2)$ for large $L$. This justifies the statistical independence of $\mathfrak{R}_{l_1,\sigma_1\sigma_1'}(t)$ and $\mathfrak{I}_{l_2,\sigma_2\sigma_2'}(t)$. By Supplementary Eqs.~(\ref{eq:45}), (\ref{eq:98}) and (\ref{eq:99}) it is equivalent to the statistical independence of $\mathfrak{R}_{l_1,\sigma_1\sigma_1'}(\boldsymbol{\varphi})$ and $\mathfrak{I}_{l_2,\sigma_2\sigma_2'}(\boldsymbol{\varphi})$.

\section{Some properties of $\boldsymbol{\check{\alpha}_m}$, $\boldsymbol{\check{\beta}_m}$ and $\boldsymbol{\check{\gamma}_m}$}
\label{sec:property_alpha_beta_gamma}

We recall that ${\check{\alpha}_m}$, ${\check{\beta}_m}$ and ${\check{\gamma}_m}$ are defined as
\beq
\check{\alpha}_m\equiv \left.\left(\begin{array}{cc}
                   \alpha_{\A\A}(k_m) & \alpha_{\A\B}(k_m)\\
                   \alpha_{\B\A}(k_m) & \alpha_{\B\B}(k_m)
                 \end{array}\right)\right.,\,
\check{\beta}_m\equiv \left.\left(\begin{array}{cc}
                   \beta_{\A\A}(k_m) & \beta_{\A\B}(k_m)\\
                   \beta_{\B\A}(k_m) & \beta_{\B\B}(k_m)
                 \end{array}\right)\right.,\,
\check{\gamma}_m\equiv \left.\left(\begin{array}{cc}
                   \gamma_{\A\A}(k_m) & \gamma_{\A\B}(k_m)\\
                   \gamma_{\B\A}(k_m) & \gamma_{\B\B}(k_m)
                 \end{array}\right).\right.
\label{eq:112}
\eeq
Combining Supplementary Eqs.~(\ref{eq:21})-(\ref{uv2}) with Supplementary Eqs.~(\ref{eq:5})-(\ref{eq:114}) and Supplementary Eq.~(\ref{eq:13}), we can show straightforwardly the following properties for the elements of these three matrices:
\begin{itemize}
  \item For $m>0$,
\beq
&&\textrm{(i)\ } \gamma_{\A\A}(k_m)=\gamma_{\A\A}(-k_m)\in \mathbb{R};    \nn\\
&&\textrm{(ii)\ } \gamma_{\A\B}(k_m)={\rm e}^{i\delta(k_m)}\,\gamma'_{\A\B}(k_m),\,\,\,\,\,\,\,\, \textrm{with } \delta(k_m)=-\delta(-k_m),\,\,\, \gamma'_{\A\B}(k_m)=\gamma'_{\A\B}(-k_m),\,\,\, \textrm{and}\, \gamma'_{\A\B}(k_m)\in \mathbb{R};\nn\\
&&\textrm{(iii)\ } \alpha_{\A\A}(k_m)=\alpha_{\A\A}(-k_m)\in \mathbb{R};\nn\\
&&\textrm{(iv)\ } \alpha_{\A\B}(k_m)={\rm e}^{i\delta(k_m)}\alpha'_{\A\B}(k_m),\,\,\, \textrm{with\ } \alpha'_{\A\B}(k_m)=\alpha'^{\,*}_{\A\B}(-k_m);  \nn\\
&&\textrm{(v)\ } \beta_{\A\A}(k_m)=-\beta_{\A\A}(-k_m)\in \mathbb{R} ;\nn\\
&&\textrm{(vi)\ } \beta_{\A\B}(k_m)={\rm e}^{i\delta(k_m)}\,\beta_{\A\B}'(k_m),\,\,\, \textrm{with\ } \beta'_{\A\B}(k_m)=-\beta'^{\,*}_{\A\B}(-k_m).
\label{eq:113}
\eeq
Recall that $\gamma'_{\A\B}(k_m)$, $\alpha'_{\A\B}(k_m)$ and $\beta_{\A\B}'(k_m)$ are respectively the off-diagonal ($\A\B$) component of $\check{\Gamma}'_m\,$, $\check{\alpha}'_m$ and $\check{\beta}'_m$.
  \item For  $m=0$, we have $k=0$ and
\beq
&&\textrm{(vii)\ }\gamma_{\A\A}(0)\in \mathbb{R}, \textrm{\ \ \ (viii)\ }\gamma_{\A\B}(0)\in \mathbb{R}, \textrm{\ \ \ (ix)\ }\alpha_{\A\A}(0)\in \mathbb{R}, \nn\\
&&\textrm{\ (x)\ }\alpha_{\A\B}(0)\in \mathbb{R}, \textrm{\ \ \ (xi)\ }\beta_{\A\A}(0)=0,  \textrm{\ \ \ \ \ (xii)\ }\beta_{\A\B}(0)\,\,\, \textrm{ purely\, imaginary}.
\eeq
\end{itemize}

\section{Proof of the identity Supplementary Eq.~(\ref{eq:105})}\label{ident}

In this Supplementary Appendix we prove the identity Supplementary Eq.~(\ref{eq:105}). Throughout this appendix we suppress all the subscripts $m$ and the corresponding Bloch momenta $k_m$ to make the formulae compact.

For $m=0$, by using the definitions of $\gamma$, $\tilde{\gamma}$, $\alpha$, $\tilde{\alpha}$ and  Supplementary Eqs.~(\ref{eq:21}), (\ref{eq:22}), (\ref{eq:23}) and (\ref{eq:24}),  we obtain
\beq
\gamma \alpha + \tilde{\gamma} \tilde\alpha=4ab\,\biggl( (au^-)^2+(b u^+)^2-\frac{1}{2} \biggr) u^+ u^-+2ab\,(a^2 u^-v^-+b^2 u^+ v^+)(u^- v^++u^+v^-)=0.
\eeq
The last expression is obtained with the help of the identities: $u^- u^+=-v^- v^+$, $(u^{\pm})^2+(v^{\pm})^2=1$ and $a^2+b^2=1$.

For $m>0$, similarly, by writing out the expressions explicitly we have
\beq
\gamma = \Bigl(|a|^2 (u^-)^2+|b|^2 (u^+)^2\Bigr)-\frac{1}{2},\,\,\,\,\,\,\,\tilde{\gamma}=|a|^2 u^- {v'}^-+|b|^2 u^+ {v'}^+,
\eeq
and
\beq
\alpha= 2 u^+ u^- (a b^*+a^*b),\,\,\,\,\,\,\,\tilde{\alpha}=(u^- {v'}^++u^+ {v'}^-) (a b^*+a^*b).
\eeq
So one can check that
\beq
\gamma \alpha + \tilde{\gamma} \tilde\alpha&=&(a b^*+a^*b)\biggl( |a|^2  \Bigl(2 (u^-)^3 u^++(u^-)^2 {v'}^- {v'}^++ u^-u^+({v'}^-)^2\Bigr)                   \nn\\
&&          +|b|^2  \Bigl(2 (u^+)^3 u^-+(u^+)^2 {v'}^- {v'}^++ u^-u^+({v'}^+)^2\Bigr)  -u^-u^+          \biggr)=0
\eeq
with the help of the identities: $u^- u^+= - {v'}^- {v'}^+$, $(u^{\pm})^2+({v'}^{\pm})^2=1$ and $|a|^2+|b|^2=1$.

\section{Time evolution of generic entanglement probes for $\boldsymbol{L\rightarrow\infty}$}
\label{sec:infinite_L}

For a generic entanglement probe $O$ such as $S$ and $S_n$, as shown in Supplementary Notes $6$ and $7$
its variance obeys the same scaling law as Supplementary Eq.~(\ref{eq:118}). (Note that the unimportant coefficients $a,b$ in Supplementary Eq.~(\ref{eq:118}) change, when $S$ is replaced by $O$.) As a result, at fixed $L_A$ a vanishing variance results in the limit $L\rightarrow\infty$, i.e.
\begin{equation}\label{eq:173}
  \lim_{T\rightarrow\infty}\int_{0}^{T}{dt\over T}\left(O(t)-\langle O\rangle\right)^2=0.
\end{equation}
Because $O(t)$ is continuous in time, $O(t)\neq \langle O\rangle$ cannot occur in a zero measure set. Therefore, $(O(t)-\langle O\rangle)^2$ must decay for sufficiently large $t$ (otherwise the limit does not exist), but this contradicts the quasiperiodic behavior of $O(t)$. So the only possibility for Supplementary Eq.~(\ref{eq:173}) to hold is that there exists a critical time $t_c$, such that
\begin{equation}\label{eq:174}
  O(t\geq t_c)=\langle O\rangle.
\end{equation}
In words, $O(t)$ is strictly a constant $\langle O\rangle$ beyond $t_c$. This kind of phenomena were first found numerically in Ref.~[1] and proven analytically in Ref.~[3], but only for the entanglement entropy $S$.
\\
\\
\textbf{Supplementary References:}
\\
\\
\noindent
[1] Calabrese, P. \& Cardy, J. L. Evolution of entanglement entropy in one-dimensional systems, \textit{J. Stat. Mech.} \textbf{2005}, P04010 (2005).
\\
\noindent
[2] Jin, B.-Q. \& Korepin, V. E. Quantum spin chain, Toeplitz determinants and the Fisher-Hartwig conjecture, \textit{J. Stat. Phys.} \textbf{116}, 79 (2004).
\\
\noindent
[3] Fagotti, M. \& Calabrese, P. Evolution of entanglement entropy following a quantum quench:
Analytic results for the XY chain in a transverse magnetic field, \textit{Phys. Rev. A} \textbf{78}, 010306 (2008).
\\
\noindent
[4] Peschel, I. \& Eisler, V. Reduced density matrices and entanglement entropy
in free lattice models, \textit{J. Phys. A: Math. Theor.} \textbf{42}, 504003 (2009).
\\
\noindent
[5] Altshuler, B. L. Fluctuations in the extrinsic conductivity of disordered conductors, \textit{Pis'ma Zh. E'ksp. Teor. Fiz.} \textbf{41}, 530 (1985) [\textit{JETP Lett.} \textbf{41}, 648 (1985)].
\\
\noindent
[6] Lee, P. A. \& Stone, A. D. Universal conductance fluctuations in metals, \textit{Phys. Rev. Lett.} \textbf{55}, 1622 (1985).
\\
\noindent
[7] Sheng, P. {\it Introduction to wave scattering, localization, and mesoscopic phenomena} (2nd Ed., Springer, Berlin,
Germany, 2006).
\\
\noindent
[8] Akkermans, E. \& Montambeaux, G. {\it Mesoscopic physics of electrons and photons} (Cambridge University Press, UK, 2007).
\\
\noindent
[9] Efetov, K. B. {\it Supersymmetry in Disorder and Chaos} (Cambridge University, Cambridge, UK, 1997).
\\
\noindent
[10] Beenakker, C. W. J. Random-matrix theory of quantum transport, \textit{Rev. Mod. Phys.} \textbf{69}, 731-808 (1997).
\\
\noindent
[11] Kac, M. {\it Statistical independence in probability, analysis and number theory} (The Mathematical Association of America, 1959).
\\
\noindent
[12] Arnold, V. I. {\it Mathematical methods of classical mechanics} (2nd Ed., Springer, New York, U.S., 1989).
\\
\noindent
[13] Samoilenko, A. M. Quasiperiodic oscillations, \textit{Scholarpedia} \textbf{2}, 1783 (2007).
\\
\noindent
[14] Modak, R., Alba, V. \& Calabrese, P. Entanglement revivals as a probe of scrambling in finite quantum systems, \textit{J. Stat. Mech.} \textbf{2020}, 083110 (2020).
\\
\noindent
[15] Talagrand, M. A new look at independence, \textit{Ann. Probab.} \textbf{24}, 1 (1996).
\\
\noindent
[16] Milman, V. D. \& Schechtman, G. {\it Asymptotic theory of finite-dimensional normed spaces} (Springer, Berlin,
Germany, 1986).
\\
\noindent
[17] Ledoux, M. {\it The concentration of measure phenomenon} (AMS, Providence, 2001).
\\
\noindent
[18] Bogolmony, E. Spectral statistics of random Toeplitz matrices, \textit{Phys. Rev. E} \textbf{102}, 040101(R) (2020).
\\
\noindent
[19] Bogolmony, E. \& Giraud, O. Statistical properties of structured random matrices, \textit{Phys. Rev. E} \textbf{103}, 042213 (2021).
\\
\noindent
[20] Haake, F. {\it Quantum signatures of chaos} (2nd Ed., Springer, Berlin, 2001).
\\
\noindent
[21] Evers, F. \& Mirlin, A. D. Anderson transitions, \textit{Rev. Mod. Phys.} \textbf{80}, 1355 (2008).
\\
\noindent
[22] Altshuler, B. L., Zharekeshev, I. Kh., Kotochigava, S. A. \& Shklovskii, B. I. Repulsion between levels and the
metal-insulator transition, \textit{Sov. Phys. JETP} \textbf{67}, 625 (1988).
\\
\noindent
[23] Hammond, C. \& Miller, S. J. Distribution of eigenvalues for the ensemble of real symmetric Toeplitz matrices, \textit{J. Theor. Probab.} \textbf{18}, 537 (2005).
\\
\noindent
[24] Bogomolny, E. B., Gerland, U. \& Schmit, C. Models of intermediate spectral statistics, \textit{Phys.
Rev. E} \textbf{59}, 1315(R), (2001).
\\
\noindent
[25] Davy, M., Shi, Z., Park, J., Tian, C. \& Genack, A. Z. Universal structure of transmission eigenchannels inside opaque media, \textit{Nature Commun.} \textbf{6}, 6893 (2015).
\\
\noindent
[26] Altland, A., Kamenev, A. \& Tian, C. Anderson localization from the replica formalism, \textit{Phys. Rev. Lett.} \textbf{95}, 206601 (2005).
\\
\noindent
[27] Fang, P., Zhao, L. Y. \& Tian, C. Concentration-of-measure theory for structures and fluctuations of waves, \textit{Phys. Rev. Lett.} \textbf{121}, 140603 (2018).
\\
\noindent
[28] Boucheron, S., Lugosi, G. \& Massart, P. \textit{Concentration inequalities - A nonasymptotic theory of independence} (Oxford University Press, UK, 2013).
\\
\noindent
[29] Vidmar, L., Hackl, L., Bianchi, E. \& Rigol, M. Entanglement entropy of eigenstates of quadratic fermionic Hamiltonians, \textit{Phys. Rev. Lett.} \textbf{119}, 020601 (2017).
\\
\noindent
[30] Kaufman, A. M. et al. Quantum thermalisation through entanglement in an isolated many-body system, \textit{Science} \textbf{353}, 794 (2016).
\\
\noindent
[31] von Neumann, J. Beweis des Ergodensatzes und des H-Theorems
in der neuen Mechanik, \textit{Z. Phys.} \textbf{57}, 30 (1929).
\\
\noindent
[32] Goldstein, S., Lebowitz, J. L., Tumulka, R. \& Zanghi, N. Long-time behavior of macroscopic quantum
systems, \textit{Eur. Phys. J. H} \textbf{35}, 173–200 (2010).
\\
\noindent
[33] D'Alessio, L., Kafri, Y., Polkovnikov, A. \& Rigol, M. From quantum chaos and eigenstate thermalization to
statistical mechanics and thermodynamics, \textit{Adv. Phys.} \textbf{65}, 239 (2016).
\\
\noindent
[34] Gogolin, C. \& Eisert, J. Equilibration, thermalisation, and the emergence of statistical
mechanics in closed quantum systems, \textit{Rep. Prog. Phys.} \textbf{79}, 056001 (2016).
\\
\noindent
[35] Borgonovi, F., Izrailev, F. M., Santos, L. F. \& Zelevinsky, V. G. Quantum chaos and thermalization in isolated systems of interacting particles, \textit{Phys. Rep.} \textbf{626}, 1 (2016).
\\
\noindent
[36] Mikeska, H.-J. \& Kolezhuk, A. K. in {\it Quantum magnetism} edited by Schollw{\"o}ck, U., Richter, J., Farnell, D. J. J., \& Bishop, R. F. (Springer, Berlin/Heidelberg, 2004).
\\
\noindent
[37] Franchini, F. {\it An introduction to integrable techniques for one-dimensional quantum systems} Lecture notes in physics Vol. \textbf{940} (Springer, Berlin/Heidelberg, 2017).
\\
\noindent
[38] Zangara, P. R. et al. Time fluctuations in isolated quantum systems of interacting particles, \textit{Phys. Rev. E} \textbf{88}, 032913 (2013).
\\
\noindent
[39] L\"{a}uchli, A. M. \& Kollath, C. Spreading of correlations and entanglement after a quench in the one-dimensional Bose–Hubbard model, \textit{J. Stat. Mech.} P05018 (2008).
\\
\noindent
[40] Giulio, G. D. \& Tonni, E.
Subsystem complexity after a global quantum quench, \textit{JHEP} \textbf{05}, 022 (2021).
\\
\noindent
[41] See for e.g. Sandvik, A. W. Computational studies of quantum spin systems, \textit{AIP Conf. Proc.} \textbf{1297}, 135 (2010).
\\
\noindent
[42] Popescu, S., Short, A. J. \& Winters, A. Entanglement and the foundations of statistical mechanics, \textit{Nat. Phys. }\textbf{2}, 754 (2006).
\\
\noindent
[43] Tian, C. \& Yang, K. Breakdown of quantum-classical correspondence and dynamical generation of entanglement, \textit{Phys. Rev. B} \textbf{104}, 174302 (2021).

\end{document}